\documentclass[twocolumn]{aastex62}

\usepackage{mathptmx}
\usepackage[T1]{fontenc}
\usepackage{ae,aecompl}
\usepackage{graphicx}	% Including figure files
\usepackage{amsmath}	% Advanced maths commands
\usepackage{amssymb}	% Extra maths symbols
\usepackage{subfigmat}
\usepackage{enumerate}
\usepackage{multirow}
\usepackage{natbib}

\usepackage{chngcntr}
\usepackage{ulem}
\usepackage{appendix}
\usepackage{CJK}
\usepackage{threeparttable}

\newcommand{\kms}{\hbox{km s$^{-1}$}}

\shorttitle{Wind absorption in \ion{C}{2}}
\shortauthors{Xu et al.}

\begin{document}
\begin{CJK*}{UTF8}{gbsn}

% Title.
\title{Probing Protoplanetary Disk Winds with \ion{C}{2} Absorption}

% Authors.
\correspondingauthor{Ziyan Xu}
\author[0000-0002-2986-8466]{Ziyan Xu (徐紫嫣)}
\affiliation{Kavli Institute for Astronomy and Astrophysics, Peking University, Yiheyuan 5, Haidian Qu, 100871 Beijing, China}
\affiliation{Department of Astronomy, Peking University, Yiheyuan 5, Haidian Qu, 100871 Beijing, China}

\author[0000-0002-7154-6065]{Gregory J. Herczeg (沈雷歌)}
\affiliation{Kavli Institute for Astronomy and Astrophysics, Peking University, Yiheyuan 5, Haidian Qu, 100871 Beijing, China}
\affiliation{Department of Astronomy, Peking University, Yiheyuan 5, Haidian Qu, 100871 Beijing, China}

\author[0000-0002-8828-6386]{Christopher M. Johns-Krull}
\affiliation{Department of Physics and Astronomy, Rice University, 6100 Main Street, Houston, TX 77005, USA}

\author[0000-0002-1002-3674]{Kevin France}
\affiliation{Laboratory for Atmospheric and Space Physics, University of Colorado, 600 UCB, Boulder, CO 80309, USA}
\affiliation{Department of Astrophysical and Planetary Sciences, University of Colorado, 389 UCB, Boulder, CO 80309, USA}
\affiliation{Center for Astrophysics and Space Astronomy, University of Colorado, 593 UCB, Boulder, CO 80309, USA}
\email{ziyanx@pku.edu.cn}

% ABSTRACT
%%%%%%%%%%%%%%%%%%%%%%%%%%%%%%%%%%%%%%%%%%%%%%%%%%%%%%%%%%%%%%

\begin{abstract}
We present an analysis of wind absorption in the \ion{C}{2} $\lambda 1335$ doublet towards 40 classical T Tauri stars with archival far-ultraviolet (FUV) spectra obtained by the Hubble Space Telescope.
Absorption features produced by fast or slow winds are commonly detected (36 out of 40 targets) in our sample. The wind velocity {of the fast wind} decreases with disk inclination, consistent with {expectations for} a collimated jet. Slow wind absorption is detected mostly in disks with intermediate or high inclination, without a significant dependence of wind velocity on disk inclination. Both the fast and slow wind absorption are preferentially detected in FUV lines of neutral or singly ionized atoms. The \ion{Mg}{2} $\lambda \lambda 2796,2804$ lines show wind absorption consistent with the absorption in the \ion{C}{2} lines.  We develop simplified semi-analytical disk/wind models to interpret the observational disk wind absorption. Both fast and slow winds are consistent with expectations from a thermal-magnetized disk wind model and are {generally} inconsistent with a purely thermal wind. Both the models and the observational analysis indicate that wind absorption occurs preferentially from the inner disk, offering a wind diagnostic in complement to optical forbidden line emission that traces the wind in larger volumes.
\end{abstract}

%%%%%%%%%%%%%%%%%%%%%%%%%%%%%%%%%%%%%%%%%%%%%%%%%%%%%%%%%%%%%%

%%%%%%%%%%%%%%%%%%
% INTRODUCTION
%%%%%%%%%%%%%%%%%%

\section{INTRODUCTION} \label{sec:intro}
Gas that accretes through a protoplanetary disk and onto the star must lose angular momentum.  The angular momentum transport and accretion are governed by different physics as a function of disk radius.
Near the star, a fast ($\sim 200$ km/s) magneto-hydrodynamic
(MHD) jet launched
within the star-disk interaction region or from the star
itself is thought to carry away much of the angular momentum \citep[e.g.][]{shu94,matt05,cranmer09}.
At larger radii ($\sim 1-10$ AU), simulations suggest that
accretion through the disk is driven by MHD winds of $\sim 30-100$ km/s \citep{ferreira06,bai13}.
Irradiation of the disk by a nearby hot star or by the central
star may lead to photoevaporation, with wind velocities
of $\sim10$ km/s \citep[e.g][]{johnstone98,alexander06,gorti09,owen10}.
These physical processes likely work together to drive disk winds throughout the disk \citep{wang19}.

The observational diagnostics of these outflow components probe the mass loss and therefore the angular momentum transport at different locations in the disk \citep[see review by][]{frank14}.

Disk winds are most commonly identified in optical forbidden line emission, especially [\ion{O}{1}] $\lambda6300$, with a low-velocity component that traces a slow disk wind and a high-velocity component that traces a fast jet \citep[e.g.][]{hartigan95,simon16,banzatti19}.  The mass flux is then measured through analyses of the emission lines, with physical conditions inferred from line ratios \citep[e.g.][]{hartigan95,natta14}.   Mass loss rates from optical forbidden lines have typically been considered uncertain to an order of magnitude, with uncertainties related to geometry, ionization, and abundances.  The recent inclusion of the [\ion{S}{2}] $\lambda 4068$ line helps to break some of the degeneracies in interpreting optical forbidden lines \citep{fang18}.

Absorption lines have the potential to provide a powerful complement to the emission lines by probing the wind in our line-of-sight to the star. Surveys of \ion{He}{1} $\lambda10830$ absorption revealed the presence of both fast and slow winds to accreting young stars \citep[e.g.][]{dupree05,edwards06,cauley14,reiter18}.  However, this \ion{He}{1} line has not been used to measure mass fluxes, in part because of the complexity of {determining the population of the meta-stable lower level.} Optical lines, including the Balmer lines and the \ion{Na}{1} D doublet, sometimes reveal wind absorption as well \citep[e.g.][]{mundt84,alencar00,cauley15}; however, the observed profiles can often be complicated.

Far-ultraviolet spectra of young stars cover many transitions across a wide range of ionization species, some of which frequently have P Cygni profiles indicating wind absorption.   In case studies, the face-on disks RU Lup and TW Hya exhibit strong, fast absorption in many neutral and singly-ionized lines \citep[e.g.][]{herczeg02,herczeg05,johnskrull07}.
In the moderately inclined disk RW Aur, blueshifted wind absorption is detected in low ionization lines with slower maximum velocities \citep{France14}.
Wind absorption has also been inferred in \ion{C}{3} and \ion{C}{4} lines based on a line profile analysis of \citet{dupree05,dupree14}, although the emission profiles may instead be intrinsically asymmetric  \citep{johnskrull07}, as also seen in X-rays \citep{argiroffi17}.

In this paper, we present an observational analysis of wind absorption in the \ion{C}{2} $\lambda 1335$ \AA\ doublet, supplemented by brief comparisons to other absorption lines.
In \S 2, we describe the observations and sample used for this survey. In \S 3, we characterize the \ion{C}{2} lines. In \S 4, we present the measurements of \ion{C}{2} wind properties. In \S 5, we then assess the presence of wind signatures in other lines in the spectra and compare our results to other lines from previous studies. In \S 6, we compare the observational results to the accretion and disk properties and discuss the implications of our results. In \S 7, we present simplified modeling of \ion{C}{2} absorption in disk winds. Our conclusions are presented in \S 8.

% TABLE OBSERVATION LOG
\begin{table*}[t]
    \centering
     \caption{Observation log \label{tab:obslog}}
    \label{tab:obslog}
    \begin{tabular}{lcccccccc}
   Star     &  Obs Date & Instrument & $\lambda$ coverage & Resolution & $t_{exp}^a$ (s)  & near-UV? & Program ID & PI \\
   \hline
    AA Tau & 2013-02-06 & COS & 1135--1790 & 15000 & 5689 & Y & 12876 & France\\
    AK Sco    & 2014-08-05 & COS &   1150--1700 & 45000 & 2917 & Y & 13372 & Gomez de Castro \\
    BP Tau  & 2011-09-09 & COS & 1135--1790 & 15000 & 5134 & N & 12036 & Green\\
    CS Cha   & 2011-06-01 & COS & 1135--1790 & 15000 & 2356 & Y & 11616 & Herczeg\\
    CV Cha   & 2011-05-23 & STIS &   1150--1700 & 45000 & 6671 & Y & 11616 & Herczeg\\
    CW Tau    & 2017-12-16 & COS & 1135--1790 & 15000 & 12546 & N & 15070 & France\\
    CY Tau   & 2000-12-06  & STIS &   1150--1700 & 45000 & 5178 & Y & 8206 & Calvet\\
    DE Tau   & 2010-08-20 & COS & 1135--1790 & 15000 & 2068 & Y & 11616 & Herczeg\\
    DF Tau  &  2010-01-11 & COS & 1135--1790 & 15000 & 4828 & Y & 11533 & Green\\
    DK Tau A & 2010-02-04 & COS & 1135--1790 & 15000 & 1585 & Y & 11616& Herczeg\\
    DM Tau   & 2010-08-22 & COS & 1135--1790 & 15000 & 3459 & Y & 11616& Herczeg\\
    DN Tau   & 2011-09-10 & COS & 1135--1790 & 15000 & 2904 & Y & 11616& Herczeg\\
    DR Tau  & 2010-02-15 & COS & 1135--1790 & 15000 & 1704 & Y & 11616& Herczeg\\
     DS Tau   & 2000-08-24  & STIS &   1150--1700 & 45000 & 5846 & Y & 8206& Calvet\\
   GM Aur   & 2010-08-19 & COS & 1135--1790 & 15000 & 2128 & Y & 11616 & Herczeg\\
    HD 104237 & 2009-12-05 & STIS &   1150--1700 & 45000 & 3279 & Y & 11616 & Herczeg\\
    HD 135344B &2010-03-14 & COS & 1135--1790 & 15000 & 2666 & N &  11828 & Brown\\
    HD 142527 &2013-03-23 & COS & 1135--1790 & 15000 & 1775 & N &13032   & Grady\\
    HD 169142  & 2013-04-05 & COS &   1150--1700 & 45000 & 4493 & N & 13032 & Grady \\
    HN Tau   & 2010-02-10 & COS & 1135--1790 & 15000 & 5725 & Y & 11616& Herczeg\\
     IP Tau   & 2011-03-21 & COS & 1135--1790 & 15000 & 3292 & Y & 11616& Herczeg\\
   KK Oph  & 2013-06-18 & COS & 1135--1790 & 15000 & 1856 & N & 12996 & Johns-Krull\\
    LkCa 15  & 2010-02-05 & COS & 1135--1790 & 15000 & 1712 & N & 11616& Herczeg\\
    PDS 66 &  2011-05-23 & STIS &   1150--1700 & 45000 & 6583 & Y & 11616 & Herczeg\\
    RECX 11  & 2009-12-12 & COS & 1135--1790 & 16000 & 3645 & Y & 11616 & Herczeg\\
    RECX 15 & 2010-02-05 & COS & 1135--1790 & 15000 & 3891 & Y & 11616& Herczeg \\
    RR Tau  & 2012-12-07 & COS & 1135--1790 & 15000 & 1856 & N & 12996& Johns-Krull\\
    RU Lup  & 2000-09-08  & STIS & 1150--1700 & 45000 & 12530 & N & 8157 & Walter \\
     RW Aur   & 2013-08-29 & COS & 1135--1790 & 15000 & 1764 & Y & 12876 & France\\
   RY Lup  &2016-03-16 & COS & 1135--1790 & 15000 & 1481 & N & 14469  & Manara\\
    SU Aur & 2011-03-25 & COS & 1135--1790 & 15000 & 1788 & Y & 11616 & Herczeg\\
    Sz 68  & 2011-03-25 & COS & 1135--1790 & 15000 & 2420 & N &14604 & Manara\\
   T Cha  & 2017-09-02 & COS & 1135--1790 & 15000 & 13051 & N &15128 & Brown \\
    T Ori  &2012-12-06 & COS & 1135--1790 & 15000 & 2420 & N & 12996 & Johns-Krull\\
    T Tau  & 2000-09-08  & STIS & 1150--1700 & 45000 & 11981 & Y & 8157 & Walter\\
    TW Hya &  2000-05-07 & STIS  &   1150--1700 & 30000 & 6583 & Y & 8041 & Linsky \\
   UX Tau & 2012-11-10 & COS & 1135--1790 & 15000 & 1628 & Y & 11616 & Herczeg\\
    V836 Tau  & 2011-02-05 & COS & 1135--1790 & 15000 & 5705 & Y & 11616& Herczeg\\
    V4046 Sgr & 2010-04-27 & COS & 1135--1790 & 15000 & 4504 & N & 11533& Green\\
    VV Ser & 2013-05-15 & COS & 1135--1790 & 15000 & 1856 & N & 12996 & Johns-Krull\\
\hline
    \multicolumn{6}{l}{$^a$Exposure time that covers \ion{C}{2} $\lambda1335$ lines}\\
    \end{tabular}
\end{table*}

% FIGURE ABSORPTION SKETCH
\begin{figure*}[t]
\plotone{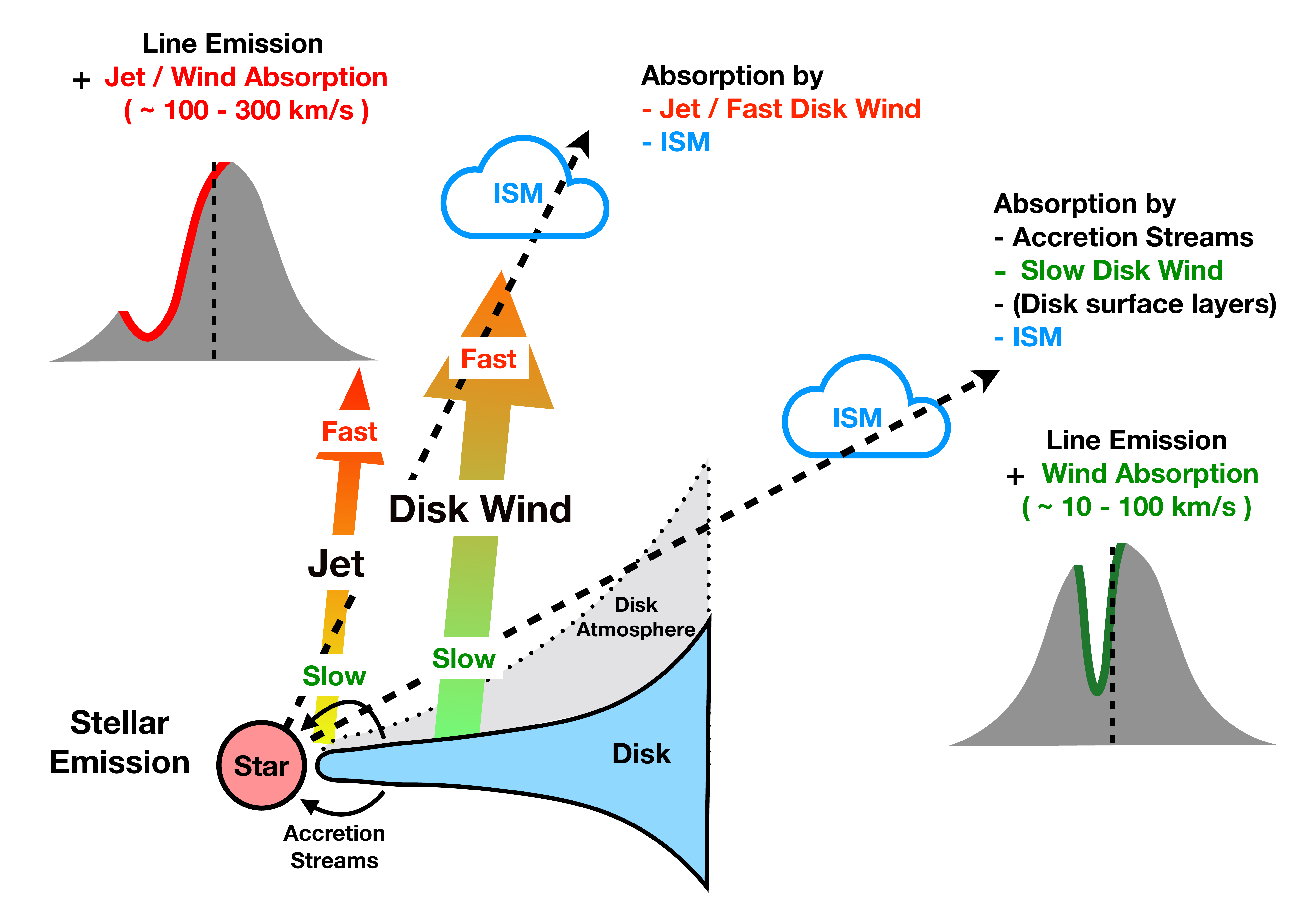}
\caption{\label{fig:absintro} Schematic illustration of the absorption components that may be detected depending on our viewing angle of the star-disk system, adapted from Figure 12 in \citet{mcjunkin14}. The blueshifted absorption components are expected to be from fast jets originated from the star-disk interaction regions, or from disk winds. The colors of green, yellow, and red in the arrows of the jet and the disk wind qualitatively represent the flow velocities from slower to faster. Dashed lines mark the lines of sight.}
\end{figure*}

%%%%%%%%%%%%%%%%%
% OBSERVATIONS
%%%%%%%%%%%%%%%%%
\section{Observations and Targets} \label{sec:method}
The far-ultraviolet spectra of accreting young stars used in this paper (see Table \ref{tab:obslog} for an observation log) were obtained by the Hubble Space Telescope (HST) across several different programs.  The selection of targets is therefore somewhat arbitrary.  From this parent sample, we select spectra of accreting young stars with sufficient $S/N$ to measure any wind absorption, if present.

%%%
\subsection{HST-COS FUV Spectra} \label{subsec:spec}

The COS G130M and G160M spectra\footnote{In some cases, only G130M spectra were obtained, covering $\sim 1100-1400$ \AA} combined cover $1100-1800$ \AA~with a resolving power of $\sim 18000$.  The circular aperture has a diameter of $\sim 2\farcs5$ and provides minimal spatial resolution.
The line spread function has significant power beyond the central core, following the aberrated point-spread function of HST.
 The COS instrument has very low background and much higher sensitivity than STIS, leading to much higher $S/N$ ratios.  The data reduction for the COS data was fully described in \citep{france12}.  The absolute wavelength calibration of COS medium-resolution spectra is typically considered $\sim 10-15$ \kms \citep{oliveira10}.  
All spectra are shifted in wavelength so that H$_2$ lines at 1338 and 1342 \AA\ are located at their theoretical wavelengths \citep{abgrall93}, in order to shift the spectrum to the stellar rest frame.  
The use of H$_2$ lines should improve this accuracy to $\sim 5$ \kms.  In a few cases the H$_2$ line emission is produced in outflows and would have a centroid of 5-10 \kms\ blueward of the stellar rest frame \citep{herczeg06,france12}.

%%%
\subsection{HST-STIS FUV Spectra}

The STIS instrument with the E140M echelle grating and $0\farcs2 \times 0\farcs2$ aperture produces spectra that simultaneously cover 1150--1700 \AA\ at $R\sim 45,000$.  When compared to COS, STIS offers higher spectral resolution at the cost of lower sensitivity and higher background rates.
 Reduced data were retrieved from the MAST archive.  Although the wavelength calibration of STIS E140M spectra is accurate to $\sim 1.5$ \kms \citep{sonnentrucker15}, we also shift these spectra to the H$_2$ rest frame for consistency with the COS spectra.

%%%
\subsection{HST-STIS NUV Spectra}

We supplement the FUV spectra with near-UV spectra for a subsample of 21 sources to analyze the \ion{Mg}{2} $\lambda\lambda2976,2804$ resonance lines.  Most E230M spectra used in this paper were obtained with the $0\farcs2 \times 0\farcs2$ aperture and cover from $\sim 2130-2940$ \AA\ at $R\sim30,000$.  Reduced STIS E230M spectra were retrieved from the MAST archive.

% TABLE WIND SUMMARY
\begin{table*}[t]
    \centering
    \caption{Summary of Wind Absorption Detections}
    \label{tab:windall}
    \begin{tabular}{lccccccccc}
Star & Incl.($^{\circ}$)     & C II Abs. Type & N I & Si II & Si III & C IV & Mg II\\
\hline
DR Tau & 5 & Fast & Y & Y & Y & Y & Y\\
TW Hya & 7  & Fast, ISM & N? & N & N & N & Y\\
HD 135344B & 12 & Fast & N? & N & N? & N? & -\\
DK Tau & 13  & Fast & N? & Y & N? & N & Y\\
HD 169142 & 13  & Fast, ISM & - & N & N & - & -\\
HD 104237 & 18$^*$ & Fast, ISM & Y & Y & Y & Y? & Y\\
RU Lup & 18  & Fast, ISM & Y & Y & Y & N? & -\\
HD 142527 & 27$^*$ &Fast & - & Y & Y & - & -\\
CY Tau & 27 & - & N & N & N & N? & -\\
T Tau & 28  & Fast & N & Y & N & N & -\\
PDS 66 & 30  & Fast & N & N & N? & N & Y\\
V4046 Sgr & 33  & Fast & N? & N & N? & N & -\\
CV Cha & 35  & Fast, ISM & N & N & N & N & Y\\
DE Tau & 35  & Fast & N & Y & N & N & Y\\
DN Tau & 35  & Fast, ISM & N? & Y? & N? & N & Y\\
UX Tau & 35 & Slow & N & N & N & N & N\\
DM Tau & 35 &  Slow & N & N & N & N & Y\\
BP Tau & 38  & Fast, ISM & N & N & N & N & -\\
V836 Tau & 43 &  Slow  & N & N & N & N & Y\\
IP Tau    & 45 & Slow  & N & N? & N & N & Y\\
SZ 68 & 48 & - & - & - & N & - & -\\
LkCa 15 & 49 &  Slow  & N? & Y? & Y? & N & -\\
GM Aur & 55 &  Fast,Slow  & N? & N? & N & N & Y\\
RW Aur & 55 & - & Y & Y & Y & N? & Y\\
RECX 15 & 60$^{*}$ &  Slow  & N & N & N? & N & Y\\
CS Cha & 60$^{*}$ & - & N & N & N & N & Y\\
SU Aur & 62$^{*}$ &  Slow  & N & Y & N & N & Y\\
CW Tau & 65$^{*}$ & Slow  & - & - & N? & - & -\\
DS Tau & 65  & Fast & N? & N & N & N & -\\
RY Lup & 68 &  Slow  & N & N? & N & N & -\\
RECX 11 & 70$^*$  & Fast & N & N & N & N & Y\\
HN Tau & 70 &  Slow & N & N & N? & N & Y\\
KK Oph & 70$^{*}$ &  Slow  & N & Y & N? & N & -\\
AK Sco & 71 &  Slow  & Y & Y & Y & N & Y\\
AA Tau & 75 &  Slow  & N & N & N & N & Y\\
DF Tau & -  &Fast, ISM & Y & Y? & N & N & Y\\
RR Tau & - & Slow  & N & Y & N & N? & -\\
T Cha & - &  Slow  & N & N & N & N & -\\
T Ori & - &  Slow  & N & N & Y & N & -\\
VV Ser & - &  {Fast}  & N? & N? & N & N? & -\\

\hline
    \end{tabular}
    \begin{tablenotes}
    \item The question marks show ambiguous identifications for wind absorption.
    \item $^{*}$: all inclinations from ALMA or SMA except where marked.

    \end{tablenotes}
\end{table*}

% TABLE FAST WINDS
\begin{table}[t]
    \centering
    \caption{Properties of Fast Winds}
    \label{tab:fastwind}
    \begin{tabular}{lcccccccc}
& Incl.     & \multicolumn{2}{c}{Wind} & Detection\\
Star & (deg) & $v_{min}$ & $v_{max}$ & Methods \\
\hline
DR Tau & 5 & -20 & -355 & sub-cont., H$_2$ \\
TW Hya & 7  & -100 & -255  & \ion{C}{2}, sub-cont., H$_2$ \\
HD 135344b & 12 & -- & -250  & \ion{C}{2}, H$_2$\\
DK Tau & 13  & -- & -230  & \ion{C}{2}\\
HD 169142 & 13  & -- & -165  & visual \\
HD 104237 & 18$^*$ & -80 & -565 & \ion{C}{2}, sub-cont., H$_2$\\
RU Lup & 18  & -140 & -300  & \ion{C}{2}, sub-cont., H$_2$\\
HD 142527 & 27$^*$ & -- & -250 &\ion{C}{2}, H$_2$\\
T Tau & 28  & -150 & -195  & \ion{C}{2}, sub-cont., H$_2$\\
PDS 66 & 30  & -- & -210  & \ion{C}{2}\\
V4046 Sgr & 33  & -- &-125  & \ion{C}{2} \\
CV Cha & 35  & -- & -280 &  \ion{C}{2} \\
DE Tau & 35  & -- &-170  &  \ion{C}{2} \\
DN Tau & 35  & -- &-265  & \ion{C}{2}  \\
BP Tau & 38  & -- &-220  & \ion{C}{2}, H$_2$ \\
GM Aur & 55  &-- & -190  & \ion{C}{2}\\
DS Tau & 65  & -10 & -100  & sub-cont. \\
RECX 11 & 70$^*$  & -- & -130  & visual \\
DF Tau & -  &-- &  -130 & visual  \\
VV Ser & -  &-- &  -300 & visual  \\
\hline
    \end{tabular}
    \begin{tablenotes}

    \item $^{*}$: all inclinations from ALMA or SMA except where marked.
        \item All of the velocities are in \kms in the stellar rest frame. ``sub-cont'': sub-continuum absorption. ``H$_2$'': the absorption of ``H$_2$'' lines. ``\ion{C}{2}'': the absorption of the blue member of the \ion{C}{2} doublet.  
    \item For fast absorption features with multiple measurements (detected in both of the doublet members or by more than one method), we adopt the maximum velocity from these measurements.
    For most of the cases, the velocity of absorption in the red component is comparable or slightly higher than that in the blue component.
    \end{tablenotes}
\end{table}

% FIGURE ABSORPTION PROFILE EXAMPLES
\begin{figure*}[t]
\plotone{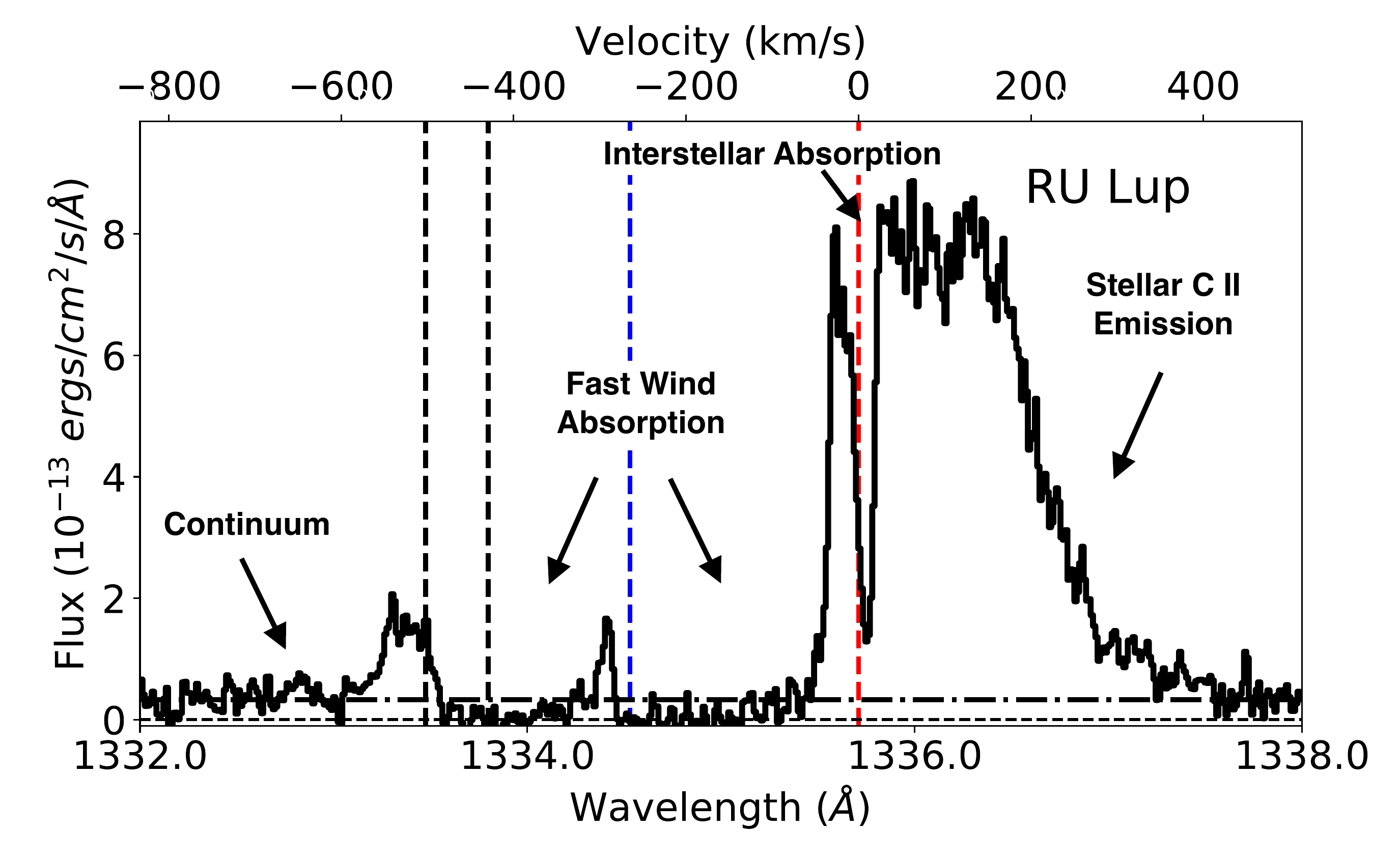}
\plotone{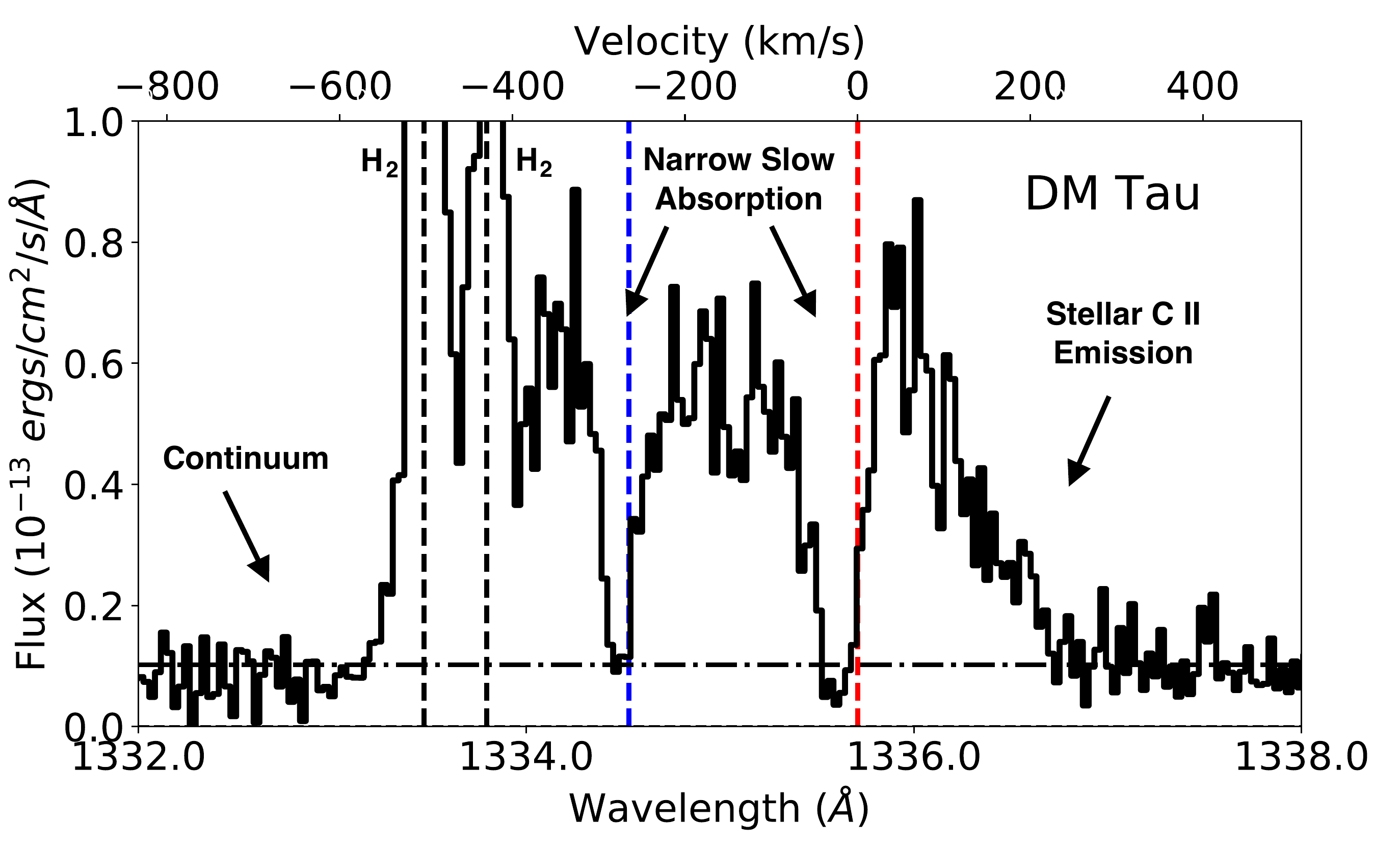}
\caption{\label{fig:lineoverview} Prototypes of the \ion{C}{2} $\lambda$1335 doublet show absorptions against the stellar line emission. The velocities are relative to \ion{C}{2} $\lambda$1335.71. Blue and red dashed lines mark the wavelengths of the doublet members, black vertical dashed lines mark the wavelengths of the H$_2$ $\lambda$1333.5 and $\lambda$1333.8 emission lines. The dashed horizontal line marks the zero-flux level, the dash-dotted horizontal line marks the flux level of continuum. Upper panel: in RU Lup, the stellar emission suffers from broad sub-continuum absorptions by the fast wind, which absorb the blue member of \ion{C}{2} doublet and the H$_2$ lines. An ISM-like feature is also detected in \ion{C}{2} $\lambda$1335.71 at about 0 \kms. Bottom panel: in DM Tau, narrow slow absorptions are detected in the \ion{C}{2} $\lambda$1335 doublet against stellar emission.}
\end{figure*}

%%%
\subsection{Target description}

Properties of the targets analyzed in this paper are described in     Table~\ref{tab:targprops}.  Spectral types or photospheric temperatures, extinctions, photospheric luminosities, and accretion luminosities are obtained from the same source, when possible.  For cool stars, the spectral type is converted to temperature using the relationship calculated by \citet{herczeg14}. Luminosities are corrected for updated distances from Gaia DR2 \citep{gaia18}.  The mass for each low-mass object is then calculated using pre-main sequence evolutionary tracks of \citet{feiden16} with no magnetic field and abundances of \citet{asplund09}.  Use of evolutionary tracks with magnetic fields or stellar spots for the low-mass stars would lead masses of low-mass stars that are larger by $\sim 50$\% \citep{feiden16,somers20}.

%%%%%%%%%%%%%%%%%%%%%%%%%%%%%
% OVERVIEW OF C II DOUBLET
%%%%%%%%%%%%%%%%%%%%%%%%%%%%%

\section{An Overview of the \ion{C}{2} $\lambda1335$ doublet} \label{sec:c2overview}

The \ion{C}{2} resonance lines at $1335$ \AA\ are typically referred to as a doublet ($\lambda1334.53$ and $\lambda1335.71$), although a weaker third line at $\lambda1335.66$ is also present. For classical T Tauri stars, the \ion{C}{2} lines are typically seen as strong emission on top of a weak continuum.  The luminosity in the \ion{C}{2} doublet correlates strongly with lines that are produced in the
accretion funnel flow and accretion shock \citep{gomezdecastro12,yang12,ardila13}, although the origin has not been confirmed with a detailed line profile analysis and modeling.  The continuum around the \ion{C}{2} lines has an uncertain origin and may result from a hot accretion component \citep{france17}.

The \ion{C}{2} emission from the star can be absorbed by different gas components along our line of sight.  Absorption studies of \ion{C}{2} have historical importance as a probe of the  local interstellar medium \citep[e.g.][]{redfield04}.  In the context of an accreting young star, the disk and different components of the wind may be present in our line of sight to the star depending on the inclination, as illustrated in Figure \ref{fig:absintro}.  Based on the framework developed to interpret \ion{He}{1} $\lambda10830$ lines \citep{edwards06}, absorption in a disk wind is expected for intermediate inclination targets. Absorption from higher inclination targets is expected to be relatively narrow and at lower velocity. For face-on disks, the emission features are absorbed mainly by the collimated fast jet/wind from star-disk interaction regions, such as X-winds \citep[e.g.][]{shu94}, or outflows driven by magnetic inflation \citep[e.g.][]{Zanni13}. The line of sight passes through jet/wind material from the flow base to the higher locations where the jet/wind accelerates.  The resulting absorption is thus broader and more blueshifted.
The wind may also emit in \ion{C}{2} lines, as blueshifted emission is detected in our sample (e.g. AA Tau in Figure \ref{fig:narrow}); scattered photons are also expected. However, the wind emission is somewhat ambiguous to identify as it blends with the \ion{C}{2} emission from the star. In this paper, we focus on \ion{C}{2} absorption in the wind.

Consistent with the expectations from \ion{He}{1}, three types of \ion{C}{2} absorption are detected in our sample: absorption in a fast wind, a broader absorption at low velocities, and a near-universal, very narrow absorption likely from the interstellar medium. Table \ref{tab:windall} summarizes the types of \ion{C}{2} absorption features detected in our sample, along with detections of wind absorption in other FUV or NUV lines. Tables \ref{tab:fastwind}, \ref{tab:narrow}, and \ref{tab:ISM-like} summarize the targets and their detected fast wind, slow wind, and ISM-like absorption features. Wind absorptions are robustly detected in the \ion{C}{2} $\lambda1335$ doublet for 36 out of 40 targets in our sample. The absence of detectable absorption does not necessarily rule out the presence of wind absorption, because the line profiles can be complicated, the line-spread function smooths out subtle features, and the signal-to-noise is often low.

Redshifted absorption has been detected in lines at other wavelengths, including H $\beta$, the Na I $\lambda$ 5895 doublet, and \ion{He}{1} $\lambda$ 10830 \citep[e.g.][]{edwards94,fischer08,alencar12}.  This downflowing gas can be generally explained by models of magnetospheric accretion flows \citep[e.g.][]{muzerolle01,kurosawa11}, with the presence or absence depending on the geometry and radiative transfer.    In \ion{C}{2} lines in our sample, redshifted absorption is not clearly detected in any spectrum.
The lack of redshifted \ion{C}{2} absorption may indicate that {singly ionized carbon is not present} in magnetospheric accretion column; however, {the lack of clear detections are challenging to interpret, and} detailed discussion of the contribution of accretion to the \ion{C}{2} lines profiles is beyond the scope of this paper.

\paragraph{Fast wind absorption:}  Broad and deep \ion{C}{2} absorption is detected in {20} of 40 targets in our sample.  The upper panel of Figure \ref{fig:lineoverview} shows the fast wind absorption of the prototype RU Lup (with disk inclination of 18$^\circ$).
The fast wind absorption features extend from low velocities to hundreds of \kms\ with sub-continuum absorption and near-zero flux. The high velocity wind is seen in absorption against the FUV continuum and  emission lines that coincide in wavelength with the velocity of absorption, including \ion{C}{2} $\lambda1334.5$ line by the $\lambda1335.7$ line and two H$_2$ $\lambda1333$ lines by the \ion{C}{2} $\lambda1334.5$ line \citep[see, e.g.,][]{herczeg06,johnskrull07}.  Details of these diagnostics will be described in \S \ref{sec:windmeasure}.

\paragraph{Narrow slow absorption:} {17} out of 40 targets in our sample have slow absorption features detected against the emission lines. {The line profiles can be well described by a Gaussian profile with line width of tens of $\kms$ to over a hundred $\kms$
(\S \ref{subsec:measurenarrow}), which is overall narrower than the fast wind absorption features detected in our sample.} The lower panel of Figure \ref{fig:lineoverview} shows a prototype of this absorption, DM Tau (with disk inclination of 35$^\circ$). These absorption lines are typically slightly blueshifted, with a central velocity of $\sim$ 10 $\kms$. 

\paragraph{ISM-like absorption:} Unresolved ISM-like absorption features are detected in 8 of our 40 targets. These profiles are initially identified as distinct from the narrow-slow absorption by eye.  This difference is then confirmed following equivalent width measurements (see \S \ref{sec:analysis}).
These ISM-like features are found mostly in targets with fast wind absorption, such as RU Lup (Figure \ref{fig:lineoverview}), since that component is well separated from the ISM component.  The \ion{C}{2} absorption from the ISM must be present in all spectra, but this absorption often blends with narrow-slow absorption and can be hard to clearly separate.

%%%%%%%%%%%%%%%%%%%%%%%%%%%%%
% MEASURING WIND PROPERTIES
%%%%%%%%%%%%%%%%%%%%%%%%%%%%%

\section{Measuring the properties of wind absorption} \label{sec:windmeasure}

\begin{figure*}[t]
\includegraphics[width=\textwidth]{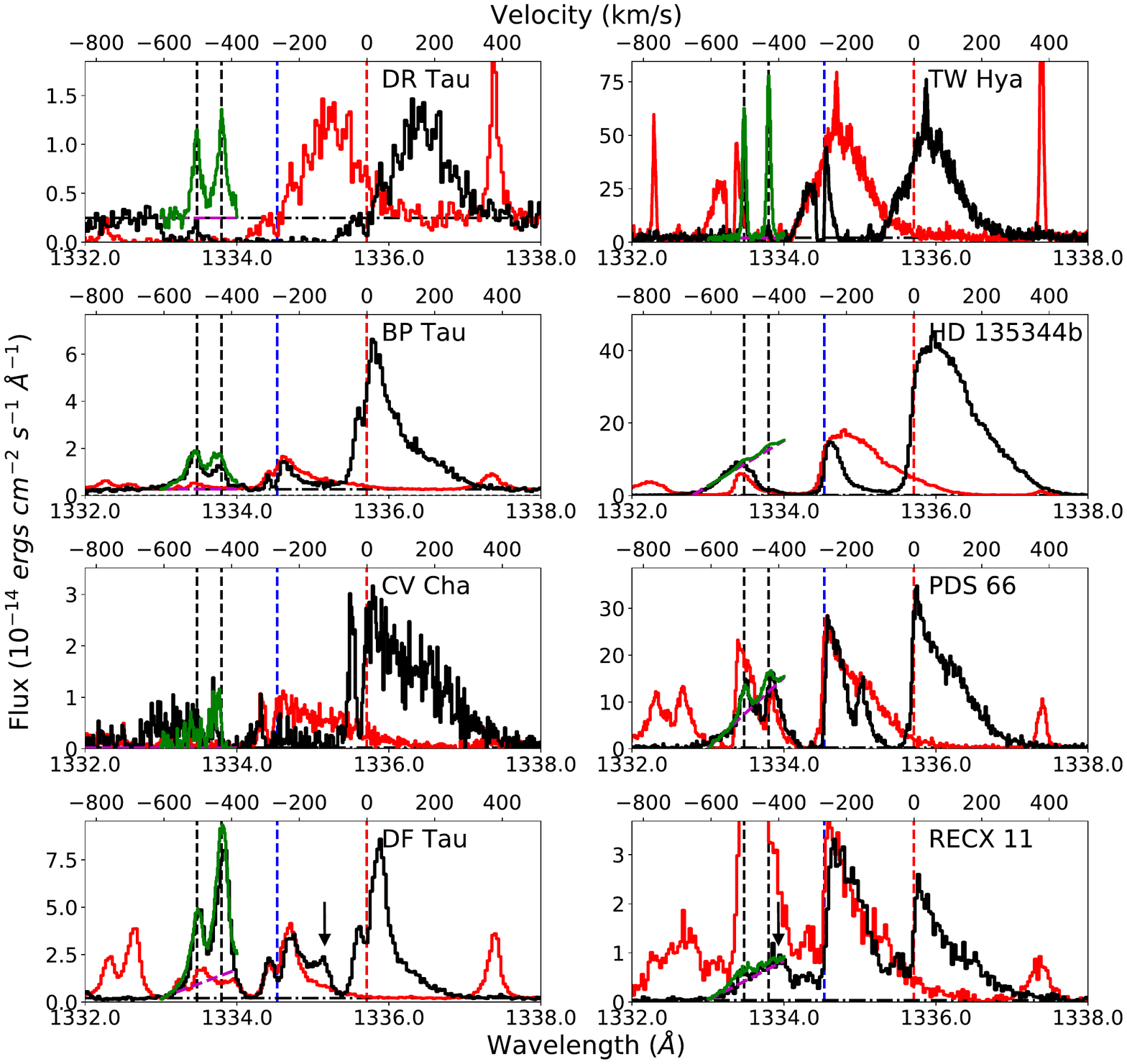}
\caption{\label{fig:fastwindspec} Representative examples of our diagnostics of fast wind absorptions. The wavelengths correspond to the black line, and the velocities are relative to \ion{C}{2} $\lambda$1335.71. Vertical dashed lines mark wavelengths of the \ion{C}{2} doublet members (blue and red) and the H$_2$ emission lines (black). The dash-dotted horizontal line marks the flux level of continuum.  The red line overplots the red member by shifting it to $\lambda$1334.53 and rescaling it to match the peak of the blue member. The green line shows the expected profile of the H$_2$ lines on the top of an expected flux (magenta slope) of the continuum or the blue side of \ion{C}{2} $\lambda$1334.53 emission. The black arrows in DF Tau and RECX 11 mark the edges of wind absorptions that are visually identified. }
\end{figure*}

In this section, we measure properties of each absorption component.    For the fast wind, the underlying line profile is uncertain and may be asymmetric, and the quantified properties of absorption features, particularly the wind velocities, are measured through three robust diagnostics (see Figure~\ref{fig:fastwindspec}): sub-continuum absorption, absorption of H$_2$ lines at shorter wavelengths, and absorption of the red wing of the blue member of the \ion{C}{2} 
doublet at 1334.53 \AA. In addition, some fast wind absorption features can be visually identified. For the narrow slow and ISM-like absorption features {(e.g. Figure~\ref{fig:narrow})}, the central velocity, FWHM, and equivalent width are easy to measure because they are less disruptive to the broader emission line profile. {The \ion{C}{2} $\lambda1335$ line profiles for all of the targets in our sample are presented in Figure~\ref{fig:spec_all}.}

% FIGURE FAST WINDS

%%%
\subsection{Fast absorption features} \label{subsec:measurefast}

Fast absorption features in \ion{C}{2} are detected in {20} targets in our sample, identified through several complementary techniques \citep[following][]{johnskrull07}. The absorption against the continuum is the most straightforward measurement, when the continuum is detected with sufficient signal. For targets with no clear sub-continuum absorption (e.g., BP Tau), the fast wind can be identified from absorption of H$_2$ lines that are located blueward of \ion{C}{2} $\lambda1334.5$, or by the absorption of the blue member of the doublet. In a few cases, subtle wind absorption features are identified by eye, guided by similar absorption components in the \ion{Mg}{2} lines.

In this subsection, we step through each of these methods to describe the fast absorption components.

%%%
\subsubsection{Sub-continuum absorption}

Continuum emission is detected to many of the targets in our sample. For six targets, the continuum is weaker on the blue side of the \ion{C}{2} lines, indicating sub-continuum absorption. An example of the sub-continuum absorption in RU Lup is shown in Figure~\ref{fig:lineoverview}. Two additional examples, DR Tau and TW Hya, are shown in Figure~\ref{fig:fastwindspec}.
The sub-continuum absorption in the red member of DR Tau starts at about $-20$ \kms\ and extends to at least $-265$ \kms, where the blue member is located. In the blue member of the doublet, the sub-continuum absorption extends to $\sim$ $-355$ \kms. For TW Hya, the sub-continuum absorption features in the doublet have similar properties, spanning from $-100$  to $-220$ \kms\ in the red member and from $-100$ to $-225$ \kms\ in the blue member.

Table \ref{tab:fastwind} lists the properties of these fast absorption lines, including the maximum and minimum velocities where the absorption goes below the local continuum ($v_{max}$ and $v_{min}$). The sub-continuum absorption for these 6 targets are all broad ($\Delta v > 100~\kms$ for most of the cases) and blueshifted (maximum velocity $v_{max}$ faster than $-100$~\kms) indicating the presence of a fast wind.

% FIGURE SLOW WINDS 

\begin{figure*}[t]
\plottwo{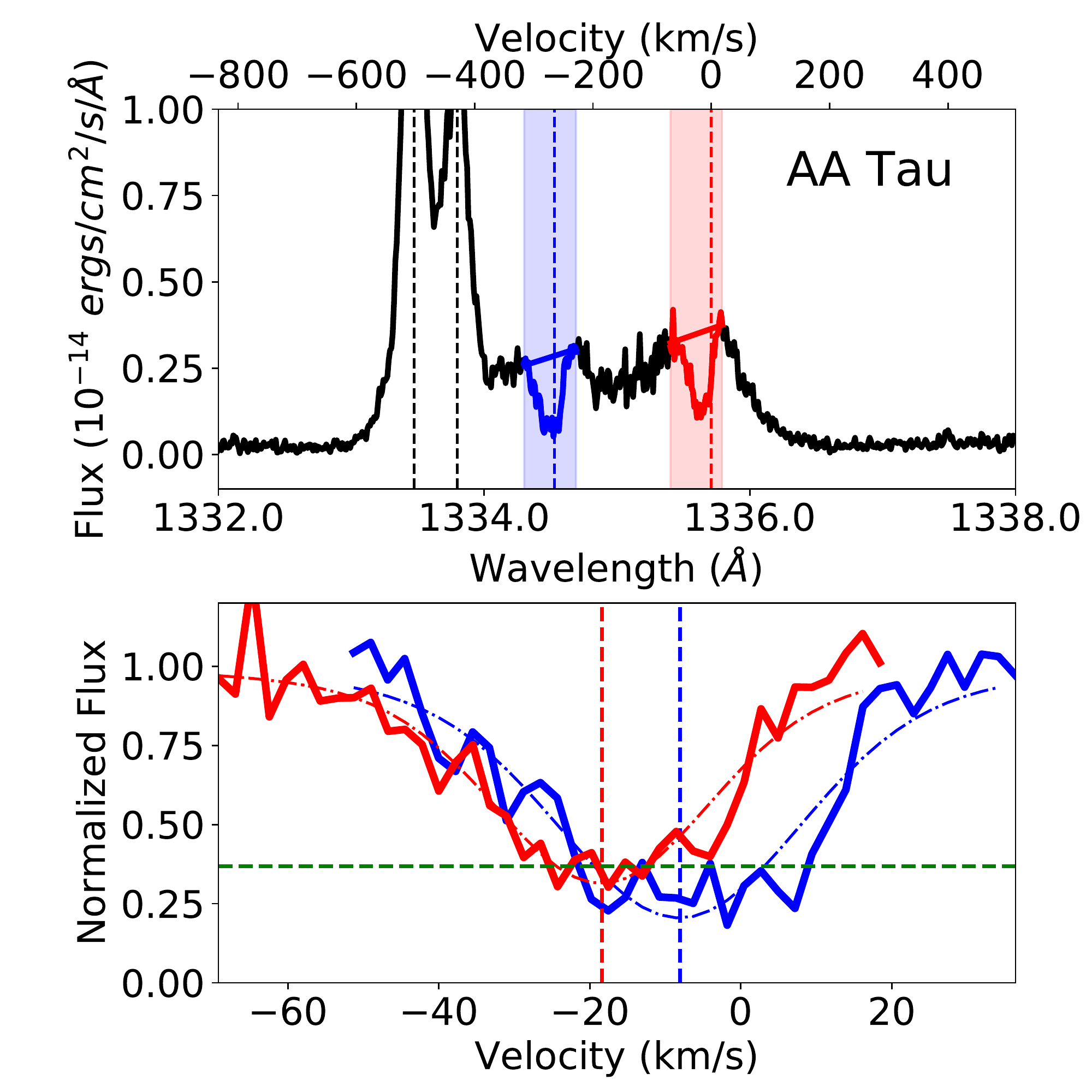}{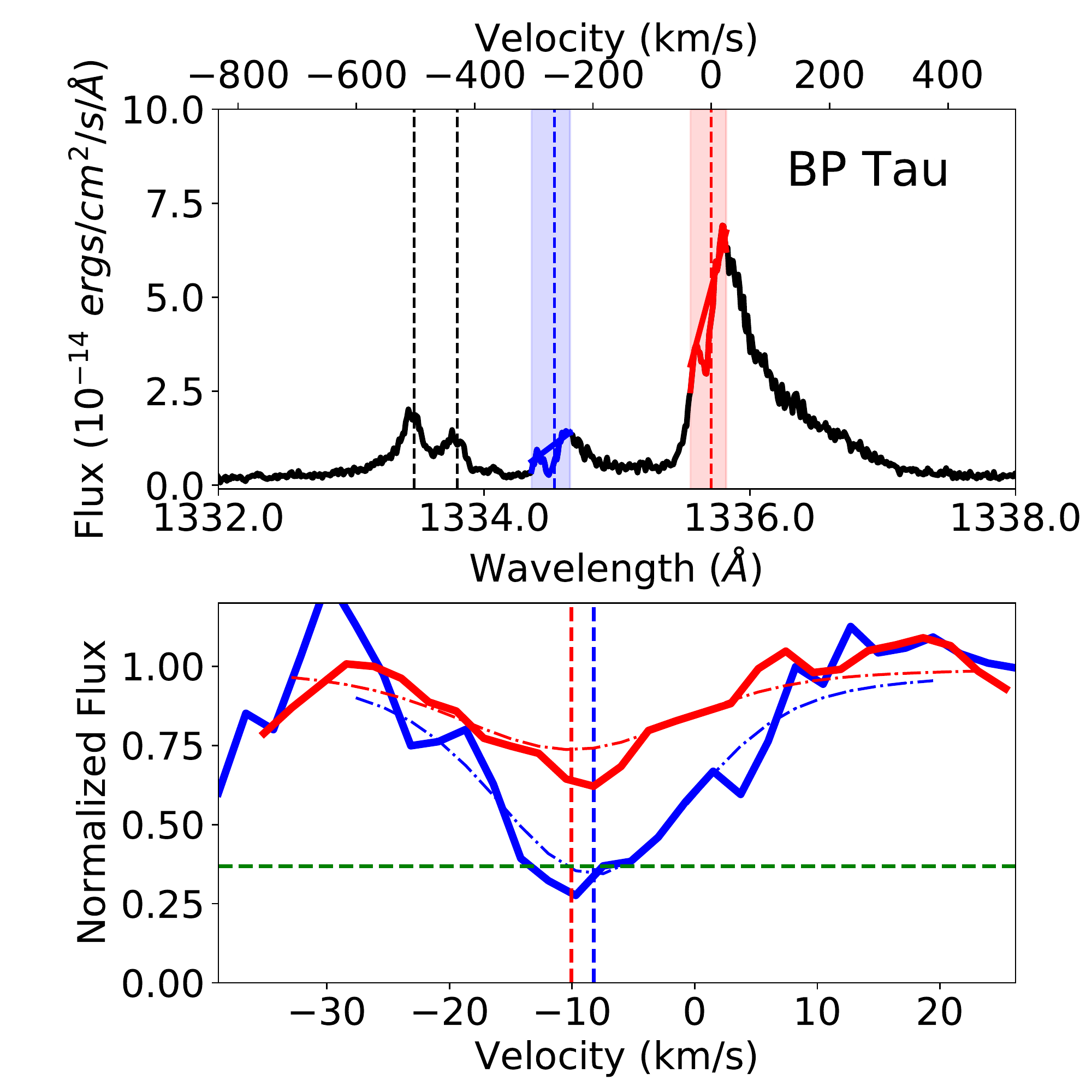}
\caption{\label{fig:narrow} Representative examples of \ion{C}{2} line profiles with narrow slow absorption (AA Tau, left) or ISM-like absorption (BP Tau, right). Upper panels: the velocities are relative to \ion{C}{2} $\lambda$1335.71. Blue and red dashed lines mark the wavelengths of the doublet members, black dashed lines mark the wavelengths of the H$_2$ emission lines. Blue and red shaded regions show the wavelength ranges of the identified absorption features, solid slopes in each region show the expected emission flux prior to absorption. Bottom panels: solid lines show the line profiles of \ion{C}{2} $\lambda$1334.53 (blue) and $\lambda$1334.53 (red) absorptions, normalized by the expected emission flux.  Dash-dotted lines show our best fit to the absorption profiles using a Gaussian profile convolved with the local line spread function. Dashed vertical lines mark the fitted line centers. Green dashed lines mark the flux corresponding to optical depths $\tau=1$.}
\end{figure*}

%%%
\subsubsection{Absorption of H$_2$ lines}

Two H$_2$ emission lines, B-X 0-4 R(1)~$\lambda1333.5$ and B-X~0-4 R(2)~$\lambda1333.8$,
are located at $-230$ and $-165$ \kms\ relative to the \ion{C}{2}~$\lambda1334.5$ line.
These H$_2$ lines originate from close to the star \citep{herczeg06}, behind the bulk of the wind along the line of sight.
The expected fluxes for these two H$_2$ lines are calculated from branching ratios\footnote{The calculated branching ratios have been confirmed to be accurate through FUV spectroscopy, including from Mira B by \citet{wood02}.} \citep{abgrall93} and
the fluxes of their P-branch counterparts from the same upper level, located at 1338.6 and 1342.3 \AA.
If the wind absorption extends beyond $-230$ \kms, then the flux in both of these H$_2$ lines are weaker than expected (for example, see DR Tau).  For other targets (e.g. TW Hya), the flux in the H$_2$ $\lambda1333.5$ line is close to expectation, indicating that the \ion{C}{2} wind at $-230$ \kms\ is optically thin, but the flux in  H$_2$ $\lambda 1333.8$ line is weaker than expected, indicating a high optical depth in the \ion{C}{2} line at $-165$ \kms.

Properties of these lines are presented in Table \ref{tab:fastwind}, including minimum and maximum velocities ($v_{min}$ and $v_{max}$) from the \ion{C}{2} $\lambda1334.53$ line where the wind is optically thick.  In eight of the 40 targets, at least one of the H$_2$ lines is weaker than expected, attenuated by \ion{C}{2} in the wind.

% FIGURE ALL PROFILES

\begin{figure*}[t]
\includegraphics[width=0.33\textwidth]{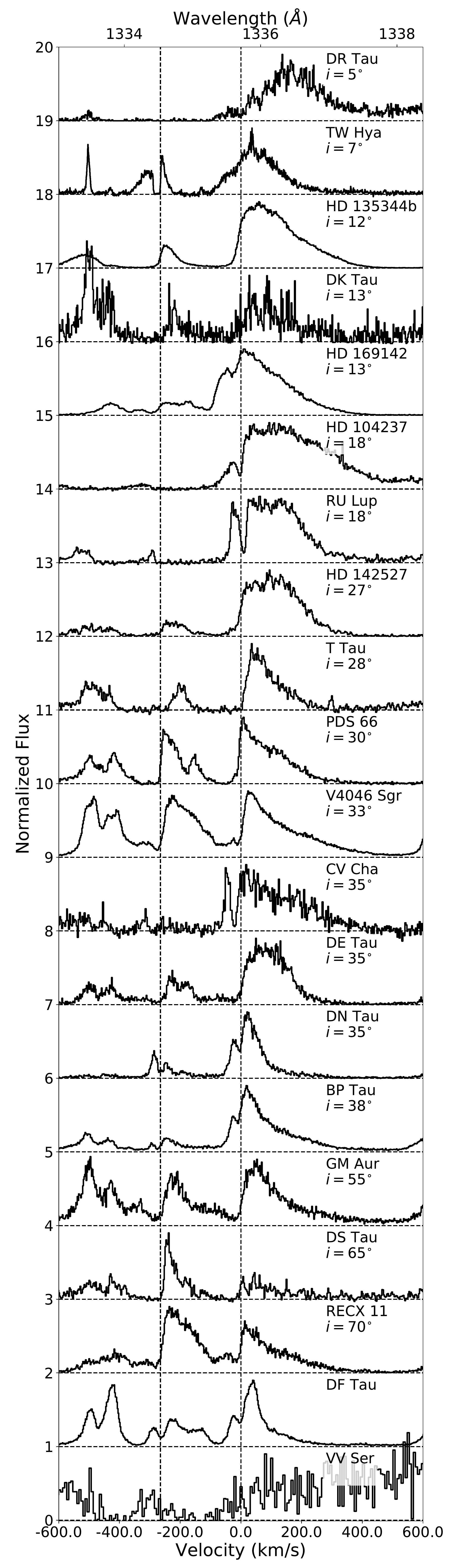}
\includegraphics[width=0.33\textwidth]{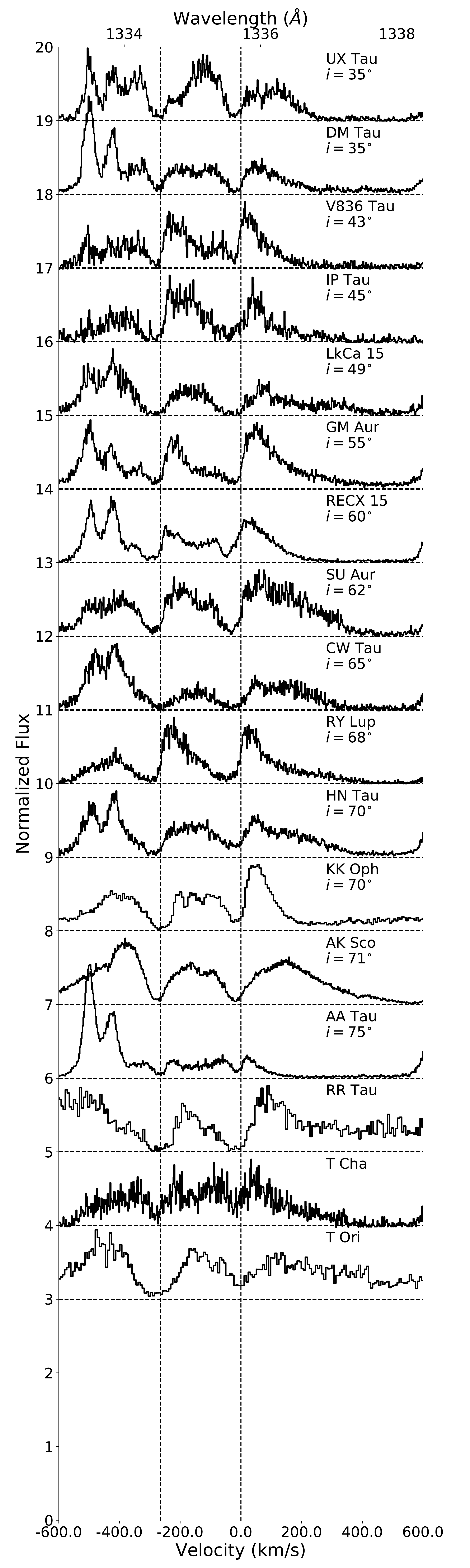}
\includegraphics[width=0.33\textwidth]{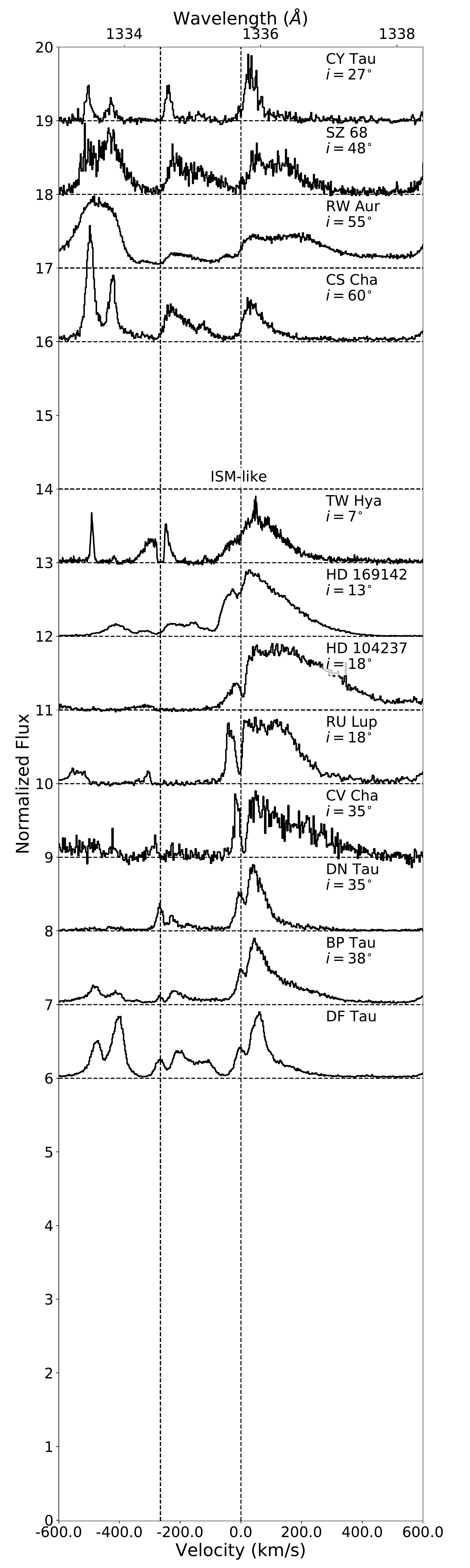}
\caption{\label{fig:spec_all} {\ion{C}{2} $\lambda$1335 line profiles for all of the targets in our sample. Dashed horizontal lines mark the zero-flux level for each profile. Dashed vertical lines mark the wavelengths of the doublet members. Left: all the targets with fast wind absorptions. Middle: all the targets with slow wind absorptions. Right: the upper part shows the targets without clear wind detection in \ion{C}{2} lines; the lower part shows targets with ISM-like absorptions, but the line profiles are presented without shifting to the stellar rest frame using H$_2$ lines (see \S \ref{subsec:spec}). Targets in each part are presented in ascending order of disk inclinations (labeled for each target, if applicable). }}
\end{figure*}

% TABLE SLOW WINDS

\begin{table}[t]
    \centering
    \caption{Properties of Slow Winds}
    \label{tab:narrow}
    \begin{tabular}{lccccccccccc}
Star & Incl.($^{\circ}$) &    & $v_{\rm max}$ & $v_{\rm cent}$ & EW & FWHM \\
\hline
         UX Tau & 35 & & -100 & -15 & 120 & 140 \\
         DM Tau & 35 & & -80 & -25 & 60 & 60 \\
         V836 Tau & 43 & & -80 & -30 & 50 & 50 \\
        IP Tau    & 45 & & -150 & -50 & 105 & 110 \\
        LkCa 15 & 49 & & -130 & -35 & 135 & 135 \\
        GM Aur & 55 & & -100 & -25 & 55 & 55 \\
        RECX 15 & 60$^{*}$ & & -100 & -45 & 55 & 55 \\
        SU Aur & 62$^{*}$ & & -130 & -35 & 95 & 95 \\
        CW Tau & 65$^{*}$ & & -150 & -55 & 115 & 115 \\
        RY Lup & 68 & & -150 & -60 & 110 & 110 \\
        HN Tau & 70 & & -120 & -30 & 80 & 95 \\
        KK Oph & 70$^{*}$ & & -150 & -25 & 90 & 65 \\
        AK Sco & 71 & & -100 & -15 & 95 & 90 \\
        AA Tau & 75 & & -50 & -15 & 35 & 35 \\
        RR Tau & - & & -150 & -25 & 160 & 170 \\
        T Cha & - & & -100 & -15 & 65 & 90 \\
        T Ori & - & & -100 & -10 & 110 & 110 \\
\hline    \end{tabular}
\begin{tablenotes}
\item The values of $v_{\rm max}$, $v_{\rm cent}$, EW and FWHM are in \kms, $v_{\rm max}$ and $v_{\rm cent}$ are in the stellar rest frame.
\item $^{*}$: less reliable inclinations that are not from ALMA or SMA.
\item {For slow absorption features detected in  both of the doublet members, we adopt the maximum velocity from the members, and average the EW and FWHM for the doublet members. For most of the cases, the velocity of absorption in the red component is comparable or slightly higher than that in the blue component.}
\end{tablenotes}
\end{table}

%%%
\subsubsection{Absorption of the short wavelength member of the \ion{C}{2} doublet} \label{subsec:redwing}

Emission in the \ion{C}{2} $\lambda1334.5$ line can be absorbed by wind in the \ion{C}{2} $\lambda1335.7$ line.  While the wavelength difference corresponds to $-265$ \kms, emission on the red wing typically extends to $\sim 100$ \kms.  The red wings of the two lines should be similar, with differences likely attributable to the absorption in the blue member. {The maximum wind velocity is determined where the absorption in the red wing of the blue member shows the greatest projected velocity from center of the red member.} In our sample, 13 targets have fast winds identified through the absorption of the red wing of the \ion{C}{2} $\lambda1334.5$ by \ion{C}{2} $\lambda1335.7$ (see Fig.~\ref{fig:fastwindspec}).

%%%
\subsection{{Slow} and ISM-like absorption features} \label{subsec:measurenarrow}

Slow absorption features are detected from {17} of 40 targets (e.g. AA Tau in Figure \ref{fig:narrow}), while 8 of 40 show ISM-like absorptions (e.g. BP Tau in Figure \ref{fig:narrow}).
{We visually identify the absorption features from the emission lines and determine the wavelength range (shown in blue and red shadows in Figure \ref{fig:narrow}) for analyzing the absorption features. Within this wavelength range, we extract the absorption line profiles by dividing against the expected line emission flux prior to absorption (shown in red and blue slopes in Figure \ref{fig:narrow}).} Although the uncertainty of the emission profile (flux) is limited within the narrow wavelength range of the absorption, the choice of the expected line emission flux prior to absorption may still affect the underlying absorption profiles.

The properties of narrow slow absorptions and ISM-like absorptions are presented in Table \ref{tab:narrow} and \ref{tab:ISM-like}, respectively. {The maximum wind velocity $v_{max}$ is visually defined as the maximum projected velocity where the absorption appears to be present. }The equivalent widths of these components are measured by summing up the absence of flux beneath the expected line profile, as estimated locally from nearby spectral regions (see Figure~\ref{fig:narrow}).  The line centers and FWHMs are measured by fitting the absorption with a Gaussian profile convolved with the local (COS or STIS) line spread function\footnote{The line spread functions (LSFs) of both COS and STIS exhibit wavelength dependent non-Gaussian wings due to mid-frequency polishing errors in the HST primary and secondary mirrors. \\ COS LSFs: \url{https://www.stsci.edu/hst/instrumentation/cos/performance/spectral-resolution} \\ STIS LSFs: \url{https://www.stsci.edu/hst/instrumentation/stis/performance/spectral-resolution}}.

In some cases, the narrow slow absorption at \ion{C}{2} $\lambda1334.53$ is blended with an additional red component, when compared to that of \ion{C}{2} $\lambda1335.71$, (e.g. AA Tau in Figure \ref{fig:narrow}).  This extra absorption feature is caused by the contribution of the interstellar absorption in the ground-state line.
This will be further discussed in \S \ref{subsec:MgII}.

% TABLE ISM

\begin{table}[!t]
    \centering
    \caption{Properties of ISM-like Absorptions}
    \label{tab:ISM-like}
    \begin{tabular}{lcccccccccc}
   & & \multicolumn{3}{c}{\ion{C}{2} $\lambda1334.53$}  &\multicolumn{3}{c}{\ion{C}{2} $\lambda1335.71$}  \\
Star &     & $v_{\rm cent}$ & EW & FWHM &$v_{\rm cent}$ & EW & FWHM\\
\hline
         TW Hya & & -10 & 30 & 25 & - & - & -\\
         HD 169142  & & - & - & - & -25 & 5 & 15\\
         HD104237  & & - & - & - & -5 & 20 & 20\\
         RU Lup   & & - & - & - & 5 & 20 & 20\\
         CV Cha  & & - & - & - & 0 & 30 & 25\\
         DN Tau  & & 0 & 15 & 10 & -5 & 10 & 5\\
         BP Tau & & -10 & 15 & 10 & -10 & 10 & 5\\
         DF Tau  & & -5 & 20 & 20 & 0 & 10 & 10\\
           \hline
    \end{tabular}
    \begin{tablenotes}
    \item{-} Values of $v_{\rm cent}$, EW and FWHM are in \kms, $v_{\rm cent}$ is in the stellar rest frame.
    \item{-} Targets are presented in ascending order of disk inclinations, except for DF Tau,the disk inclination of which has not been measured with ALMA.
    \end{tablenotes}
\end{table}

The narrow slow absorptions and ISM-like absorptions are classified separately by eye based on the absorption properties.  The line widths and equivalent widths in the slow narrow absorptions are higher than those of the ISM-like absorptions (shown in the left panel of Figure \ref{fig:narrowplots}). The narrow slow absorptions are detected mostly in targets
with intermediate or high inclinations (higher than 35$^\circ$, see Figure~\ref{fig:narrowplots}) and have line center
velocities of tens of \kms.  On the other hand, the ISM-like absorptions are
detected only in disks with lower inclinations, with centroid
velocity within -25 \kms\ of the stellar radial velocity, slower than that of narrow slow absorptions\footnote{We also present ISM-like absorption line profiles without shifting to the stellar rest frame in the right panel of Figure~\ref{fig:spec_all}, showing that ISM-like absorptions detected in targets in Taurus (DN Tau, BP Tau, and DF Tau) have very similar heliocentric velocities.}. For inclined disks, ISM-like absorptions are often blended with narrow slow absorption (\S \ref{subsec:MgII}).  {The ISM absorption is always spectrally unresolved at the COS resolution.}

{The absorption features are classified as (narrow) slow absorption and distinguished from fast wind absorption by eye, instead of a fixed criterion of velocity or width. As introduced in \S \ref{sec:c2overview}, the most notable characteristic for the slow absorption is that the line profiles can be described by a single (Gaussian) component. The center velocity and line width for the slow absorption are thus measured based on fitting this single component. The FWHM of slow absorption ranges from 35 to 295 \kms (Table \ref{tab:narrow}), generally narrower than the width of fast absorption ($v_{min}$ to $v_{max}$ ranges from 90 to 485 \kms, see Table \ref{tab:fastwind}). }
{The $v_{max}$ for slow winds ranges from 50 \kms to 150 \kms, generally lower than the $v_{max}$ for fast winds, ranging from 100 \kms to 565 \kms. For slow winds, $v_{max}$ can be affected by the line broadening or uncertainties of the emission line profile. Although $v_{cent}$ may be a better indicator for the bulk wind velocity for some objects with slow winds, we use $v_{max}$ for all winds for consistency. The use of different indicators of $v_{max}$ and $v_{cent}$ brings in uncertainties in slow wind velocity of $\lesssim$ 100 \kms, as shown in Figure \ref{fig:velerror}.}
{The difference and similarity of fast and slow winds can be seen in Figure~\ref{fig:spec_all}, where \ion{C}{2} line profiles for all of the targets in our sample are presented separately according to our wind classification.}

% FULL FIGURE SET SHOULD BE INSERTED HERE

% FIGURE IONIZATION EXAMPLE

\begin{figure*}[t]
\plottwo{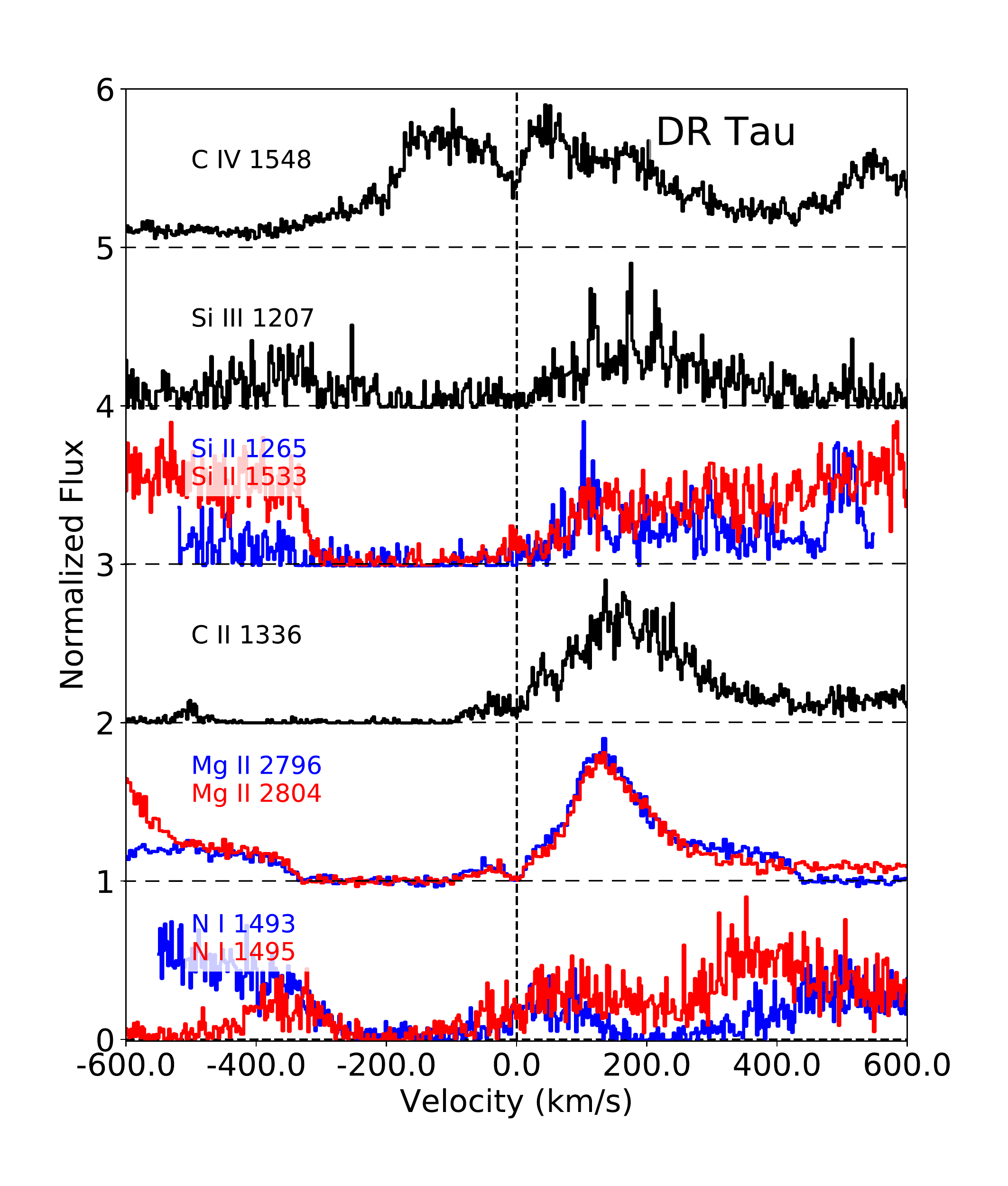}{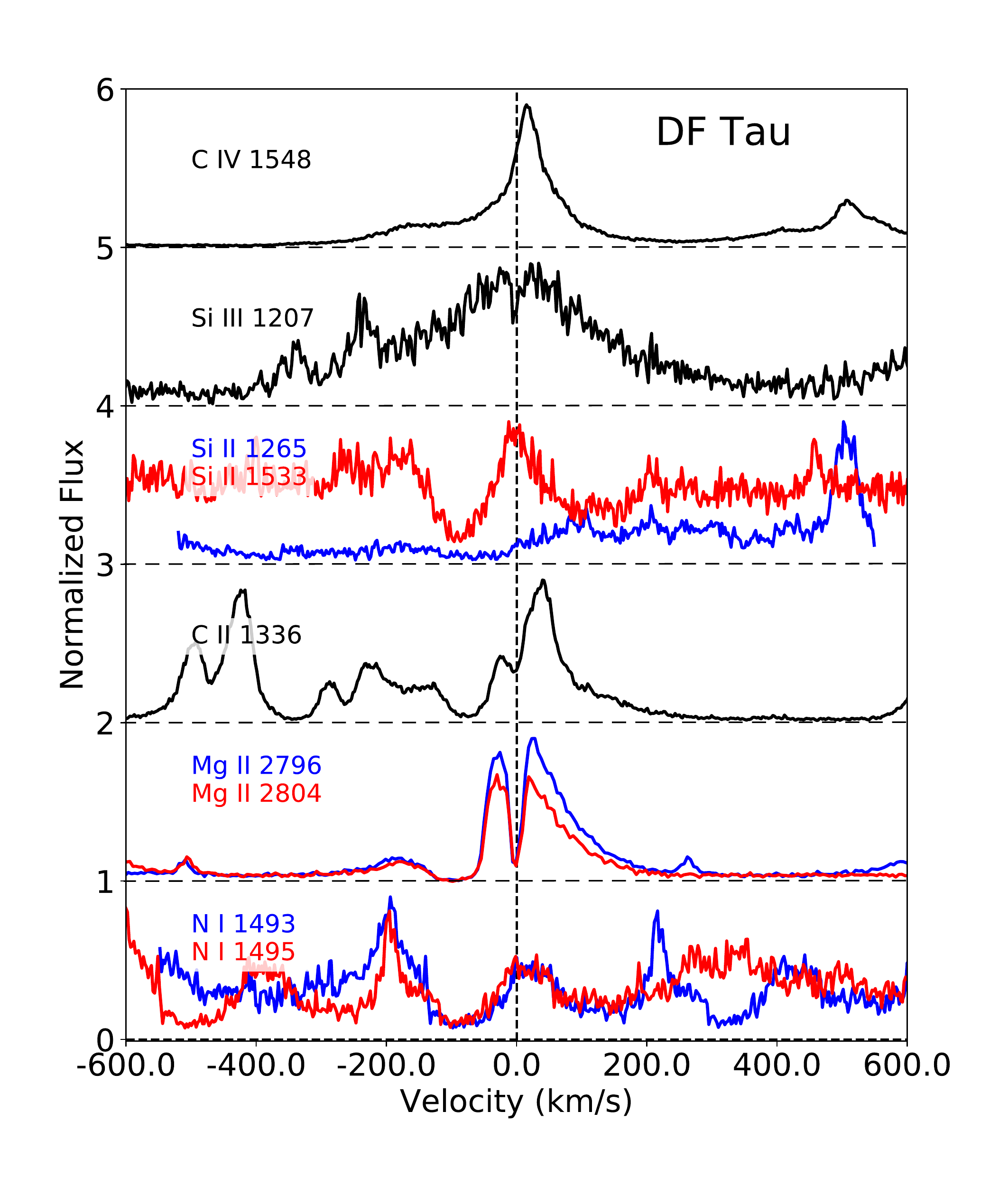}
\plottwo{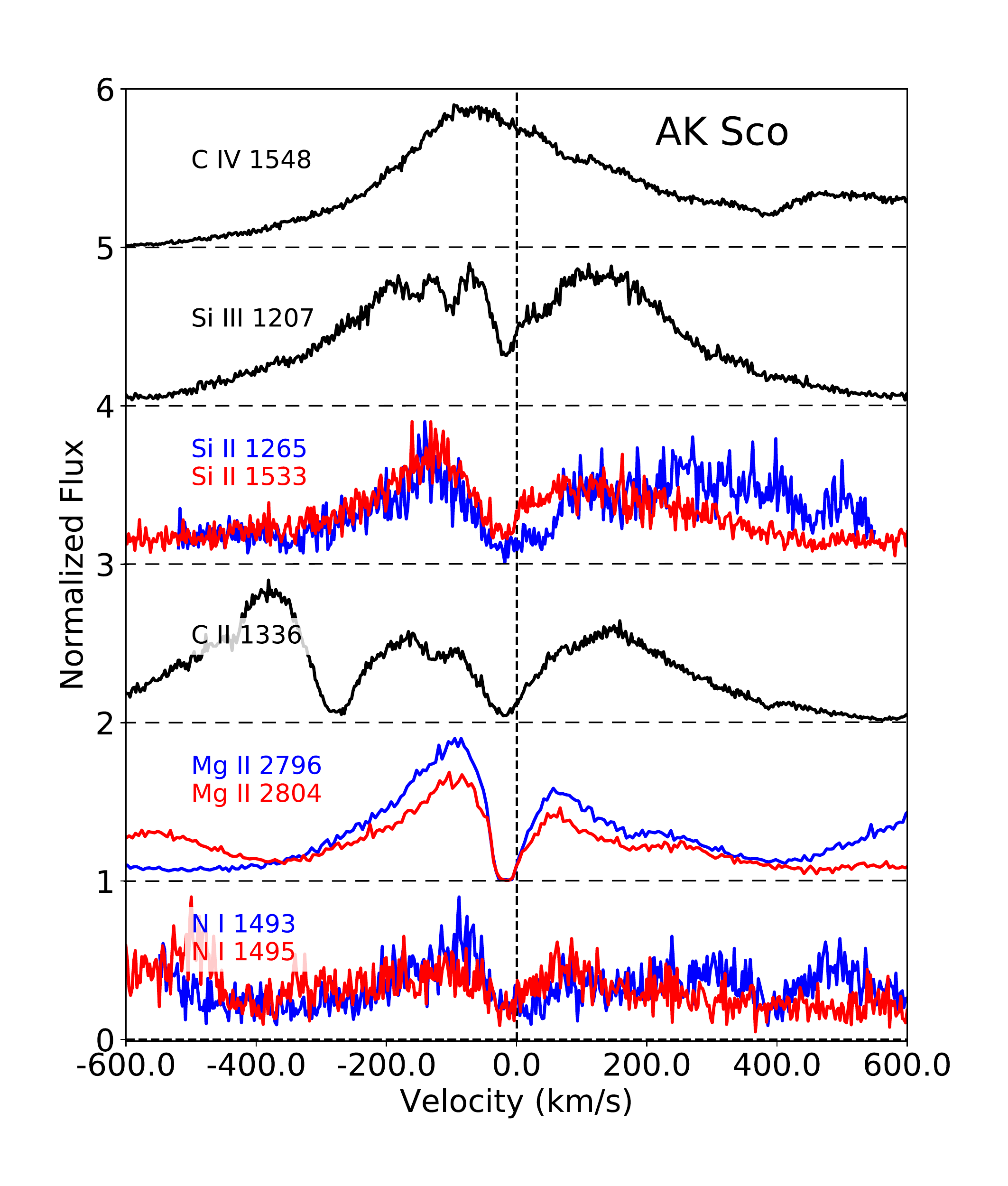}{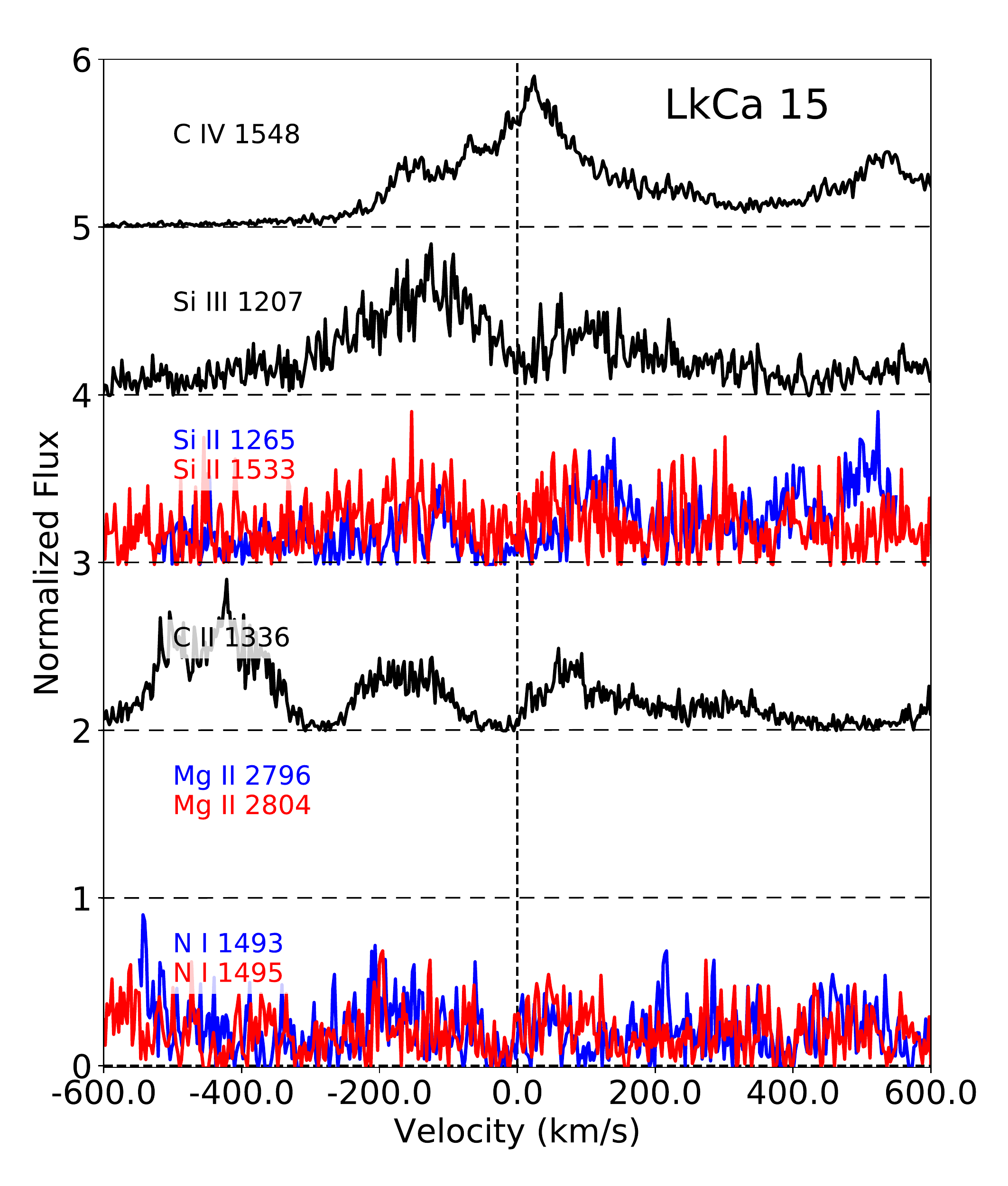}
\caption{\label{fig:ionization} FUV line profiles of different ionization states and \ion{Mg}{2} NUV line profiles, if present. Dashed horizontal lines mark the zero-flux level for each line. The dashed vertical line shows zero-velocity. Upper panels show two fast wind cases, bottom panels show two slow wind cases. The absorption is generally detected in low-ionization species. The detection of wind absorptions in \ion{Mg}{2} NUV lines are similar to those from \ion{C}{2} lines. {The complete figure set for all the targets (40 images) is available in the online journal.}}
\end{figure*}

%%%%%%%%%%%%%%%%%%%%%%%%%%%%%
% COMPARISON TO OTHER LINES
%%%%%%%%%%%%%%%%%%%%%%%%%%%%%

\section{Comparisons of \ion{C}{2} absorption to other lines} \label{sec:comparisons}

In this section, the \ion{C}{2} wind absorption is compared to other tracers of winds from young stars; each wind diagnostic has relative strengths and weaknesses.
This section begins with a brief analysis of many other FUV lines that could have absorption profiles.  We then compare in more detail the \ion{C}{2} absorption with absorption in the near-UV \ion{Mg}{2} h \& k doublet and the \ion{He}{1} $\lambda10830$ line and emission in the  [\ion{O}{1}] $\lambda6300$ line.

%%%
\subsection{FUV lines of different ionization states} \label{subsec:ionization}

The launching location of the wind can be informed by assessing the ionization structure of the absorbing gas.  The far-ultraviolet includes many lines other than \ion{C}{2} that would be detectable in the wind, if sufficiently abundant.  The ionization structure of the absorbing wind is uncertain because the best-studied wind absorption line, \ion{He}{1} $\lambda10830$, has a lower level that may be excited thermally ($\sim 25000$ K) or by X-rays \citep{kwan07}.  A description of the ionization in the wind  helps to break this degeneracy.

Previous evaluations of FUV absorption have characterized wind absorption in TW Hya, RU Lup and RW Aur \citep{herczeg05,dupree05,johnskrull07,France14}.  In addition, \citet{cauley16} searched for wind absorption in \ion{C}{4} in several Herbig AeBe stars, with only a single detection.

Table \ref{tab:windall} summarizes detections of FUV wind absorptions in our sample for different ionization states. Figure \ref{fig:ionization} shows line profiles for two fast wind and two slow wind cases, selected here for high S/N (figure set of line profiles for all of the targets is available in the online journal).  For both the fast and slow winds, the absorption is detected in low-ionization species, including \ion{N}{1}, \ion{O}{1}, \ion{Si}{2}, and in some cases \ion{Si}{3}.  Wind absorption is detected in \ion{Si}{4} and \ion{C}{4} only from HD 104237 \citep[as identified by][]{cauley16}.  In one other case, DR Tau, \ion{C}{4} self-absorption is present at line center of the blue component, either because the line is optically thick or because the circumstellar medium is ionized.

% FIGURE C II VS. Mg II SPECTRA

\begin{figure*}[t]
\plottwo{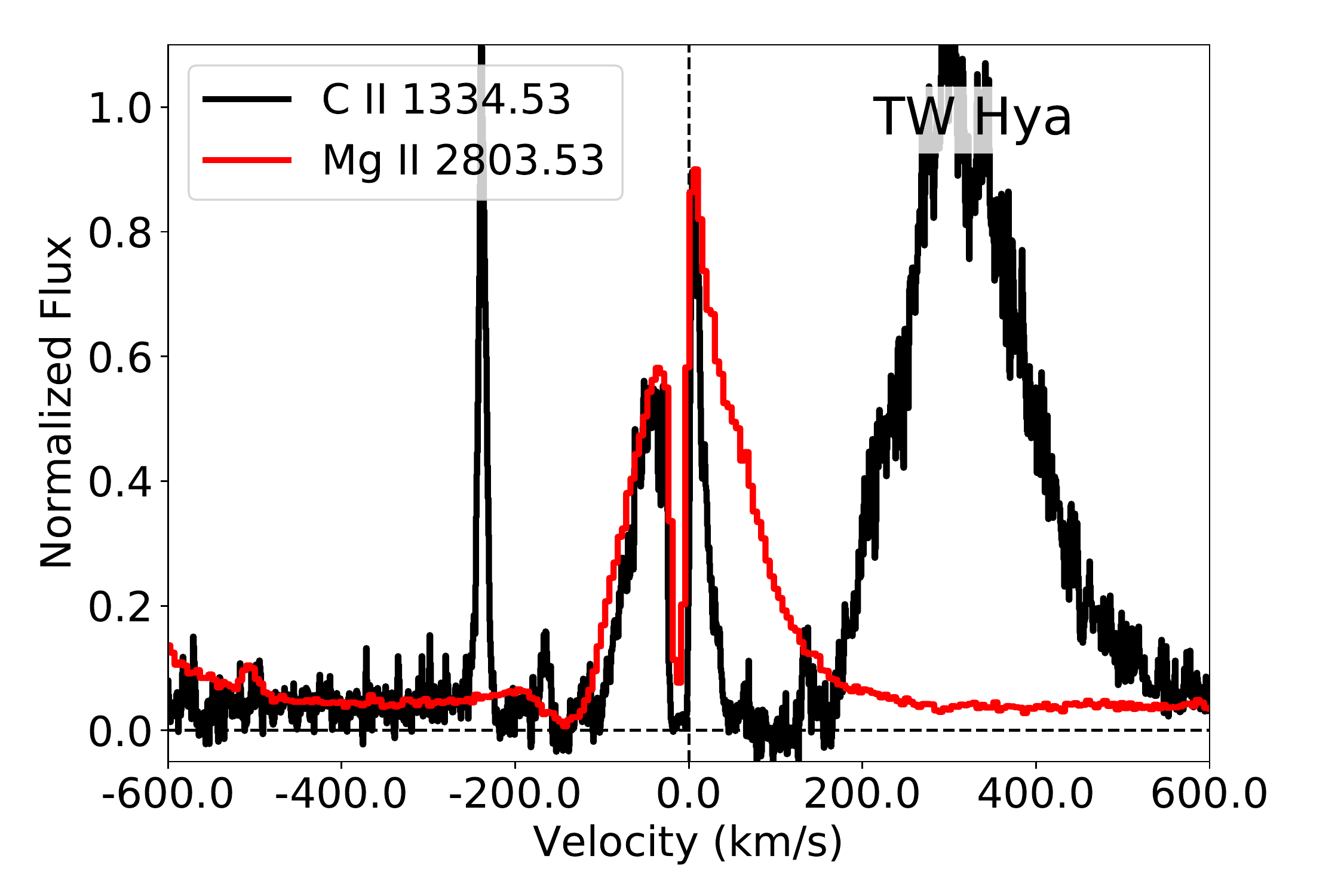}{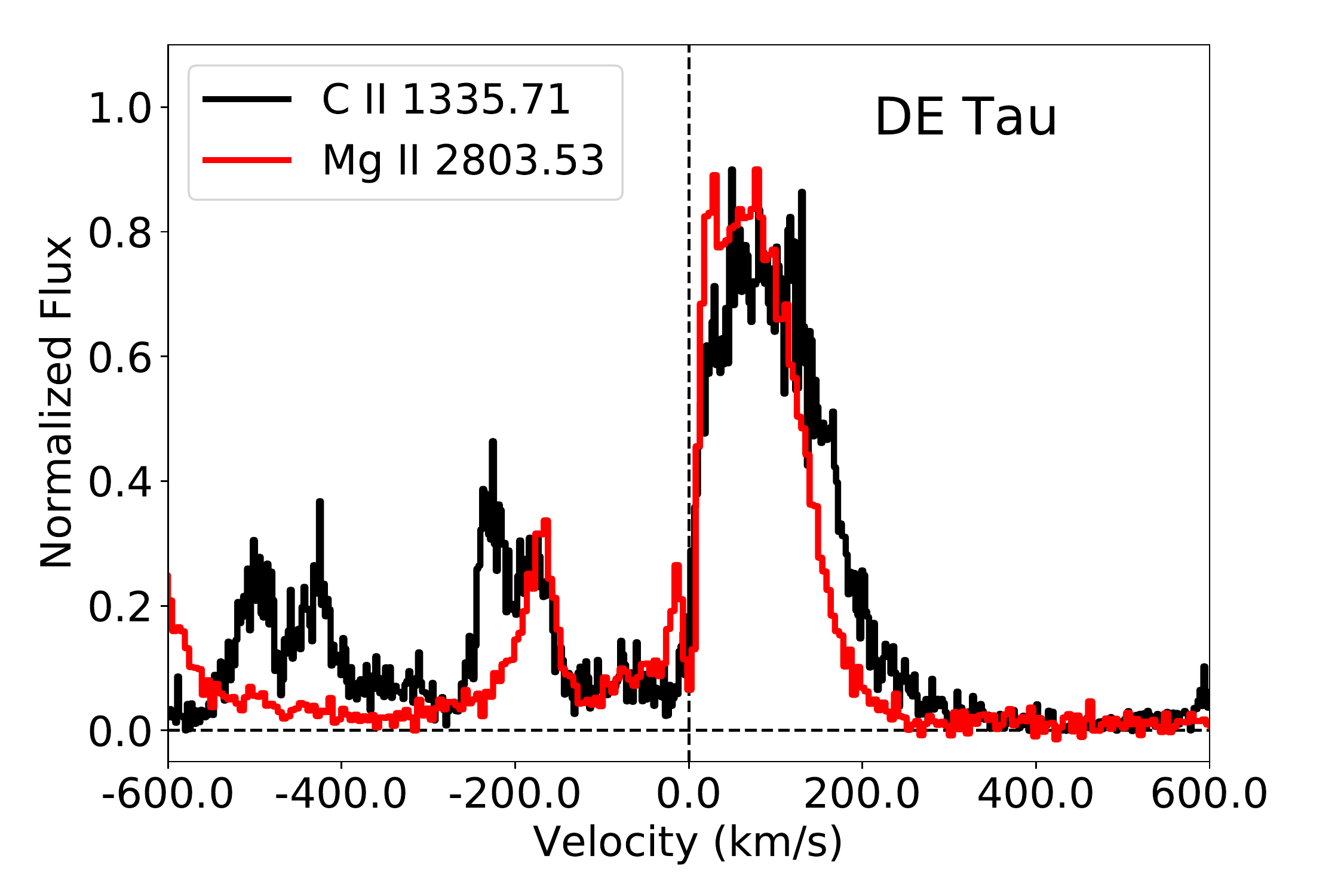}
\plottwo{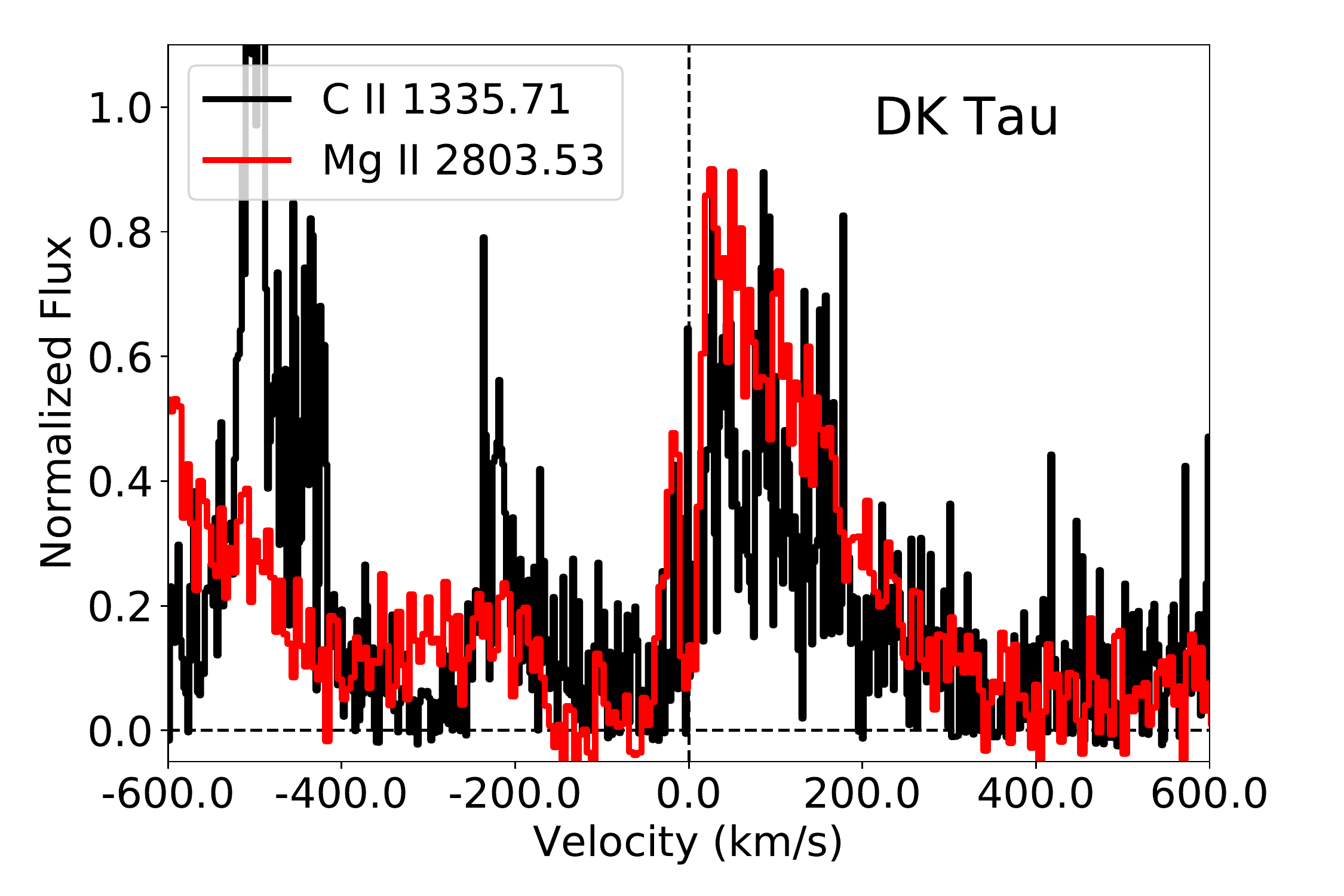}{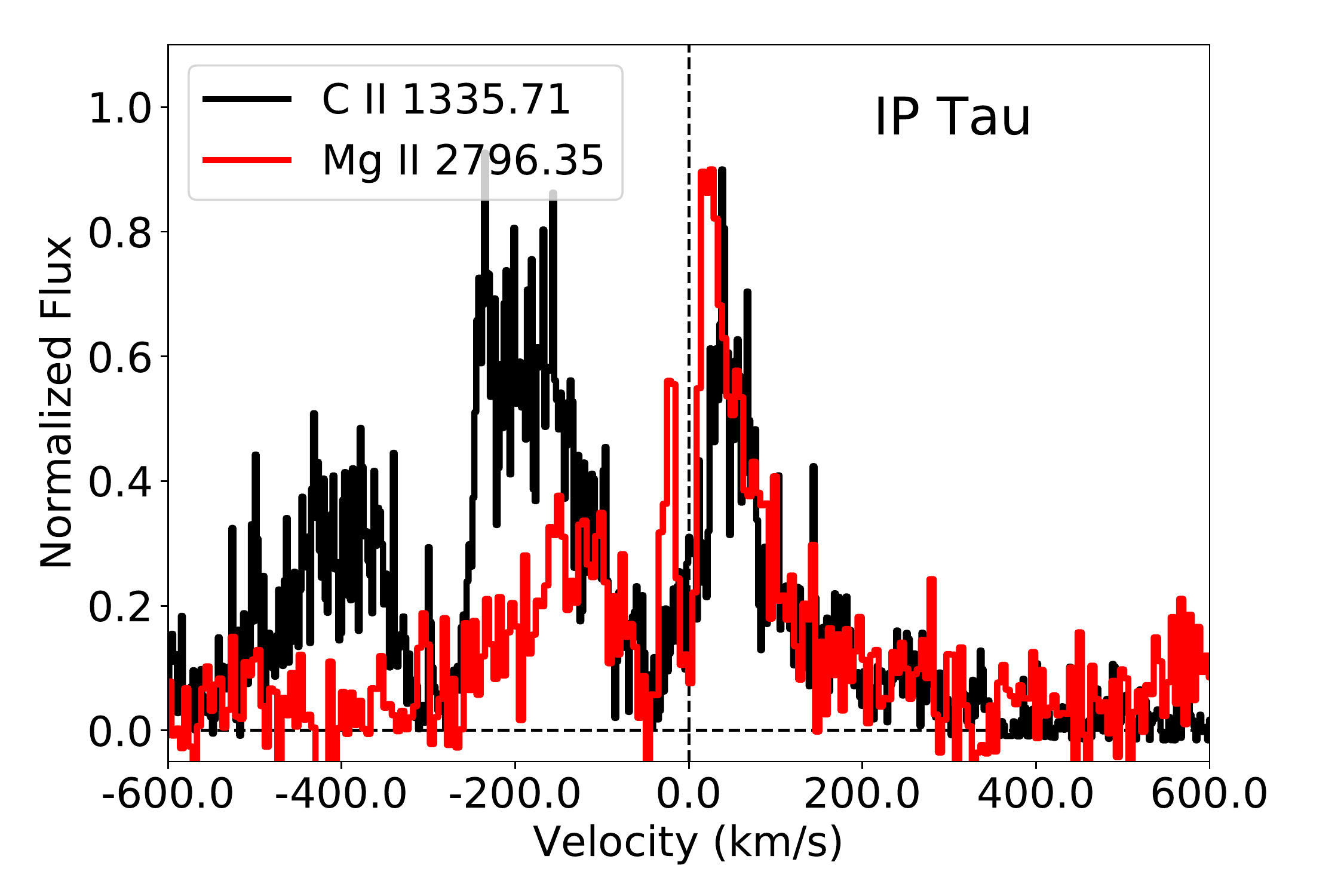}
\plottwo{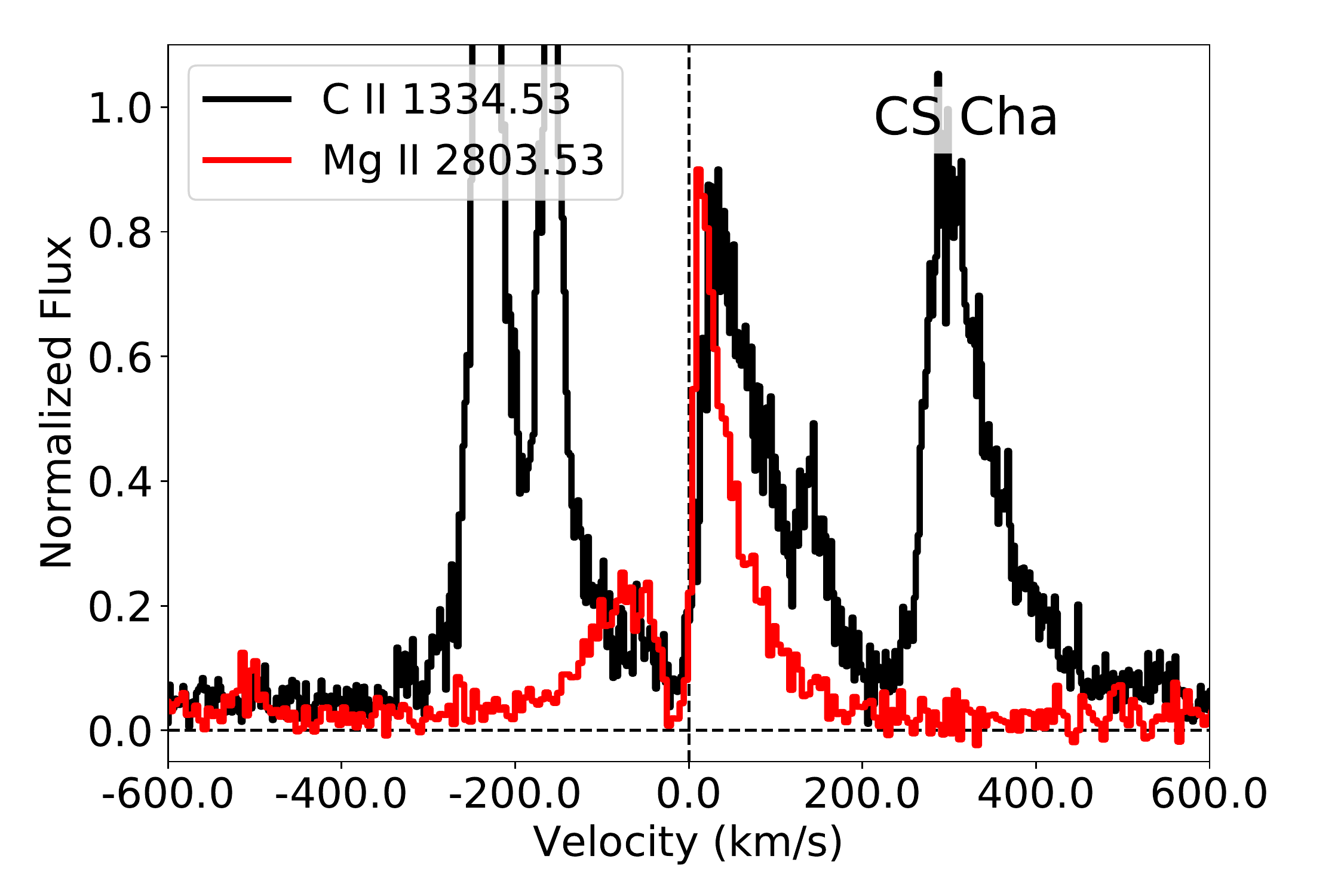}{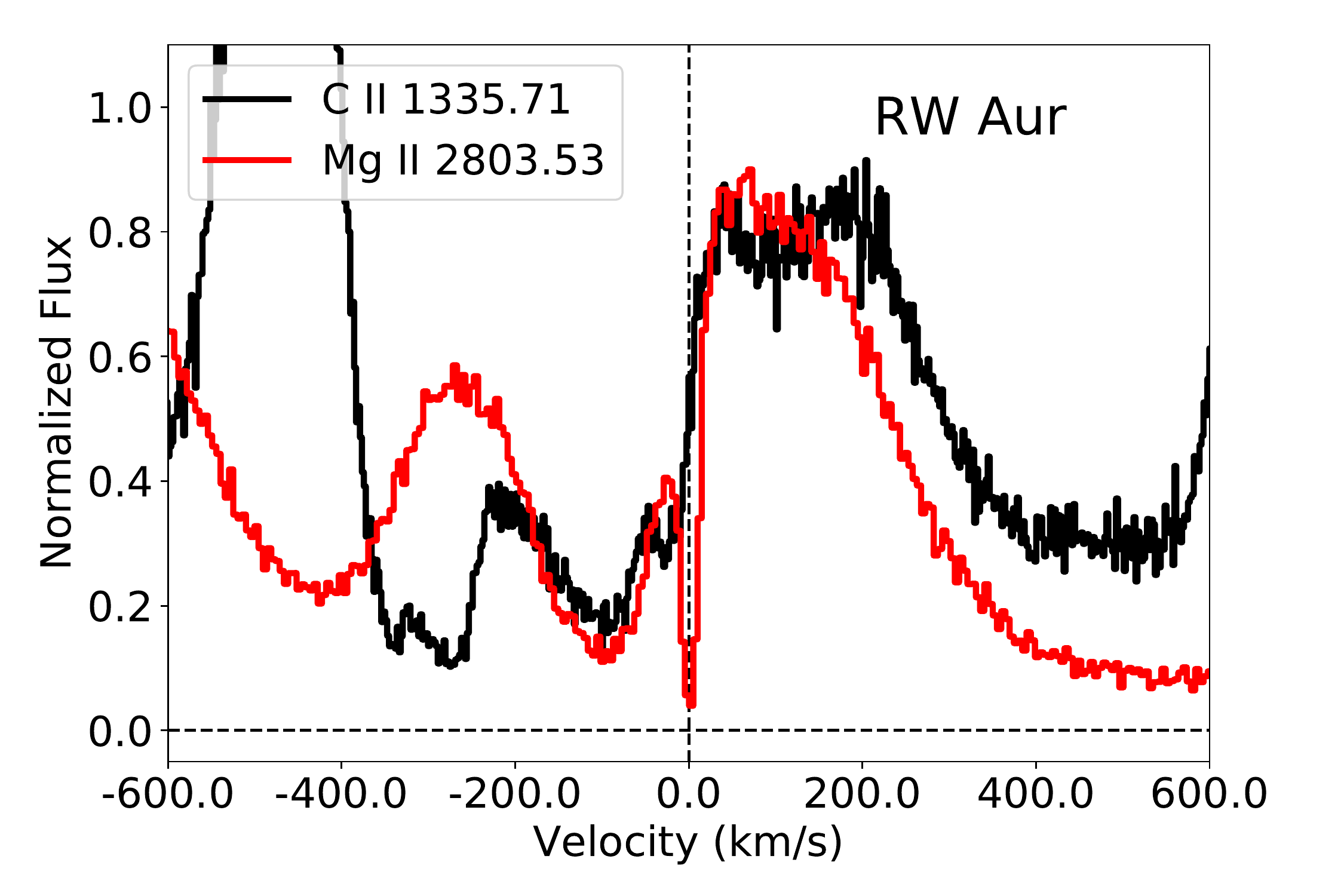}
\caption{\label{fig:mgiicompare} Line profiles of \ion{C}{2} $\lambda$1335 (black) and \ion{Mg}{2} $\lambda\lambda$2796.35,2803.53 (red). The horizontal dashed line shows zero-flux level, and the vertical dashed line shows zero-velocity. The line profiles are scaled to have the peaks of \ion{C}{2} and \ion{Mg}{2} lines match. The shape of \ion{Mg}{2} lines are overall similar to \ion{C}{2} profiles. TW Hya shows an example of the consistency of fast wind absorption between \ion{C}{2} and \ion{Mg}{2} lines.  The higher $S/N$, higher spectral resolution and larger line separation in \ion{Mg}{2} lines help identify otherwise ambiguous wind absorptions (DE Tau, DK Tau, CS Cha and RW Aur). The higher resolution in the \ion{Mg}{2} lines also resolve low velocity or ISM-like components that could be blended with faster components in \ion{C}{2} lines (e.g. IP Tau).}
\end{figure*}

% FIGURE C II VS. Mg II

\begin{figure*}[t]
\plottwo{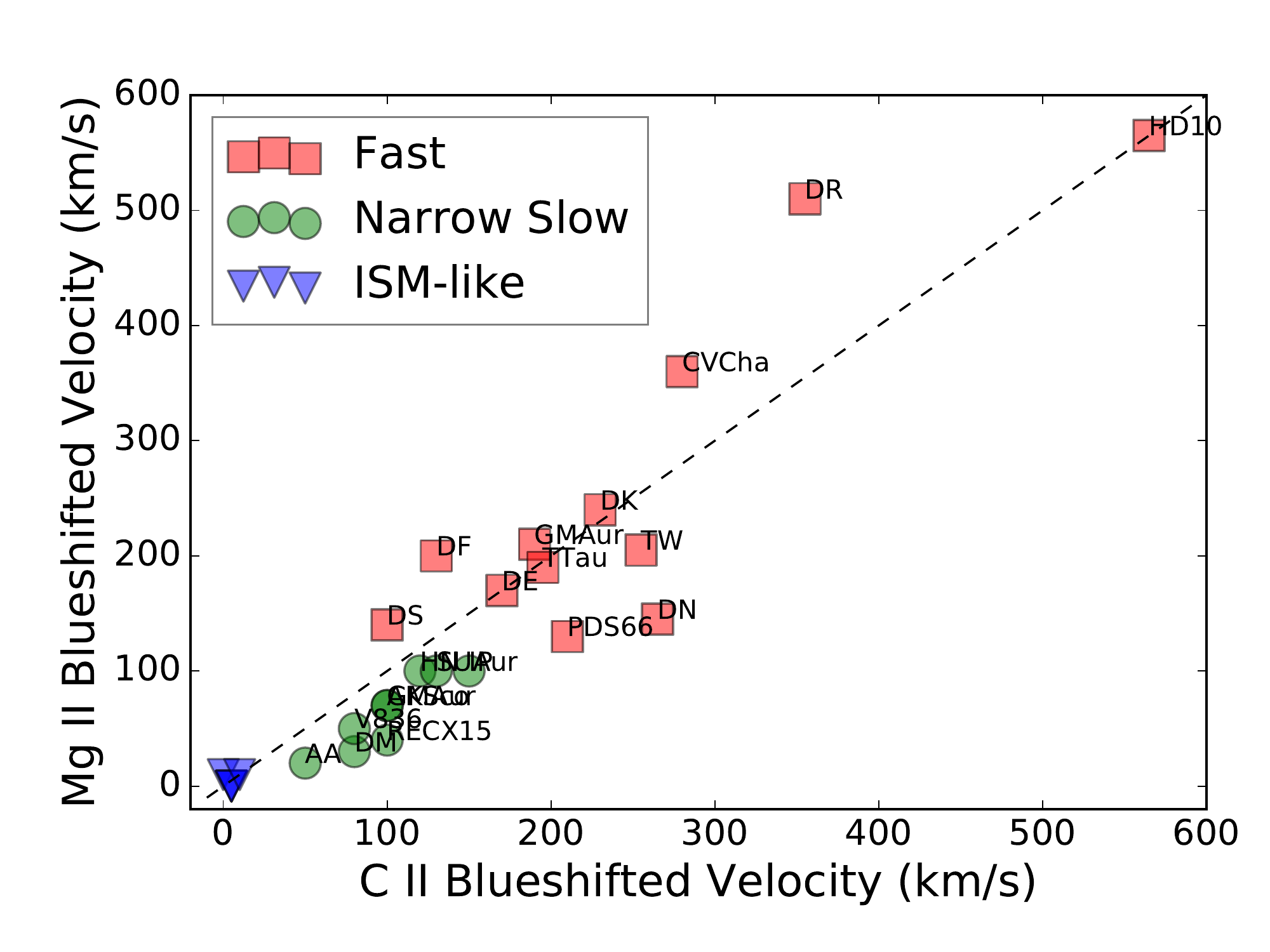}{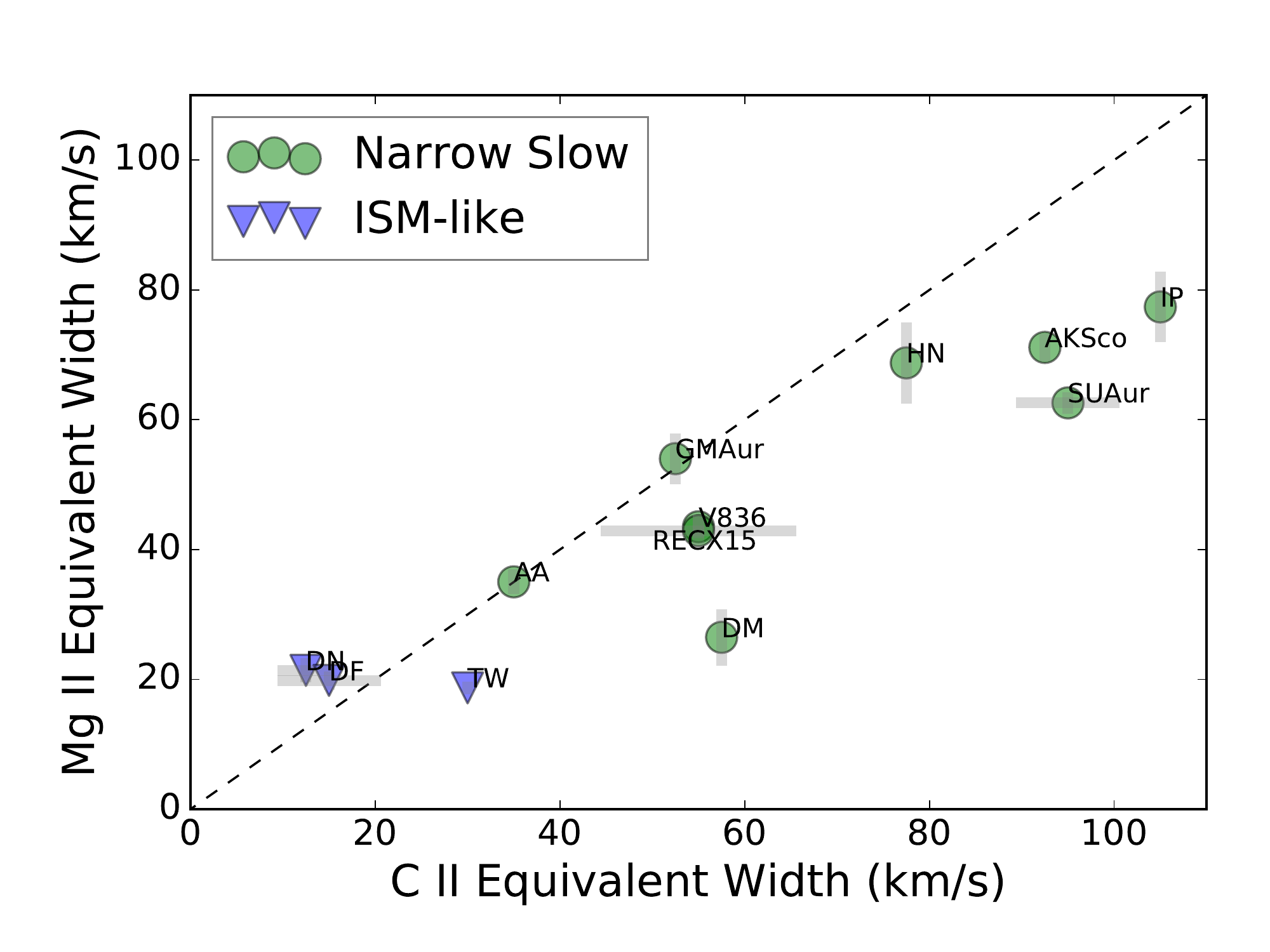}
\caption{\label{fig:mgiivscii} Correlations between \ion{C}{2} and \ion{Mg}{2} absorption line properties. The black dashed lines mark where the \ion{C}{2} and \ion{Mg}{2} line properties are equal. Left: the wind detection and velocities in \ion{Mg}{2} lines are similar to those from \ion{C}{2}. {The velocities are indicated by $v_{max}$ for fast winds (Table \ref{tab:fastwind}), and slow winds (Table \ref{tab:narrow}),  and by $v_{cent}$ for ISM-like absorptions (Table \ref{tab:ISM-like}).} Right: the equivalent widths (EW) of narrow slow (and ISM-like) absorptions in \ion{C}{2} and \ion{Mg}{2} are overall consistent, with the EW from \ion{Mg}{2} slightly slower. The data points mark the averaged value of the EWs measured from two members in the \ion{C}{2} or \ion{Mg}{2} doublets, the grey bars link the absorption EWs of the two members. }
\end{figure*}

Wind absorption is detected in \ion{Si}{3} $\lambda1206.5$ for 7--8 targets (AK Sco, DR Tau, HD 104237, HD 142527, RU Lup, RW Aur, T Ori, and a tentative detection of LkCa 15),
and in \ion{N}{1} $\lambda\lambda 1492,1494$ for 6 targets (AK Sco, DF Tau, DR Tau, HD 104237, RU Lup, and RW Aur). These \ion{N}{1} lines have lower levels excited 2.4eV above the ground state. The detection of \ion{Si}{3} and \ion{N}{1} lines could be explained by {a cool wind with ionization dominated by radiation} \citep{johnskrull07}.

These results are consistent with previous conclusions that fast winds have low ionization and are therefore not tracing hot coronal winds \citep{herczeg05,johnskrull07}.  Our results further establish that the ionization structure of the slow wind is qualitatively similar to the fast wind.

%%%
\subsection{Wind absorption in \ion{Mg}{2} NUV lines} \label{subsec:MgII}

The shape of the \ion{Mg}{2} $\lambda\lambda 2796.35, 2803.53$ resonance lines are usually similar to the \ion{C}{2} line profiles, with strong emission produced by accretion processes and absorption from winds and the interstellar medium \citep[see also analysis of \ion{Mg}{2} lines by][]{ardila02,Lopez-Martinez15}.
Wind features are typically easier to measure in \ion{Mg}{2}  than in \ion{C}{2} because of brighter continuum emission, higher $S/N$, and higher spectral resolution when compared to COS.  However,
fewer sources have been observed in \ion{Mg}{2} than in the FUV\footnote{Most sources with observations of both \ion{C}{2} and \ion{Mg}{2} lines had those lines observed within hours.} and any future model analysis of \ion{Mg}{2} lines would suffer from additional abundance uncertainties. 

Absorption in \ion{Mg}{2} is detected in fast and slow winds and in the ISM, using similar methods as those described for \ion{C}{2}.
An example of the comparison of fast wind absorption between \ion{C}{2} and \ion{Mg}{2} lines is presented in Figure \ref{fig:mgiicompare} for TW Hya, showing the consistency between \ion{C}{2} and \ion{Mg}{2} wind absorptions: the blue edge of fast wind absorption in \ion{Mg}{2} $\lambda2803.5$ is located at a similar velocity as the sub-continuum absorption as well as the absorbed H$_2$ $\lambda$1333.8 emission line.
For DR Tau and HD 104237, the maximum velocity is faster than 500 \kms, similar to \ion{C}{2}.  

Overall, all targets in our sample observed in \ion{Mg}{2} with significant $S/N$ show wind absorption.  The identification and classification of fast wind absorption and their maximum velocities are similar to those detected from \ion{C}{2}, as is shown in the left panel of Figure \ref{fig:mgiivscii}.

The higher spectral resolution in the \ion{Mg}{2} lines (compared to COS spectra) helps us to confirm the detection of otherwise ambiguous absorption in \ion{C}{2} lines of DE Tau, DK Tau, DF Tau, and GM Aur.
The fast absorption in \ion{C}{2} lines of PDF 66 and RECX 11 also shows weak counterparts in their \ion{Mg}{2} lines. Similarly, wind absorption to CS Cha and RW Aur is clearly detected in \ion{Mg}{2} but is not claimed as a detection in \ion{C}{2} because the absorption is only marginally detected.  From visual inspection these differences are dominated by the lower sensitivity (lower $S/N$ and spectral resolution) in \ion{C}{2} and are not significant discrepancies.  Some temporal variations are expected but are {likely} not significant in these specific comparisons.

For some targets, multiple low velocity components are detected, usually with a wind component centered near $-30$ \kms\ and an ISM component centered near $0$ \kms.  For example, in the IP Tau spectrum the broader  \ion{C}{2} absorption is consistent with a blend of the two resolved \ion{Mg}{2} absorption components (Figure \ref{fig:mgiicompare}).

%%%
\subsection{\ion{He}{1} $\lambda10830$}

Absorption in the \ion{He}{1} $\lambda10830$ line is an excellent probe of wind dynamics.  Seven targets analyzed here overlap with the \ion{He}{1} survey of \citet{edwards06}.  The classification of disk and stellar winds from \ion{C}{2} and \ion{Mg}{2} absorption are broadly consistent with the classification from \ion{He}{1}.

DR Tau has wind absorption that extends to $\sim 500$~\kms\ in \ion{C}{2} and \ion{Mg}{2} but only $\sim400$~\kms\ in \ion{He}{1}, with the difference indicating either variability in the wind velocity or higher optical depths in \ion{C}{2} and \ion{Mg}{2}.  Other discrepancies between wind properties of \ion{He}{1} and \ion{C}{2} lines are introduced by differences in sensitivity.  For example, SU Aur has weak absorption in \ion{He}{1} that extends to high velocities and would be undetectable with the existing far-ultraviolet spectrum.  CY Tau has disk wind absorption seen in \ion{He}{1} is not detected in \ion{C}{2} because of a lack of sufficient signal.  Spectra at 1 $\mu$m usually have high $S/N$ in the continuum, against which even 1\% absorption features may be identified.
On the other hand, for \ion{C}{2}, the continuum is usually detected at much lower $S/N$, forcing us to use a combination of indicators that are sensitive only to relaively strong absorption.

%%%
\subsection{Comparison with [\ion{O}{1}] line emission} \label{subsec:oi}

Disk winds have been most commonly studied in emission in optical forbidden lines. High velocity components (HVC) of the emission line trace the jet that is launched in the inner disk, while low velocity components (LVC) are attributed to slower disk winds from the outer regions \citep[see, e.g., also][]{hartigan95,bacciotti02,simon16,fang18,banzatti19}.
We analyze cases of fast wind / HVC and slow wind / LVC seperately, by comparing \ion{C}{2} wind absorptions with wind components detected in [\ion{O}{1}] emission lines \citep{banzatti19}.

\begin{figure}[t]
\plotone{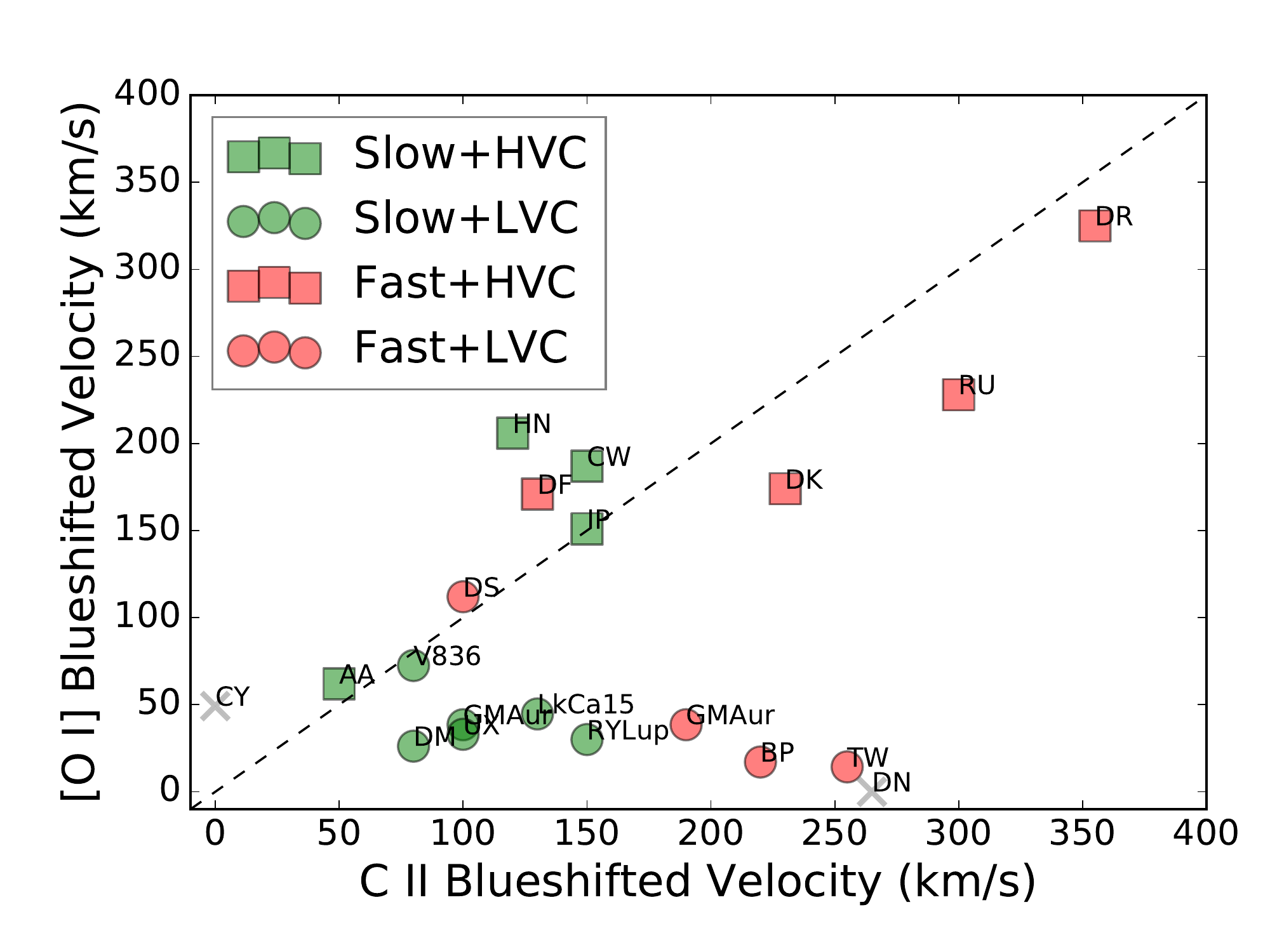}
\caption{\label{fig:oivscii} Correlation of wind velocities between \ion{C}{2} absorption (our work) and [\ion{O}{1}] emission \citep{banzatti19}. {The \ion{C}{2} velocities are indicated by $v_{max}$ for fast winds (Table \ref{tab:fastwind}), and slow winds (Table \ref{tab:narrow}), and by $v_{cent}$ for ISM-like absorptions (Table \ref{tab:ISM-like}). The [\ion{O}{1}] velocities are maximum wind velocities determined by summing the line centroid velocity and the FWHM.} The black dashed line marks where the \ion{C}{2} and [\ion{O}{1}] wind velocities are equal. ``Slow'' and ``fast'' in the legend stand for the wind type identified in \ion{C}{2} lines. ``HVC'' and ``LVC'' stand for the types of components identified in [\ion{O}{1}] emission. The grey crosses mark CY Tau, where LVC is detected in [\ion{O}{1}], but no clear wind absorption is identified in \ion{C}{2} lines; {and DN Tau, where fast wind absorption is detected in \ion{C}{2}, but no wind reported in [\ion{O}{1}]}. Overall, the wind velocities measured from [\ion{O}{1}] emission are lower than that measured from \ion{C}{2} absorption.}
\end{figure}

Figure \ref{fig:oivscii} and Table \ref{tab:OI} show that the wind velocities measured from [\ion{O}{1}] emissions are generally lower than that from \ion{C}{2} absorption, especially for {targets with no [\ion{O}{1}] HVC detected}.   Absorption is expected to be more sensitive to jets originated from the innermost disk regions. Particularly for face-on targets, the line of sight is highly aligned with the collimated jet and better captures the physical properties of the material in the jet. In contrast, emission is more powerful for slow disk winds at outer regions since the flux in emission lines increases as the emitting radius in the disk (hence the emitting area) increases.

% FIGURE C II VS. O I

Eight targets in our sample have an HVC in [\ion{O}{1}] $\lambda$6300 emission. From these targets, the three\footnote{DR Tau, DK Tau, and RU Lup; the inclination of the DF Tau disk has not been measured with ALMA.} with disk inclinations known to be less than 45$^{\circ}$ have fast wind absorption in \ion{C}{2}, consistent with the assumption of the collimated jet from close to the star. {Overall, for targets with HVC, the $v_{max}$ values of \ion{C}{2} wind are comparable to $v_{[O\ I]}$, the maximum velocities of components in [\ion{O}{1}], as shown in Figure \ref{fig:oivscii} (all of the squares). In \citet{banzatti19}, properties of [\ion{O}{1}] components are tabulated with centroid velocities and FWHM values. In order to compare our $v_{max}$ of \ion{C}{2} absorptions to [\ion{O}{1}], we approximate the [\ion{O}{1}] maximum wind velocity $v_{[O\ I]}$ by summing the line centroid velocity and the FWHM.}

Five targets with fast wind absorption, TW Hya, DN Tau, BP Tau, GM Aur and DS Tau, do not show a high-velocity component in [\ion{O}{1}] emission\footnote{Although BP Tau has redshifted HVC emission}.  The absence of jet emission despite the presence of fast winds may indicate that the mass flux or density in the wind is lower than the sources with jet emission.

For the ten targets with narrow slow wind absorption in \ion{C}{2}, their [\ion{O}{1}] emission components generally have relatively low velocities. Six of these targets\footnote{UX Tau, DM Tau, V836 Tau, LkCa 15, GM Aur, and RY Lup} have only low-velocity components detected in [\ion{O}{1}] emission, and $v_{LVC}$ is generally lower than $v_{wind}$ from \ion{C}{2} lines.  {Four of these targets (IP Tau, CW Tau, HN Tau and AA Tau) have high-velocity components with velocities within $\sim$ 200 \kms, generally comparable with their $v_{wind}$ measured from \ion{C}{2} lines}\footnote{In \citet{banzatti19}, HVC is defined by centroid velocity higher than 30\kms, which is different than our definitions of fast or slow winds based on line profiles.}. 

The slow \ion{C}{2} absorption detected in targets that also have HVCs are relatively broad.
For IP Tau and HN Tau, \ion{Mg}{2} lines in higher resolution spectra (Figure \ref{fig:mgiicompare}) show that their slow wind absorption is blended with the ISM absorption. The velocities of the slow wind absorption are reasonably consistent with the corresponding HVC velocities.
Similarly, the absorption feature in CW Tau is one of the broadest in all of the slow wind absorption we detected, which is possibly the mixture of wind components and the ISM absorption. The underlying wind component is potentially faster than measured, possibly explaning the inconsistency to the high velocity of the [\ion{O}{1}] HVC.

In \citet{banzatti19}, the HVCs are interpreted as emission from jets.  For disks that are viewed as highly inclined, like AA Tau, the jet is moving almost in the plane of the sky, so the high-velocity component is projected to a low velocity.  In general, all of these nine targets with slow wind absorption in \ion{C}{2} have moderate or high inclinations ($i>35^{\circ}$), and the HVC/jet emission is projected to lower velocities. We present models in \S \ref{sec:model} to further discuss the projected fast wind (and jet) as an explanation of the line profiles of \ion{C}{2} slow wind absorption, especially for those with relatively broader line profiles. LVCs in optical forbidden lines trace MHD winds from the inner disk \citep{simon16,banzatti19}.  Most disks with inner dust cavities are lack of secure inner MHD wind detection, likely indicating evidence for an evolution of disk winds (see also \citealt{pascucci20}). For \ion{C}{2} absorption, no clear dependence of wind detection upon disk evolutionary stage is seen in our sample.  Further discussions on wind origin with models in \S \ref{windwhere} indicate that wind absorption occurs close to the star. 

Recently, \citet{weber20} computed emission line profiles from a photoevaporative wind model and an MHD wind model separately, and concluded that the combination of line profiles from photoevaporative and MHD wind models would be plausible for reproducing the observed components. With our thermal-magnetic wind model, we will interpret the absorption line profiles and further discuss the wind physics in \S \ref{thermormag}.

% TABLE C II VS. O I

\begin{table}[t]
    \centering
     \caption{Comparison of [\ion{O}{1}] Emission to \ion{C}{2} Absorption}
    \label{tab:OI}
    \begin{tabular}{lccccc}
    Star     &   Incl.    &  $v_{wind}$   &   Wind   &  $v_{[O\ I]}$   \\
             &   (deg)  &  (km/s)  &    Type     &      (km/s) \\
          \hline
    DR Tau   &   5    &  -355   &   fast   &      -325     \\
   TW Hya    &   7    &  -225   &   fast   &      -14.3    \\
   DK Tau    &   13    &  -230   &   fast  &      -174  \\
   RU Lup    &   18    &  -300   & fast  &      -228   \\
   CY Tau    &   27    &  -   &   -   &     -49.2     \\
   UX Tau    &   35    &  -100   &   slow   &    -33.0   \\
   DN Tau    &   35    &  -85$^a$   & fast   &     -  \\
   DM Tau    &   35    &  -80   &   slow   &     -26.2  \\
   BP Tau    &   38    &  -165   &   fast  &     18.9   \\
   V836 Tau  &   43    &  -80   &   slow   &      -72.5 \\
   IP Tau    &   45    &  -150   &   slow   &      -151 \\
   LkCa 15   &   49    &  -130   &   slow   &    49.3    \\
   GM Aur    &   55    &  -210$^b$   &  fast  &   -38.5  \\
   DS Tau    &   65    &  -100   &   fast  &    112.0  \\
   CW Tau    &   65$^*$    &  -150   &   slow   &      -187 \\
   RY Lup    &   68    &  -150   &   slow   &      30.0 \\
   HN Tau    &   70    &  -120   &   slow   &      -206 \\
   AA Tau    &   75    &  -50   &   slow   &      -62 \\
   DF Tau    &   -    &  -130   &  fast  &      -171 \\
   \hline
    \end{tabular}
    \begin{tablenotes}
    \item  $v_{wind}$: maximum wind velocity measured from absorption. Taken maximum among the two C II members, different observations, different components (fast or slow) and different lines including \ion{Mg}{2}. {Indicated by $v_{max}$ for fast winds (Table \ref{tab:fastwind} or in \ion{Mg}{2}) and for slow winds (Table \ref{tab:narrow}).}
    \item Wind type: from our work.
    \item $v_{[O\ I]}$: {maximum wind velocity defined by summing up the line centroid velocity and the FWHM} of the bluest component taken from [\ion{O}{1}] 6300~\AA~from \citet{banzatti19}, classified into HVC when centroid velocity $<-30$~\kms.
    \item $^a$:  $-145$ \kms\ in \ion{Mg}{2}
    \item $^b$:  Measured from \ion{Mg}{2}.
    \item $^*$: less reliable inclinations that are not from ALMA or SMA.

    \end{tablenotes}
\end{table}

% FIGURE VELOCITY VS. INCLINATION

\begin{figure*}[t]
\plotone{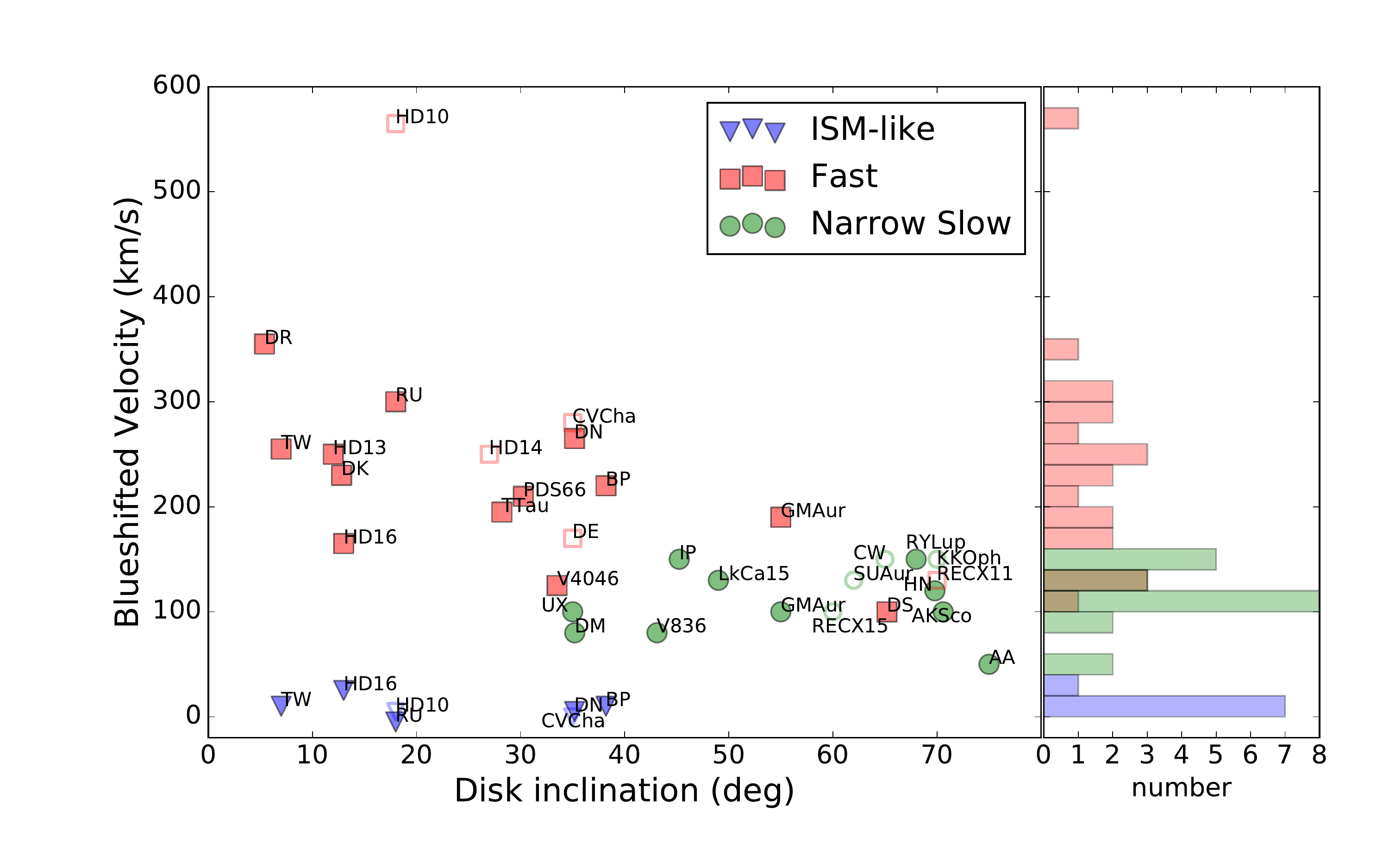}
\caption{\label{fig:windvel} Wind velocities measured from \ion{C}{2} lines, as a function of disk inclinations. {The wind velocities are indicated by $v_{max}$ for fast winds (Table \ref{tab:fastwind}) and for slow winds (Table \ref{tab:narrow}), and by $v_{cent}$ for ISM-like absorptions (Table \ref{tab:ISM-like}).} Open markers show targets with less reliable inclinations that are not from ALMA or SMA. {The histogram to the right presents velocity distribution for all of the targets, including disks without applicable inclinations.}  Fast wind absorption (red) is more likely to be detected at low disk inclinations, with wind velocity decreasing toward higher inclination, while slow wind absorption (green) tends to appear in more inclined disks. }
\end{figure*}

% FIGURE EQUIVALENT WIDTH

\begin{figure*}[t]
\plottwo{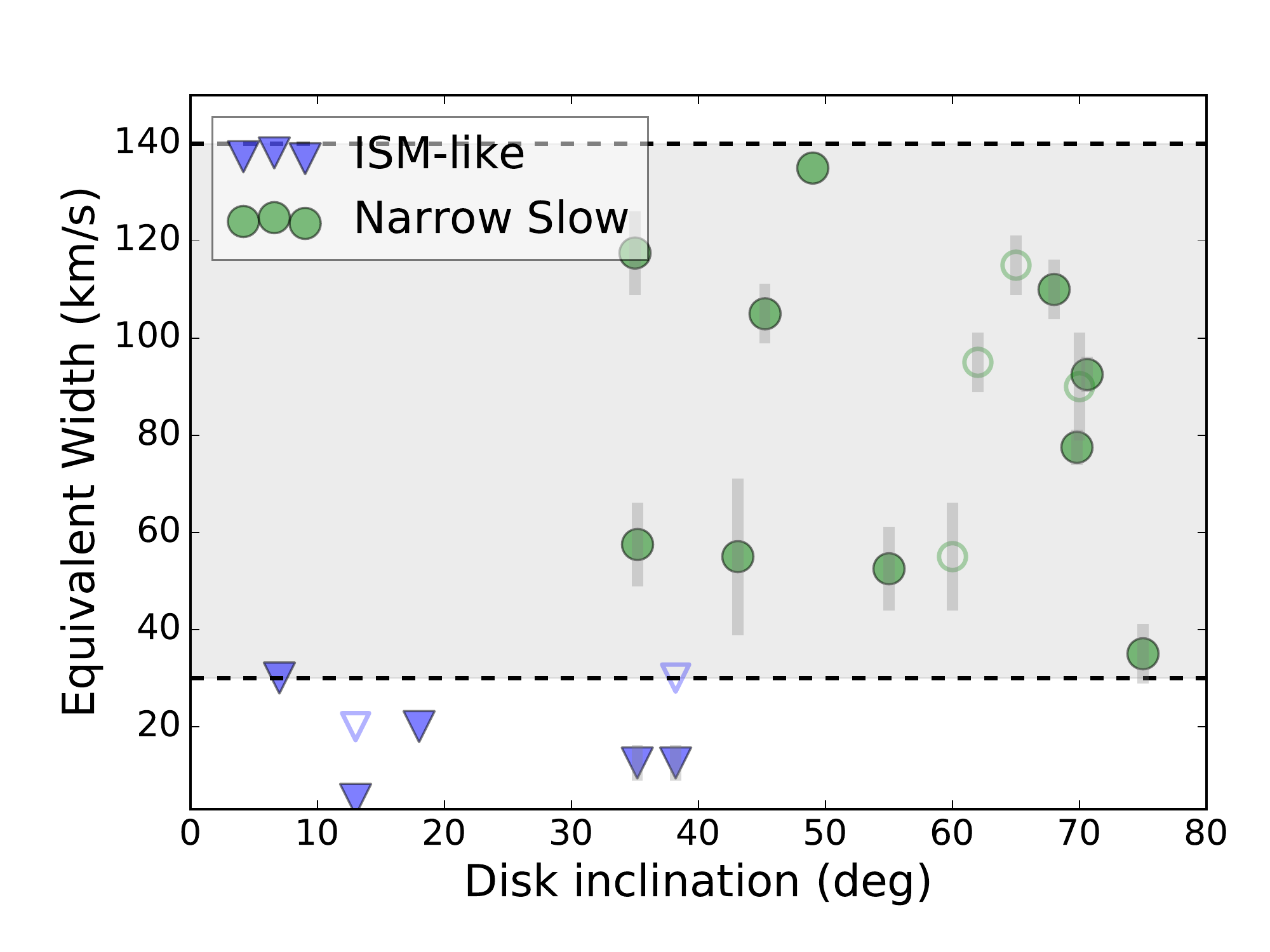}{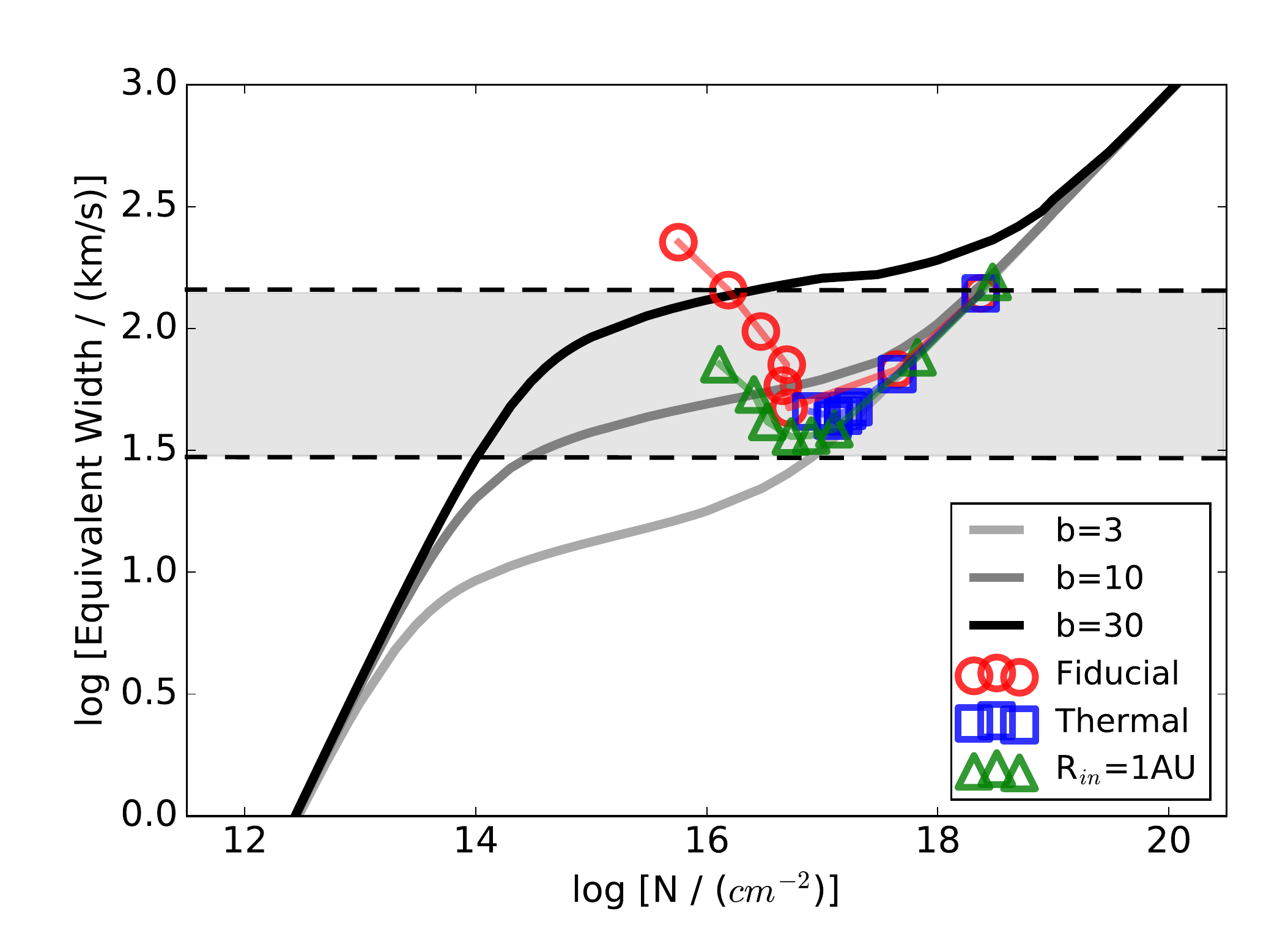}
\caption{\label{fig:narrowplots}  Left: The equivalent width (EW) of slow wind and ISM-like absorptions measured from the \ion{C}{2} lines, as a function of disk inclination. Open markers show targets with less reliable inclinations that are not from ALMA or SMA. The data points mark the averaged value of the EWs measured from two members of the \ion{C}{2} doublet, the grey bars link the absorption EWs of the two members. Right: the curve of growth of \ion{C}{2} lines, with different Doppler broadening parameters $b$ in \kms. The X-axis denotes \ion{C}{2} column density. The horizontal dashed lines and the grey shaded region show the range of EW measured from slow wind absorptions, the same as the dashed lines and the shaded region in the left panel. Data points in the right panel present what the models in \S \ref{sec:model} predict for the EW and C$^+$ column density along the line of sight assuming inclinations of 25$^{\circ}$, 30$^{\circ}$, 35$^{\circ}$, 40$^{\circ}$, 45$^{\circ}$, 50$^{\circ}$, 60$^{\circ}$, and 70$^{\circ}$, respectively. Models predict C$^+$ column density increasing with disk inclination. Overall, the predicted equivalent widths are reasonably consistent with the observations of slow wind absorption.}
\end{figure*}

% FIGURE ACCRETION RATE

\begin{figure}[t]
\plotone{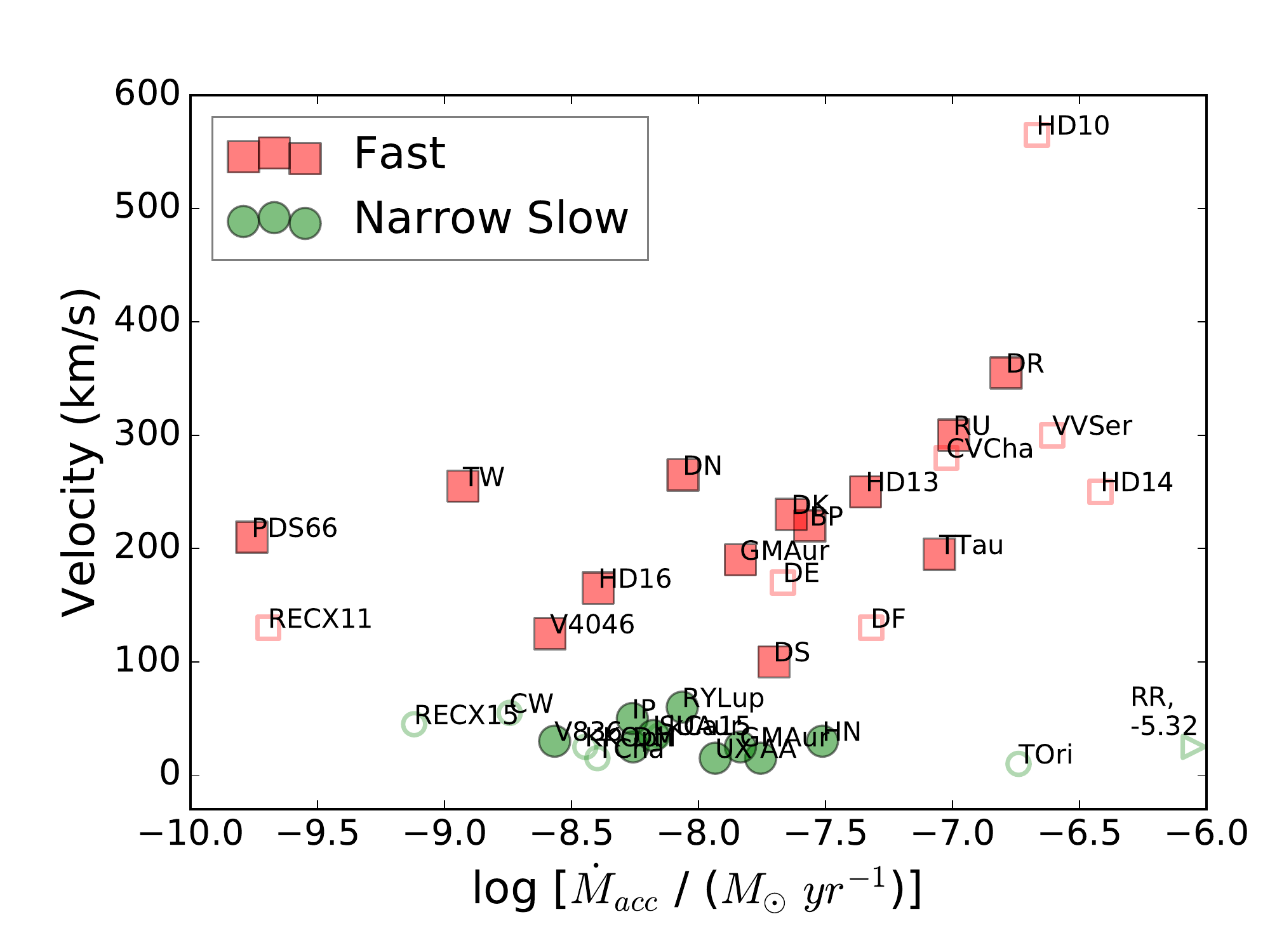}
\caption{\label{fig:mdot_v} Wind velocity from \ion{C}{2} lines as a function of accretion rate. The highest wind velocity measured from \ion{C}{2} is 565 \kms (HD 104237). The highest accretion rate in our sample is from RR Tau (4.8$\times 10^{-6}~M_{\odot}~yr^{-1}$), exceed the plot boundary, and is marked at $10^{-6}~M_{\odot}~yr^{-1}$ in the figure. No significant correlation is found overall or for the slow wind, and a weak correlation may present for the fast wind.}
\end{figure}

% FIGURE MODELS 10 AU

\begin{figure*}[t]

\includegraphics[width=0.33\textwidth]{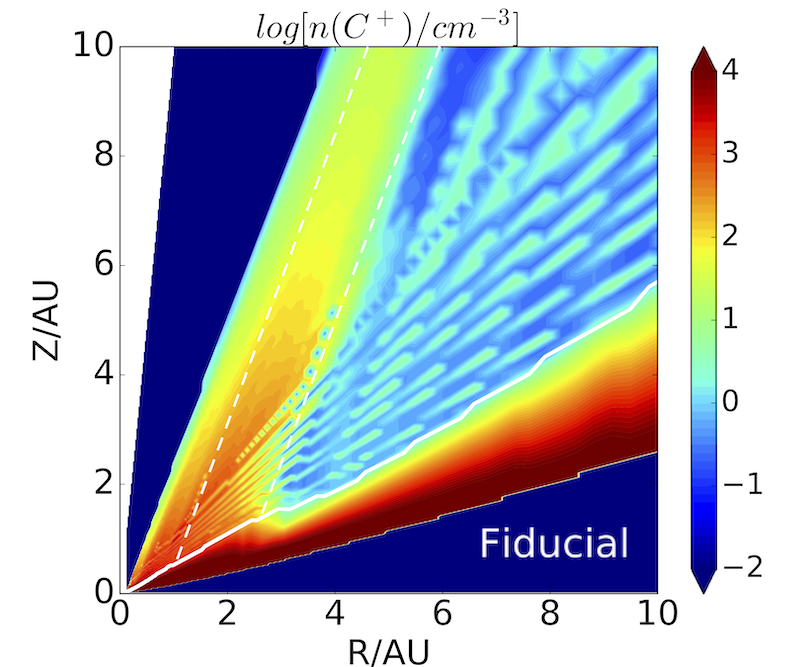}
\includegraphics[width=0.33\textwidth]{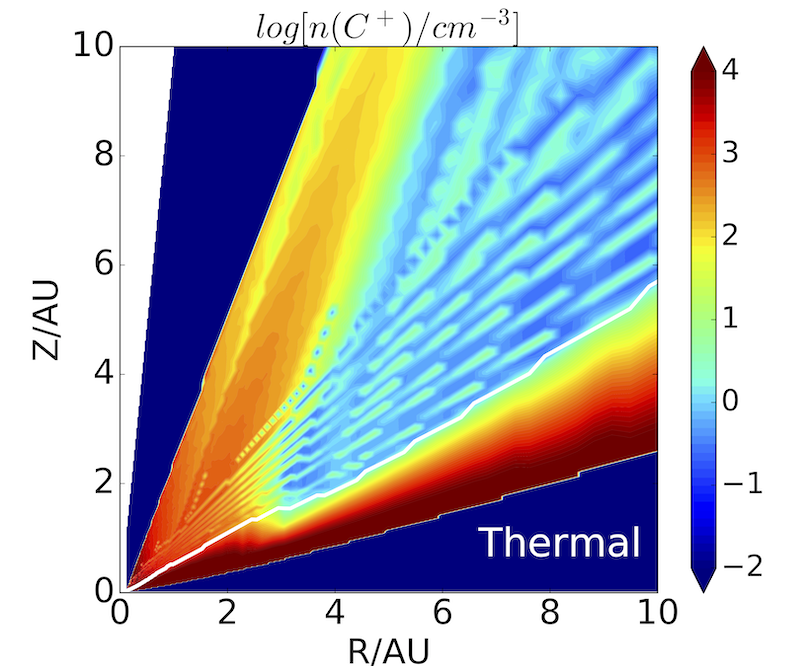}
\includegraphics[width=0.33\textwidth]{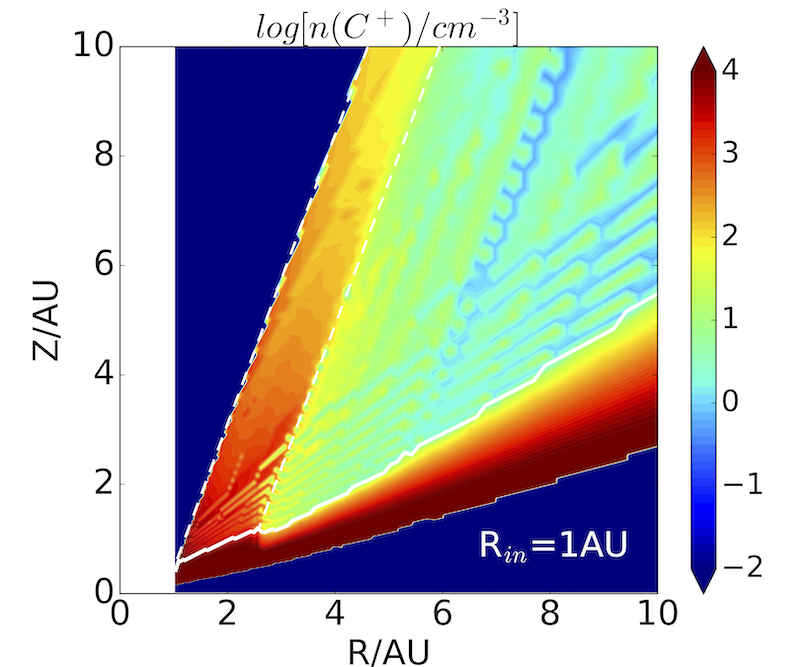}
\includegraphics[width=0.33\textwidth]{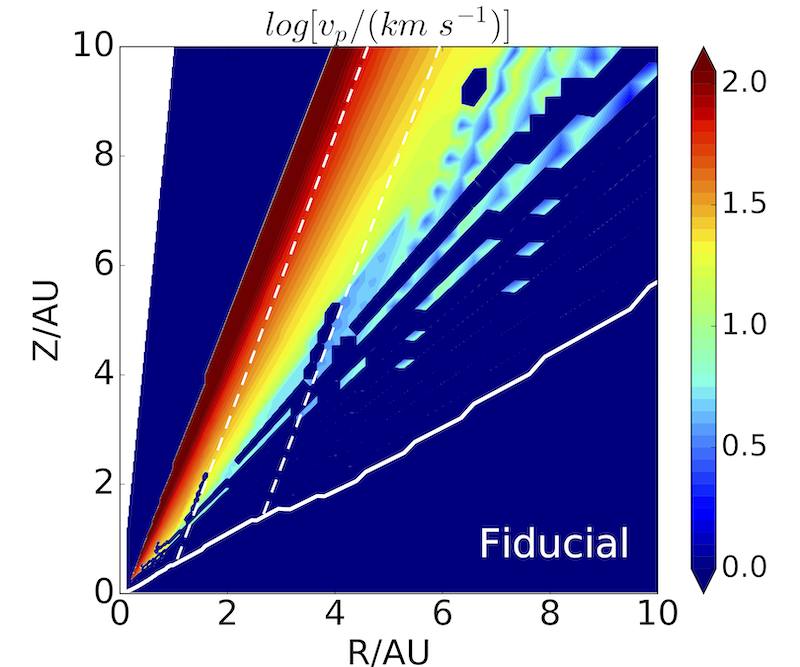}
\includegraphics[width=0.33\textwidth]{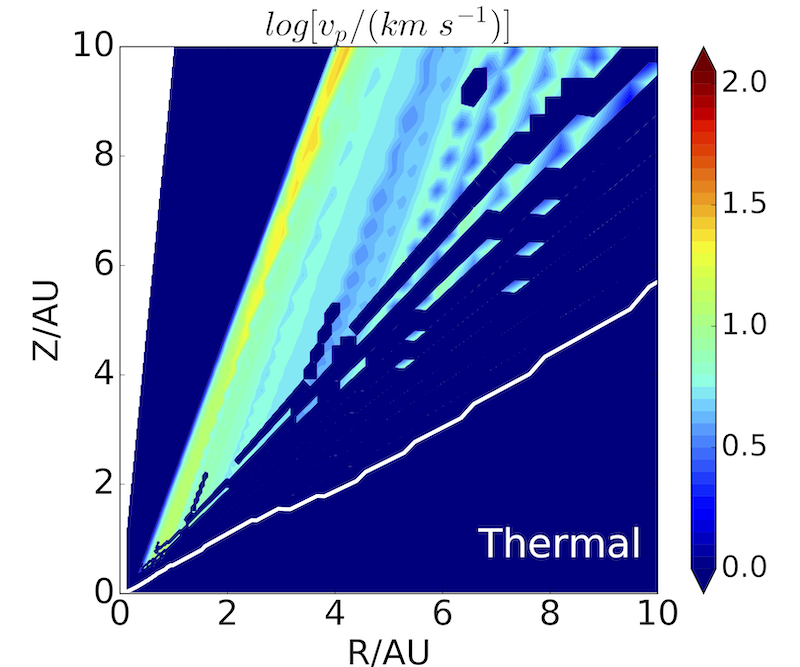}
\includegraphics[width=0.33\textwidth]{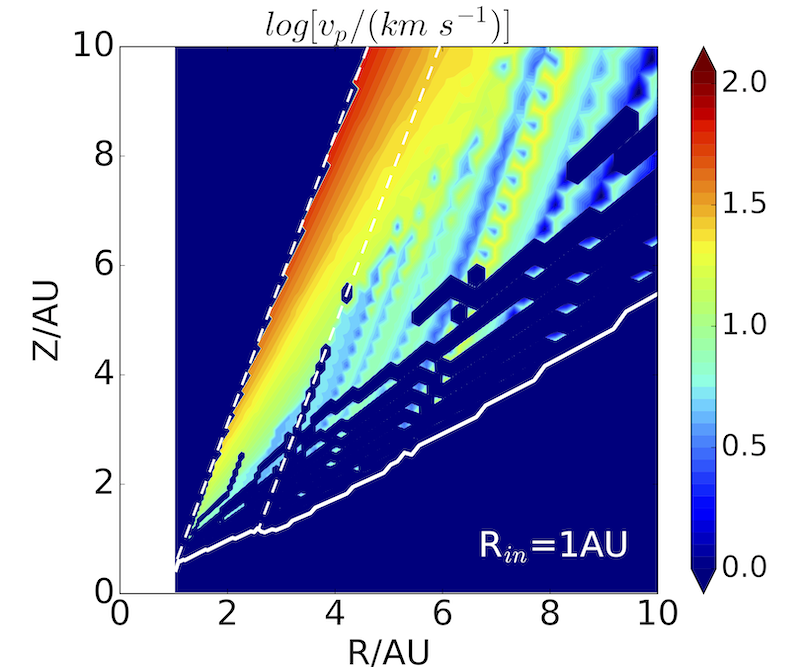}
\caption{\label{fig:modeldiskinner} C$^+$ number density (top row) and poloidal wind velocity (bottom row) structure in the innermost region for the fiducial model, the pure thermal model, and the magnetized model with an inner cavity at 1 AU. The white solid line marks the flow base, which is slightly higher than the layer of $A_V=1$, and marks the boundary between the disk atmosphere and the wind. The white dashed lines in left and right panels mark the location of the dense flow generated in the model with an inner cavity, for comparison to the fiducial model.}
\end{figure*}

%%%%%%%%%%%%%%%%%%%%%%%
% WIND INTERPRETATION
%%%%%%%%%%%%%%%%%%%%%%%
\section{Interpretation of wind components} \label{sec:analysis}

As presented in \S \ref{sec:c2overview}, blueshifted wind absorption features in the \ion{C}{2} doublet are detected and measured in a vast majority (36 of 40) targets in our sample. In this section, we present an analysis of wind properties based on the measurements from \ion{C}{2} absorption features.

%%%
\subsection{Wind velocity}

A summary of wind velocity measurements is presented in Table \ref{tab:fastwind} and \ref{tab:narrow}.
Figure \ref{fig:windvel} shows the wind velocities measured from \ion{C}{2} absorption as a function of inclination of the outer disk, where available from ALMA and SMA studies. Although we visually classify absorption as fast or slow wind, {these different types of wind absorption appear to be part of the same trend with inclination.} Fast winds have {maximum velocities $v_{max}$ (Table \ref{tab:fastwind})} ranging from 100 to 600 \kms, while {maximum velocities $v_{max}$ of slow winds (Table \ref{tab:narrow}) range from 50 to 150 \kms}\ (and ISM-like absorption occur within $v_{cent}$ of 20~\kms, see also the right panel of Figure \ref{fig:windvel}).

All disks in our sample likely have fast and slow winds; the four {disks without significant wind detection} are likely caused by low S/N and complicated line profiles. {Estimating the emission line profile prior to absorption using Gaussian fitting to the red side of the profile, as we will present in \S \ref{sec:model} for model-predicted line profiles, would suggest wind absorption for these four targets, but the imprecision of such a Gaussian fit precludes further analysis and measurements of wind properties, including wind velocities.}

The component that we detect depends on the inclination. The fast wind absorption tends to appear at low inclinations, likely tracing the highly collimated wind/jet launched near the star; while slow wind absorption appears for objects with disks viewed at higher inclinations. This slow wind absorption likely traces the same fast collimated wind/jet, but projected to a lower velocity. Slow wind absorption also appears at intermediate inclinations, likely tracing a slower wind launched at larger radii, and is not sensitive to the wind opening angle. A further interpretation of the wind velocity by disk/wind models is presented in \S \ref{sec:model}.

%%%
\subsection{Column Density Estimate of the Slow Wind} \label{subsec:columndensity}

The equivalent widths for narrow slow and ISM-like absorption profiles are shown in the left panel of Figure \ref{fig:narrowplots}, for targets with reliable disk inclinations from previous ALMA/SMA observations.  Slow wind absorption is detected in more edge-on disks and has equivalent widths that range from 30 to 200 \kms. ISM-like absorption is detected towards disks with low inclinations, with equivalent widths that range from 5 to 30 \kms.   Within each category, the equivalent width is independent of disk inclinations, which is consistent with the expectations for optically thick absorption lines. The ranges of equivalent width for slow wind and ISM-like absorption features is overplotted with the curves of growth of the \ion{C}{2} line at 1335.71 in the right panel of Figure \ref{fig:narrowplots}, assuming Doppler parameter $b$ of 3, 10 and 30 \kms.

Most equivalent widths for the slow wind
absorptions are located in the flat portion of the curves of growth, suggesting a wide range
of column density from $\sim 10^{14}$ to $\sim 10^{18}$~cm$^{-2}$. Furthermore, the FWHM of the
absorption lines are comparable to the equivalent widths for most of the targets, ranging from
30 to more than 200 \kms. One of the possible explanations for this relatively high line width is that the absorption features are the mixture of components with different velocities  (e.g. wind absorption blended with ISM absorption, as is demonstrated in \S \ref{subsec:measurenarrow}, also see further discussion in \S \ref{subsec:MgII}) instead of a single component. The effective line width is then broader than the intrinsic Doppler $b$ parameter, leading to additional uncertainties in column density estimation.

%%%
\subsection{Wind velocity versus accretion rate}

If a high accretion rate leads to a smaller disk truncation radius, then the fast wind may be launched closer to the star.  Higher accretion rates should also correspond to higher mass loss rates. In principle, these properties could increase the launch velocity and optical depth of the fast wind.

Figure \ref{fig:mdot_v} shows that the wind velocity measured from \ion{C}{2} (highest is 565 \kms) is not significantly correlated with accretion rate, with
{a Spearman coefficient of 0.32, and a p-value of 0.056.} A weak correlation may be present for the fast wind, with a Spearman coefficient of 0.56, and a p-value of 0.013.\footnote{For the slow wind, the Spearman coefficients is -0.46, and p-value is 0.069. The weak negative correlation is likely due to the scatter in slow wind velocities; excluding RECX 15 and CW Tau brings down the Spearman coefficients for the slow wind to -0.30, with a p-value of 0.298.} 
The velocity of absorption in \ion{He}{1} $\lambda10830$ is also not correlated with accretion rate \citep{edwards06}, {although \citet{Lopez-Martinez15} measured correlations between line broadening and both wind terminal velocity and accretion rate in \ion{Mg}{2 }. } At low accretion rates, PDS 66 and TW Hya both have strong absorption at high velocities.
While weak correlations may be present in velocity versus accretion rate, we would require much higher significance to claim a statistical correlation because of confounding variables (inclination, stellar mass), uncertainties in classification, and the influence of outliers.
Indeed, despite the lack of a significant overall correlation, the wind velocity is fastest for HD 104237, DR Tau and RU Lup, three stars in our sample with high accretion rates.

In contrast to the absorption, the presence and velocity of the high velocity component in \ion{O}{1} emission both correlate with accretion rate \citep[e.g.][]{banzatti19}.  Stars with low accretion rates, such as TW Hya and DN Tau, lack any detectable signature of a high velocity component in [\ion{O}{1}] emission, despite the presence of fast wind in absorption.
These two stars may have what could be interpreted as jets, but with low enough mass loss rates and densities that the [\ion{O}{1}] emission is not detectable.

The lack of correlation between wind velocity and accretion rate is evaluated here using the observed velocities.  We also find a lack of any correlation if the velocities are deprojected for inclination.  Whether a  deprojection is appropriate depends on the wind geometry.

%%%%%%%%%%%
% MODELS
%%%%%%%%%%%
\section{Modeling \ion{C}{2} absorption in disk winds} \label{sec:model}

In order to better interpret the origins of the disk wind absorption, we develop disk wind models to compare the observed and expected absorption line profiles. The wind models are built on the semi-analytical photoevaporative model of \citet{gorti09}, which describes how irradiation of the disk leads to photoevaporation of the disk surface.  We modify this model by using simplified chemistry and by adding the effect of magnetic fields to generate a thermal-magnetic wind. The absorption line profiles are calculated for different disk inclinations, which changes our line of sight through the wind. Our approach is somewhat simplistic and designed to qualitatively interpret line profiles to understand general wind properties. The models are designed to be informative and are not intended for detailed line fitting.

In \S \ref{windmodel}, we briefly introduce our fiducial disk/wind model. In \S\ref{windwhere}, we measure in the models where the absorption occurs along our line of sight to the emitting gas.  In \S\ref{thermormag}, the velocity of the absorption profile is used to demonstrate that both the slow and the fast winds are magnetized. 

%%%
\subsection{The fiducial disk/wind model}
\label{windmodel}

The fiducial disk/wind model describes the physical and chemical structure of the disk and the wind.
This model is based on a steady-state semi-analytical model of a photoevaporative wind generated by FUV and X-ray irradiation of the disk by the central star \citep{gorti09}, with a simplified approach for disk chemistry following \citet{gorti15}. The primary input parameters in the model are the stellar mass, FUV luminosity, X-ray luminosity, and the inner and outer disk radii (Table~\ref{tab:model}). The gas and dust are treated separately.

The fiducial disk/wind model solves for thermal balance and simple chemistry.   A vertical hydrostatic equilibrium is imposed to separately calculate the density and gas and dust temperatures as a function of spatial location in the disk.

% FIGURE MODEL COLUMN DENSITY

\begin{figure}[t]
\includegraphics[width=0.48\textwidth]{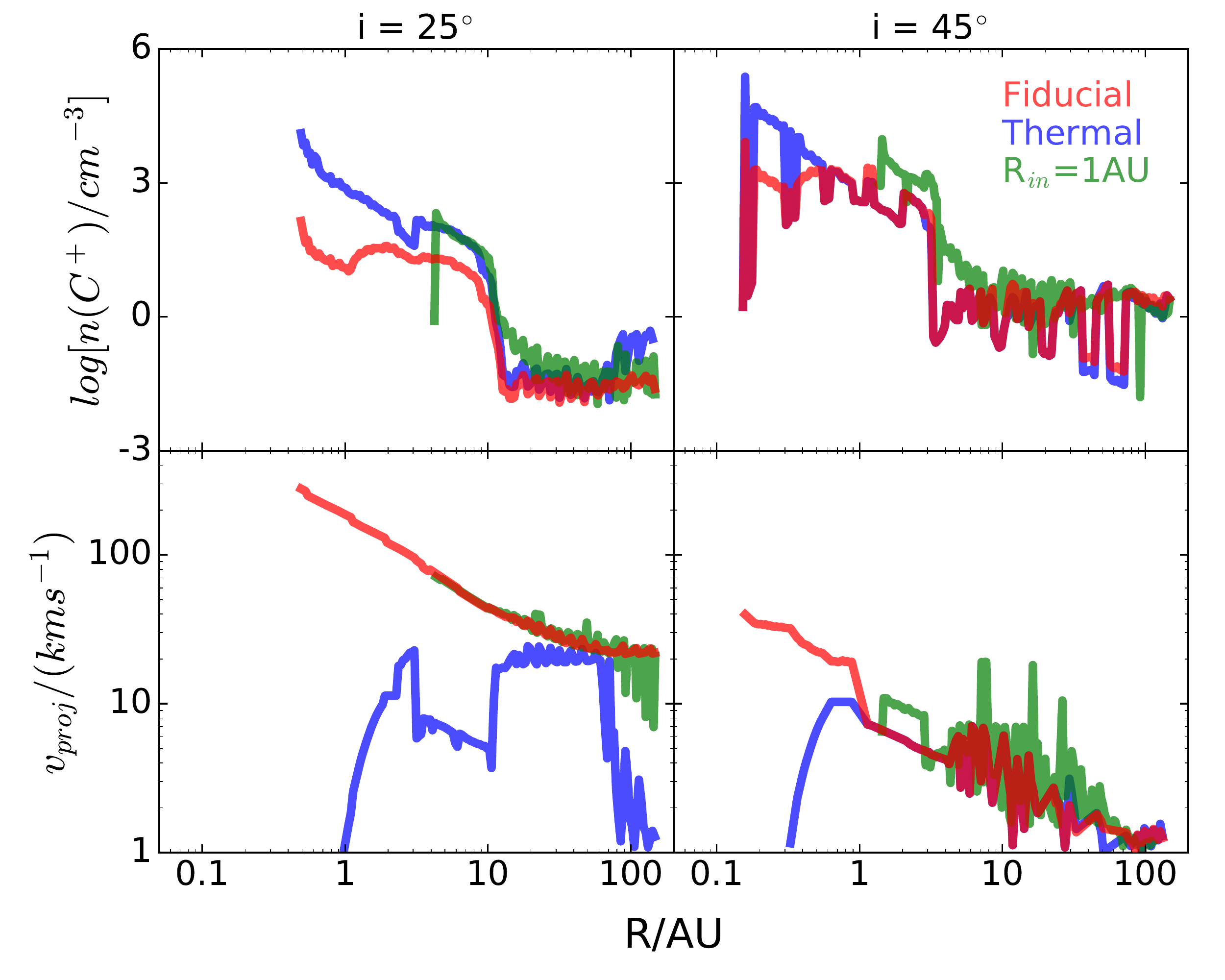}
\caption{\label{fig:ncol_r} Model-predicted radial profiles of the C$^+$ number density and the projected wind velocity along the line of sight, assuming disk inclinations of $i=25^{\circ}$ and $i=45^{\circ}$. The inner radii for the profiles are larger than the disk inner radii in the model, as the X-axis marks the radial location along the line of sight, which passes through the flow at larger $z$ and $r$ than at the flow base.}
\end{figure}

The gas is heated by X-rays, grain photoelectric heating, and collisions with warmer dust grains. Accretion heating is ignored for simplicity, as we are primarily interested in the disk evolution stage where the accretion rate is relatively low, and because accretion heating plays more important role in the midplane than in the disk upper layer. Gas cooling mainly occurs in resonant radiative transitions from atoms, ions and molecules, including Ly$\alpha$, H and He recombination, fine structure and metastable lines of dominant species of \ion{C}{1}, \ion{C}{2}, \ion{O}{1}, \ion{O}{2}, \ion{Si}{1}, \ion{Si}{2}, \ion{S}{1}, \ion{S}{2}, \ion{Fe}{1}, \ion{Fe}{2}, \ion{Ne}{2}, \ion{Ar}{2}, and CO rotational lines. Cooling due to collisions with dust grains is also considered when the gas is heated to temperatures higher than the dust. The ionization structure for these dominant species, and thus the gas cooling due to these lines, is determined by simple chemical assumptions, as described below.

The abundances of neutral carbon and C$^+$ are determined by considering carbon to be singly ionized if the photoionization rate of neutral carbon is higher than the C$^+$ recombination rate. The ionization structure for Fe, S and Si follows carbon ionization and are set to be singly ionized where carbon is in C$^+$. The Ne$^+$ and Ar$^+$ abundances are calculated following \citet{glassgold07} and \citet{gorti08}, where Ne and Ar are doubly ionized by X-rays and then recombine with electrons or transfer charge with H to form singly ionized species.
Hydrogen is considered atomic if the formation rate of H$_2$ (on dust grains or via three-body reactions) is lower than the destruction rate (by UV and X-ray photodissociation or by reactions with O or C$^+$).  For atomic hydrogen, the abundance of H$^+$ is calculated with the ionization rate from FUV and X-ray emission \citep{glassgold07} and the recombination rate of H$^{+}$. The (single) ionization fractions of He and O follows the ionization fraction of hydrogen. CO is present where H is atomic, and the abundance of neutral oxygen is set by assuming oxygen that is not in CO is in gas phase. The detailed gas-phase abundances of species adopted in the model are shown in Table~\ref{tab:model}.

The dust physics is calculated independently of the gas. We consider a range of grain size distribution and conduct dust radiative transfer with a modified two-layer approach first described by \citet{chiang97}.
The model then calculates the density and temperature of gas and dust as well as the ionization structure as a function of spatial location throughout the disk.

% FIGURE MODEL PROFILE

\begin{figure*}[t]
\plotone{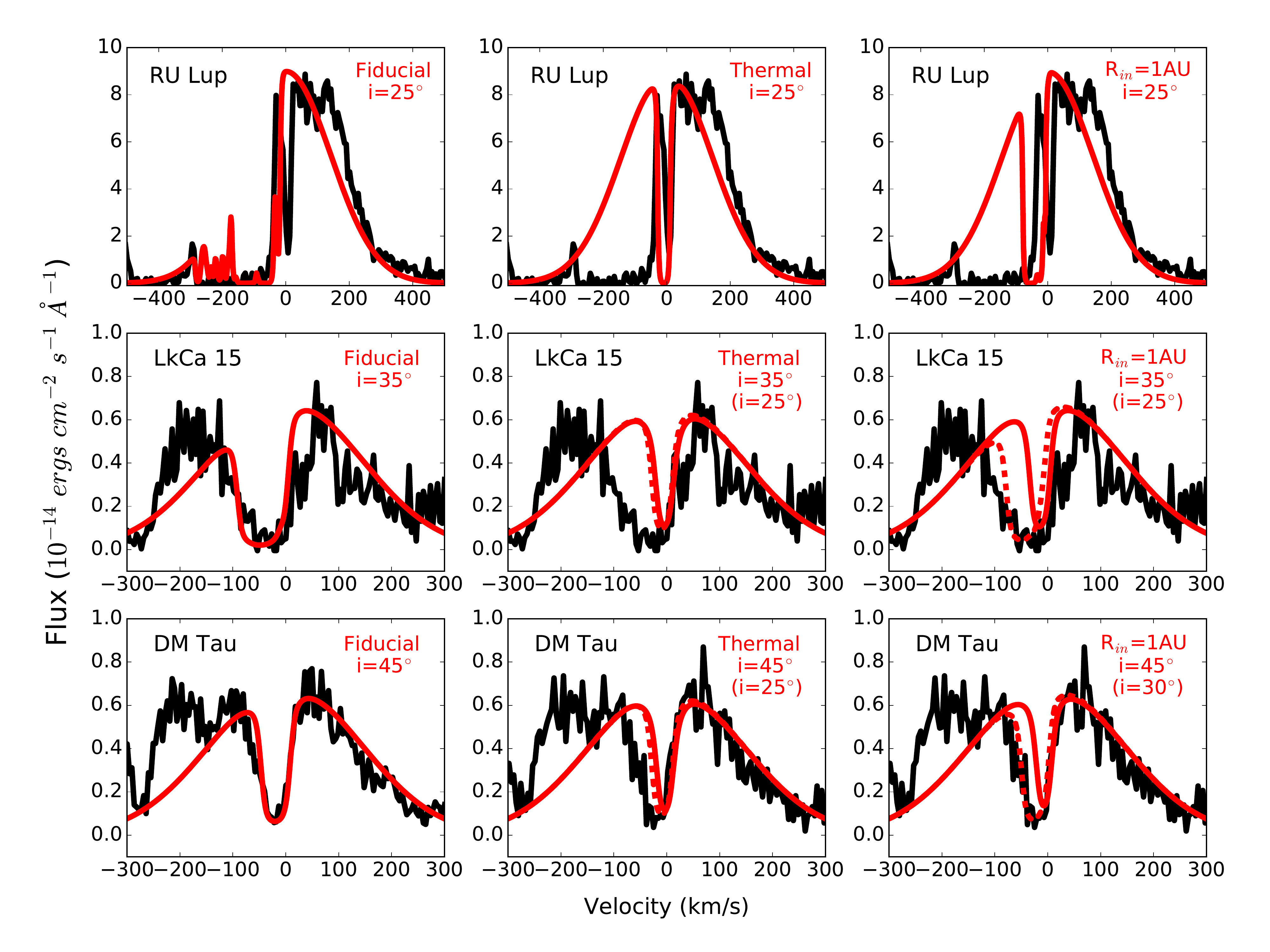}
\caption{\label{fig:modelprofile} Model-predicted absorption line profiles (red) overplotted on observations (black) of a fast wind case (RU Lup, top row), a slow wind case with relatively broad absorption (LkCa 15, middle row), and a slow wind case with one of the narrowest absorptions (DM Tau, bottom row). All of the three cases are well-explained by the fiducial model. Model-predicted line profiles at additional inclinations (red dashed lines, corresponding to inclinations in parentheses in the legend) show that slow wind absorptions can be explained by the magnetized model with inner cavity at higher inclinations. None of the winds can be explained by the pure thermal model.}
\end{figure*}

Fiducially, the disk/wind model is thermal-magnetic, where the wind is launched thermally at the flow base, setting the wind mass loss rate, but a magnetic torque is also considered once the wind is launched.
Photoevaporative flows are driven by gas thermal pressure gradients, with gas density and temperature distribution calculated in the model.
Following \citet{gorti15}, the potential photoevaporation mass loss rate $\dot{\Sigma}_{PE}$ at each spatial location is calculated slightly differently than \citet{gorti09}. The gas thermal speed and additional support by angular momentum of Keplerian rotation determine a critical radius $r_{crit}$ where the flow is unbound to the stellar gravity. Within $r_{crit}$, the flows are assumed to be launched subsonically, with Mach number dependent on the radial location in the disk (see eqn.~3 in \citealt{gorti15}). Beyond $r_{crit}$, the flow is assumed to be sonic. At a given radius, the model estimates the potential $\dot{\Sigma}_{PE}$ for each height $z$, and determines the flow base at the height where the potential $\dot{\Sigma}_{PE}$ is maximum. This flow base is slightly higher than the layer of $A_V=1$ and marks the boundary between the disk atmosphere and the wind. 

The flow structure is then solved assuming streamlines with a fixed opening angle of 70$^{\circ}$ relative to the disk midplane, and we assume the streamlines in the model to also follow the magnetic field lines.  
An opening angle of less than 60$^{\circ}$ is conventionally stated for launching a magneto-centrifugal wind \citep{blandford1982} and applies to a cold wind without the effect of thermal pressure. For a thermal-magnetic wind, there is no distinct critical angle for wind launching.  Larger opening angles relative to the disk correspond to larger temperature gradient along the field line, which helps overcome the gravitational potential and generates more effective wind acceleration.
The poloidal magnetic field is assumed to be strong enough near the flow base that it forces the flow to corotate with the flow base and accelerates the wind centrifugally up to the Alfv\'{e}n point. Beyond the Alfv\'{e}n point, $v_{\phi}$ is calculated assuming angular momentum conservation throughout the streamline. The poloidal flow velocity $v_p$ in the model is determined by solving for the energy conservation in a frame corotating with the field lines.
The Alfv\'{e}n radii $R_A$ adopted in the model is prescribed according to the radial profile of $R_A$ calculated from the fiducial model of \citet{bai2016}. The Alfv\'{e}n radius is normalized to the wind launching radius $R_0$ and is $R_A\sim 10~R_0$ in the inner 0.1 AU of the disk and $R_A \sim 1.6~R_0$ at tens of AU in the disk.

{The streamlines in our model have an opening angle of 70$^\circ$.  A disk viewed pole-on would therefore not have any wind in our line of sight.  Since stars with face-on disks, such as DR Tau and TW Hya, have absorption from a fast wind, the fast wind from the inner disk must be well collimated.  Our assumption in this section is that the structure of this well-collimated wind is approximated by the structure of the innermost stream line.}

The disk/wind model developed here is thermal-magnetic, with input parameters presented in Table \ref{tab:model}. For comparisons, we also calculate a thermal-magnetic model with a disk inner cavity of $R_{in}=$ 1~AU and a pure thermal wind model. We will discuss these models in \S \ref{windwhere} and \S \ref{thermormag}.

%%%
\subsection{Where in the wind does the absorption occur?}
\label{windwhere}

The mass loss in the models is concentrated in the innermost $\sim 2$ AU, as shown in Figure \ref{fig:modeldiskinner}.
The concentration of the wind in the inner disk is also seen in the radial profile of C$^{+}$ number density along different lines of sight (Figure \ref{fig:ncol_r}). For a moderately inclined disk of $i=45^{\circ}$, the line of sight passes through the flow base, and the C$^{+}$ number density drops off beyond 3~AU. For a low disk inclination of $i=25^{\circ}$, the C$^{+}$ number density drops off at a larger radius of $\sim 10$ AU, because the line of sight passes through the flow at much larger $z$ and $r$ than the flow base.  For both of the cases, the projected wind velocity is higher towards the innermost regions where the mass loss is concentrated, due to the higher Keplerian velocity at the flow base and the stronger magnetic field to accelerate the wind.

% FIGURE MODEL VELOCITY

\begin{figure*}[t]
\plottwo{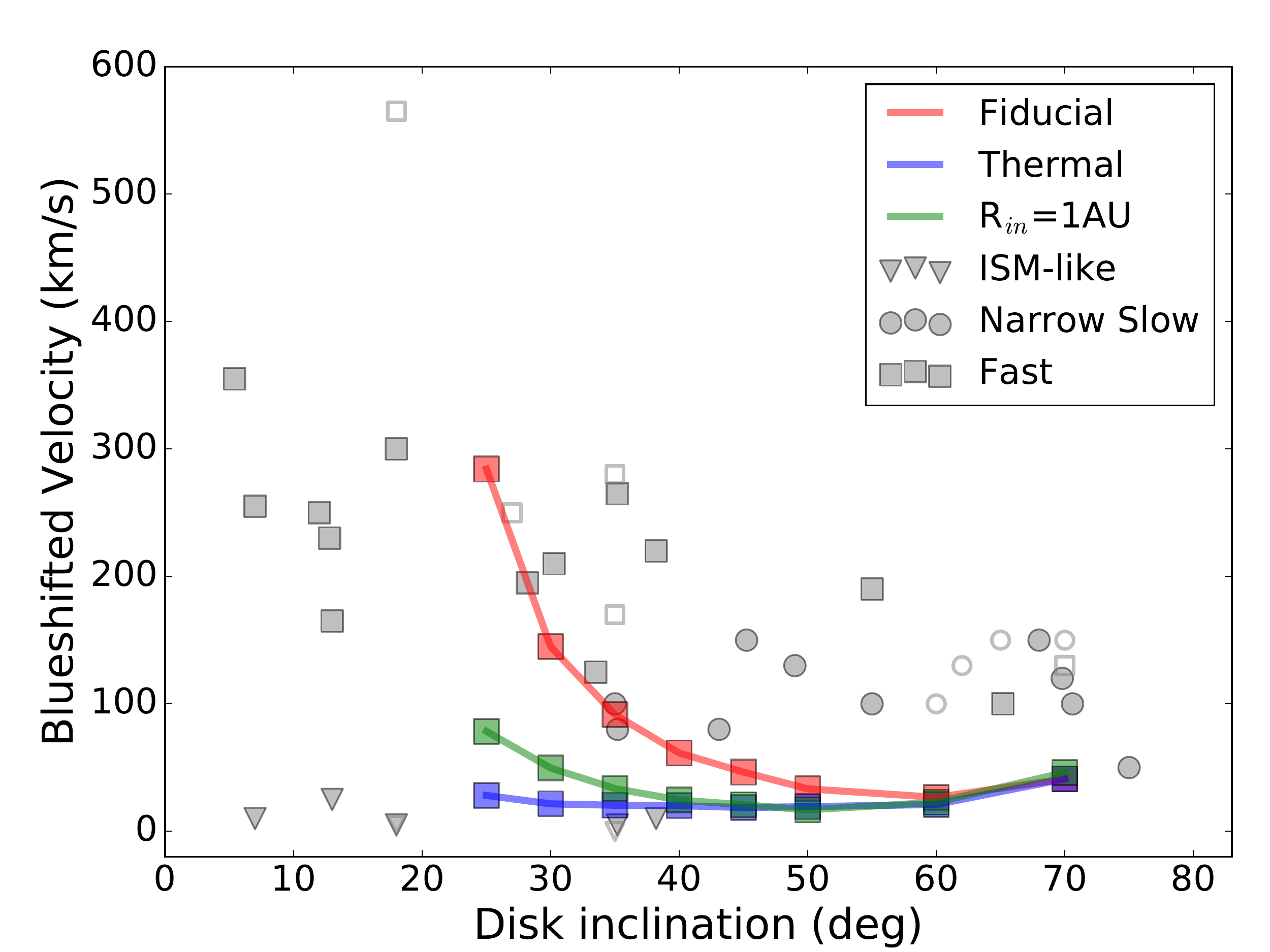}{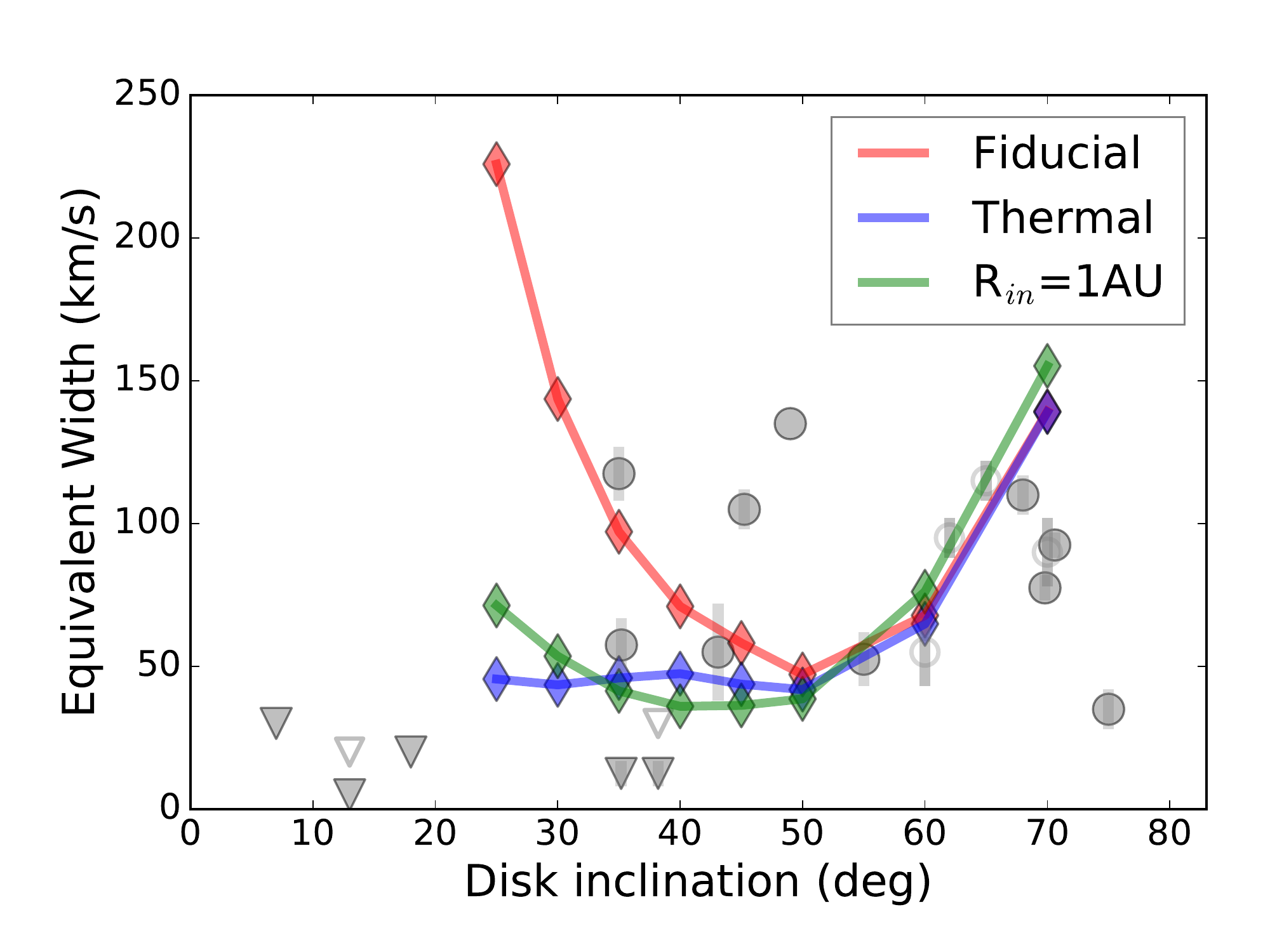}
\caption{\label{fig:modelv} Colored lines and markers show the model-predicted wind properties in \ion{C}{2} as a function of disk inclination. {The wind velocity is defined as the maximum wind velocity for optical depth $\tau>1$  (squares).} Left panel: wind velocity as a function of disk inclination.  The wind velocities measured from the observations are over-plotted for comparison, same as Figure \ref{fig:windvel}, but in grey markers. Right: the equivalent width of wind absorption as a function of disk inclination. The grey markers show the equivalent widths that are measured from the observation, as is presented in the left panel of Figure \ref{fig:narrowplots}.}
\end{figure*}

In order to further investigate the importance of the innermost regions, we compute a model with a disk inner cavity of $R_{in}=$1~AU for comparison. Overall, the model with an inner cavity gives flow structures similar to the fiducial model, with the absence of the innermost region, where the fastest wind is generated in the fiducial model (Figure \ref{fig:ncol_r}). The model with an inner cavity gives a slightly higher mass loss rate of 4$\times 10^{-9}~M_{\odot}~yr^{-1}$, compared to 2.5$\times 10^{-9}~M_{\odot}~yr^{-1}$ from the fiducial model. The enhancement in mass loss rate is likely due to the inner rim of the disk being directly exposed to the stellar irradiation at a larger radius, leading to more efficient evaporation. This enhancement in mass loss is reflected in Figure \ref{fig:modeldiskinner}, where in the presence of the inner cavity, the flow generated at $\sim$2~AU (between white dashed lines) has higher density (and similar velocity) compared to the fiducial model. In contrast, in the fiducial model the wind generated in the innermost 1~AU attenuates the incident photons and shadows the outer regions, leading to lower density in the flow region $\gtrsim$ 2~AU. This effect of mass loss rate enhancement in the presence of an inner cavity relies strongly on the fact that the wind is thermally-launched in the model and may not apply with magnetically driven wind considered. 

Figure~\ref{fig:modelprofile} compares the model-predicted line profiles with observations. {The line profiles are generated by the combination of accretion-related \ion{C}{2} emission from the stellar surface and the disk wind absorption. The emission is represented by a Gaussian profile with parameters obtained by fitting the profile to the red side of the observed line. This Gaussian profile is meant to guide the interpretation and show the general consistency of the wind model, instead of providing a robust fit to the emission line component of the profile. The wind absorption profile is then calculated against the Gaussian emission profile} by deploying an intrinsic Voigt profile of optical depth as a function of velocity ($\tau_v$) at each point considering thermal and collisional broadening, summing up $\tau_v$ along the line of sight to the center of the star, and multiply $e^{-\tau_v}$ with the Gaussian emission profile. The line profiles are convolved with the local (COS or STIS) line spread functions. The fast wind absorption in RU Lup is consistent with the fiducial model at a low inclination (25$^{\circ}$, nearly aligned with the streamlines in the models), while the model with an inner cavity predicts a much slower wind in absorption. Similarly, for LkCa 15 with $i=35^{\circ}$, the slow wind absorption is reasonably consistent with the fiducial model, while the model with an inner cavity produces a much slower and narrower absorption. Even if the inner disk is inclined and has $i=25^{\circ}$, the absorption profile in the model with cavity is still narrower than the observation. For DM Tau, where one of the narrowest and slowest winds in our sample is detected, the line profile is consistent with both the fiducial model at $i=45^{\circ}$ and the model with a 1 AU cavity at $i=30^{\circ}$.

Figure \ref{fig:modelv} generalizes this comparison. As the inclination increases, the models with and without the innermost regions produce more similar wind velocities and equivalent widths because the line-of-sight passes through the flow base.
The fast wind velocity from observations is generally comparable to that predicted by the fiducial model, but much faster than the model with a cavity. The observed slow wind is overall slightly faster than both of the magnetized models (with or without a cavity) at corresponding inclinations, but the wind velocity can be produced by both of the models at lower inclinations ($i<45^{\circ}$). However, at $i<45^{\circ}$, the equivalent widths predicted by the model with a cavity only cover the narrowest winds in our sample (e.g. DM Tau). For slow winds with broader absorptions (e.g. LkCa 15), only the fiducial model explains both the wind velocity and the equivalent width.

The comparison of models and observations indicates that the fast wind absorption traces the innermost disk regions (within 1 AU in our model). Most of the slow wind absorptions more likely trace the projected fast wind seen at high inclination from the innermost regions, while the narrowest slow wind absorption may indicate the presence of  a cavity in the inner disk. 

Although the wind is generated thermally, following \citet{gorti09}, our modified wind model generates winds from regions much closer to the star than in \citet{gorti09}. The wind launching at the innermost regions can be explained by the lack of coolants due to simplified chemistry, which leads to an over-heating of the inner disk. In addition, the flow structure in our model is simplified by assuming streamlines, and the wind properties measured at different inclinations in the model suggest that the absorption line profile is sensitive to geometry. Despite these limitations, our simplified model captures the essential physics in thermal-magnetic winds. In particular, the mass loss at the innermost regions in our model is qualitatively consistent with the expectation of a fast wind (and jet) from the inner disk\footnote{For the fastest absorptions observed at the lowest inclinations (e.g. DR Tau) that would likely require a jet intersecting the star along a nearly pole-on line of sight, the absorptions can be interpreted as a disk wind extended to within the $R_{in}$ of our model, with flow structure similar to the pole-launched jet.}, generated by the strong magnetic field. Furthermore, the comparison of the fiducial model and the model with an inner cavity shows that the absorption lines are sensitive to the inner disk and can potentially provide information for the radial distribution of wind mass loss rate.  

%%%
\subsection{Thermal or Magnetized Wind?}
\label{thermormag}

Figure \ref{fig:modeldiskinner} shows that the magnetized wind is much faster than the pure thermal wind. As the wind accelerates, the flow in the fiducial model is more diffuse than in the pure thermal model. The differences of the flow structures in thermal and magnetized winds are also reflected in Figure \ref{fig:ncol_r}, where a purely thermal model predicts a denser innermost region. In addition, the maximum wind velocity along the line of sight is $\sim$20~\kms\ in the pure thermal model, which is much slower than that in the fiducial magnetized wind model.

The line profile comparisons indicate that both the slow and the fast winds are thermal-magnetized (Figure~\ref{fig:modelprofile}). The pure thermal model predicts wind absorption that is much slower and narrower than the observations, even for DM Tau, one the narrowest slow winds. Figure \ref{fig:modelv} shows further comparisons of wind properties, where the thermal model predicts wind velocities that are generally lower than the fast and slow winds detected in our sample. At lower inclinations where the line of sight mainly passes through the wind, our pure thermal model predicts equivalent widths that are lower than the observations. Both the fiducial and the pure thermal models predict higher equivalent widths as the inclination increases.  The line of sight goes completely under the wind base at $i>=70^{\circ}$, where the absorption line is broadened by the high column density.

The right panel of Figure \ref{fig:narrowplots} compares the curve of growth with the equivalent width for different inclinations in the models. The predicted equivalent widths match the observations of slow wind absorption reasonably well, with the caveat that the observed equivalent widths have a lot of scatter. At intermediate inclinations corresponding to slow winds, the column densities in the line of sight are $\sim 10^{17}$ cm$^{-2}$. At low inclinations, the equivalent widths and column densities remains mostly the same in the pure thermal model, while the fiducial model predicts broader absorption with lower column densities due to the velocity gradient along the line of sight. At high inclinations, all of the models predict line broadening towards higher column densities, consistent with the damped portion in the curve of growth.  In order to further distinguish the line broadening mechanisms of the narrow slow components, higher resolution spectroscopy is needed to resolve the velocity structure in the wind absorption, and to investigate the relative contributions of thermal and magnetized winds.

{The conclusion that the fast and slow absorption features are both produced by a thermal-magnetized wind does not necessarily apply to interpretations of the low-velocity component of optical forbidden line emission, although \citep{whelan21} reach a similar conclusion for [O I].  The \ion{O}{1} emission is likely emitted at larger radii than is traced by the slow wind absorption.}

%%%%%%%%%%%%%%%
% CONCLUSIONS
%%%%%%%%%%%%%%%
\section{Conclusions and Perspectives} \label{sec:conclusions}
In this paper, we survey \ion{C}{2} absorption in 40 classical T Tauri stars, of which 36 show wind absorption in either a fast or a slow component.  The four non-detections are ambiguous and do not indicate that a wind is absent.  We find the following results:

\begin{itemize}
  \item For the fast wind, the velocity decreases with disk inclination, consistent with a collimated jet.

  \item The slow wind absorptions are detected mostly in targets with intermediate or high inclinations, without significant dependency of wind velocity or equivalent width on disk inclination.

  \item Wind absorption in both the slow and fast components is preferentially detected in lines of neutral or singly ionized atoms, is rarely detected in \ion{Si}{3}, and is only detected in \ion{C}{4} in one case.  Wind absorption in \ion{Mg}{2} is consistent with that in \ion{C}{2}.

  \item The fast and slow winds are both consistent with a magnetized wind, which may be enhanced by photoevaporation, and are inconsistent with expectations from a purely thermal wind.
\end{itemize}

These results lead to a morphological picture consistent with that developed for \ion{He}{1} $\lambda 10830$ absorption by \citet{edwards06}.  The strength of the \ion{He}{1} line is in the high S/N at high spectral resolution, thereby providing a much more powerful diagnostic of the wind dynamics.  In contrast, the FUV lines have an advantage in diagnosing the ionization levels in the wind.

In the disk models, the wind absorption occurs close to star, providing a diagnostic of the launch region, while emission in [\ion{O}{1}] and other optical forbidden lines should trace the wind in large volumes.  The optical forbidden line emission traces the density and mass in the wind, so some components can be absent; the absorption traces line-of-sight components and indicate that the fast (high-velocity) and slow (low-velocity) winds are universal, with detections that depend only on the geometry.

At this stage, the use of far-ultraviolet lines as a probe of mass loss rate is limited by the  optical depth in most detected lines.  The  excited \ion{N}{1} $\lambda1243,1492,1494$ lines may provide an optically thin diagnostic with deep spectra.  A comprehensive evaluation of mass loss should account for both absorption and emission components, including non-detections of different species.  Future investigations will be facilitated by the FUV spectral survey of classical T Tauri stars in the ULLYSES DDT Program \citep[PI Roman Duval;][]{roman20,proffitt21} the accompanying archival program ODYSSEUS (HST AR 16129, PI Herczeg and Co-PI Espaillat), and the ODYSSEUS archival program (HST AR 16129), and the PENELLOPE VLT Large Program \citep{manara21}.

\acknowledgements{We thank Suzan Edwards for her early support of this project and energetic and constructive guidance.  We thank the referee for insightful comments that improved the quality of the paper.  We also appreciate valuable discussions with Uma Gorti, Ilaria Pascucci, and Xuening Bai for various incarnations of the paper.   We thank the DAO of Tao team (HST GO-11616) and other teams for obtaining the data that was used for this paper.  

ZX and GJH acknowledge the support of National Key R\&D Program of China No. 2019YFA0405100 and NSFC project 11773002.  This work is based on observations made with the NASA/ESA Hubble Space Telescope, obtained from the data archive at the Space Telescope Science Institute. STScI is operated by the Association of Universities for Research in Astronomy, Inc. under NASA contract NAS 5-26555.  These observations are associated with programs HST-GO 8041, 8157, 8206, 11533, 11616, 11828, 12036, 12996, 13032, 14604, 14469, 15070, and 15128.
}

\bibliographystyle{apj}
\bibliography{ms}

\clearpage
\begin{appendix}

%%%%%%%%%%%%%%%%
\section{Tables and Figures}
\setcounter{table}{0}
\renewcommand\thetable{\Alph{section}.\arabic{table}}
\setcounter{figure}{0}
\renewcommand\thefigure{\Alph{section}.\arabic{figure}}

% TABLE TARGET PROPERTIES
\begin{table*}[t]
\footnotesize
    \centering
    \numberwithin{table}{section}
     \caption{Properties of Targets}
    \label{tab:targprops}
    \begin{tabular}{lccccccccccc}
    Star&distance& SpT &$T_{{\rm eff}}$&$A_V$&$L_{\star}$&$L_{acc}$&$R_{\star}$&$M_{\star}$&$\dot{M}_{acc}$ & Incl. &Reference  \\
        &(pc)&   &(K)&(mag) &($L_{\odot}$)&($L_{\odot}$)&($R_{\odot}$)&($M_{\odot}$)&($M_{\odot}~yr^{-1}$) & (deg) &   \\
    \hline
    AA Tau &137&K7&4020&1.9&0.96&0.139&2.02&0.63&1.76E-08& 75&1,28\\
    AK Sco AB    &141& F5 &6500&1.06& 26.3 & 3.8 & -- & --& -- & 70.6 &4,27\\
    BP Tau     &129&K7&4020&1.1&0.85&0.238&1.9&0.65&2.74E-08&38.2&1,18\\
   CS Cha     &176&K2&4710&0.8&1.75&0.054&1.99&1.32&3.21E-09&60$^{d}$&1,36\\
    CV Cha     &193&K0&4870&1&2.55&1.63&2.24&1.52&9.47E-08&35$^{d}$&1,37\\
    CW Tau     &132&K3&4545&2.4&0.45&0.042&1.08&0.99&1.81E-09& 65$^{d}$&5,6,29\\
    CY Tau     &129&M1&3720&0.32&0.39&0.034&1.5&0.54&3.72E-09& 27&7,23\\
    DE Tau     &127&M2&3560&0.9&0.66&0.097&2.14&0.38&2.15E-08&35$^{d}$&1,34\\
    DF Tau      &124&M2&3560&0.6&1.14&0.143&2.8&0.33&4.78E-08&-&8\\
    DK Tau      &130$^a$&K7&4020&1.3&0.86&0.2002&1.91&0.65&2.32E-08&12.8&1,18\\
    DM Tau      &145&M2&3560&1.1&0.39&0.0359&1.64&0.42&5.53E-09&35.2&2,26\\
    DN Tau    &128&M0&3900&0.9&1.25&0.045&2.45&0.5&8.69E-09&35.18&1,25\\
    DR Tau   &196&K5&4210&1.4&0.78&2.11&1.66&0.85&1.62E-07&5.4&1,18\\
    DS Tau   &159&K5&4210&0.34&0.74&0.27&1.62&0.87&1.98E-08&65.19&7,25\\
   GM Aur      &160&K5&4210&0.6&1.29&0.13&2.14&0.75&1.46E-08&55&2,20\\
    HD 104237   &108& A7 &8000&0&22.7&4.42&2.48&2.01&2.15E-07&18$^{d}$&9,38\\
    HD 135344b    &136& F8 & 6375&0.23&6.7&0.86&2.12&1.58&4.55E-08&12&9,20\\
    HD 142527   &157& F6 & 6550&0.78&20.5&6.12&3.52&2.22&3.82E-07&27$^{d}$&10,32\\
    HD 169142   &114& &7500&0&4.69&0.135&1.28&1.69&4.03E-9& 13&4,21,40\\
    HN Tau   &144$^a$&K5&4210&1.1&1.27&0.276&2.12&0.75&3.08E-08&69.8&1,18\\
    IP Tau    &131&M0&3900&1.1&0.61&0.0505&1.71&0.62&5.49E-09&45.24&1,25\\
    KK Oph   &194$^a$& A6 &8500&2.7&4.21&0.248&0.95&2.59&3.59E-09& 70$^{d}$&9,33\\
    LkCa 15    &159&K5&4210&1.1&1.03&0.0697&1.91&0.79&6.64E-09&49&1,20\\
    PDS 66   &99&K1&4790&0.2&1.19&0.0037&1.58&1.32&1.75E-10& 30.26&1,24\\
    RECX 11     &99&K6&4105&0&0.484&0.0032&1.38&0.86&2.02E-10& 70$^{d}$&1,3,35\\
    RECX 15 &99$^b$&M3.25&3355&0&0.077&0.00705&0.82&0.3&7.60E-10& 60$^{d}$&1,3,35\\
     RR Tau     &779& A0 &10000&1.58&107& 203 &3.45&5.8&4.80E-06&-&11\\
    RU Lup     &160&K7&4020&0&1.49&0.57&2.52&0.56&1.01E-07& 18&12,22\\
    RW Aur A  &163$^a$&K3&4545&0.5&0.68&0.563&1.33&1.12&2.64E-08&55.1&1,18\\
    RY Lup    &159&K2&4710&0.4&1.86&0.141&2.05&1.32&8.63E-09& 68&13,22\\
    SU Aur     &158& G4 &5945&0.9&8.8&0.127&2.8&2&7.01E-09& 62$^{d}$&14,39\\
    Sz 68     &154 &K2&4710&1&5.83&0.072&3.63&1.26&8.18E-09& 48.1&13,30\\
    T Cha      &110& G8 & 5520& 0.086 &3.15& &1.94&1.65& 4E-09 & -&15\\
    T Ori    &408& A3 &9000&1.5&39& 5.94 &2.57&3.3& 1.82E-07 & -&9\\
    T Tau    &144&K0&4870&1.8&6.8&0.943&3.7&1.55&8.87E-08&28.2 &14,18\\
    TW Hya    &60&K7&4020&0&0.215&0.03&0.96&0.96&1.18E-09& 7&2,19\\
    UX Tau   &146$^a$& K2 &5520&1.8&2.95&0.253&1.88&1.61&1.16E-08& 35&16,20\\
    V836 Tau     &170&K7&4020&1.5&1.47&0.0157&2.5&0.57&2.71E-09&43.1&1,18\\
    V4046 Sgr AB  &72& K5 &4250&0&0.51&0.049&1.32&0.98&2.60E-09& 33.5&17,31\\
    VV Ser &431$^c$& A5 & 14000& 5.35 & 890& 5.88 & 5.07 & 4 & 2.47E-07 & -&4,40\\
\hline
\multicolumn{12}{l}{Empirical measurements from 1:  \citet{ingleby13}; 2:  \citet{manara14};  3:  \citet{rugel18}; 4:  \citet{alecian13};} \\
\multicolumn{12}{l}{ 5:  \citet{valenti93}; 6:  \citet{herczeg14};  7: \citet{gullbring98}; 8: \citet{herczeg08}; } \\
\multicolumn{12}{l}{9:  \citet{fairlamb15}; 10:  \citet{mendigutia14}; 11:  \citet{mendigutia11}; 12: \citet{alcala14}; }\\
\multicolumn{12}{l}{13:  \citet{alcala17}; 14:  \citet{calvet04}; 15: \citet{schisano09}; 16: \citet{espaillat10}; 17: \citet{donati11}}\\
\multicolumn{12}{l}{Disk inclination measurements from 18: \citet{long19}; 19: \citet{andrews16}; 20: \citet{andrews11};}\\
\multicolumn{12}{l}{21: \citet{fedele17}; 22: \citet{andrews18}; 23: \citet{tripathi17}; 24: \citet{avenhaus18}; 25: \citet{long18};}\\
\multicolumn{12}{l}{26: \citet{kudo18}; 27: \citet{czekala15}; 28: \citet{andrews07}; 29: \citet{pietu14}; 30: \citet{kurtovic18};}\\
\multicolumn{12}{l}{31: \citet{rosenfeld12}; 32: \citet{boehler17}; 33: \citet{kreplin13}; 34: \citet{johnskrull01};}\\
\multicolumn{12}{l}{35: \citet{lawson04}; 36: \citet{Espaillat07,Espaillat11}; 37: \citet{hussain09}; 38: \citet{grady04};}\\
\multicolumn{12}{l}{39:\citet{akeson02}; 40: \citet{salyk13};  $^d$: Less accurate inclination.}\\
\multicolumn{12}{l}{Distances from Gaia DR2 parallaxes; $^a$: weighted average of from parallaxes of binary; $^b$:  distance to cluster}\\
\multicolumn{12}{l}{from \citet{cantat18} or $^c$ from \citet{ortiz17} and \citet{herczeg19}.}\\
\end{tabular}
\end{table*}

% TABLE MODEL PARAMETERS
\begin{table}[]
    \centering
    \caption{Fiducial Model Input Parameters}
    \label{tab:model}
    \begin{tabular}{lcr}

Parmeter &  ~~~~~~~~~~~~~~~~~~~~~~~  & Value\\
\hline
         Stellar mass  && 0.5 $M_{\odot}$\\
         FUV luminosity  && 1.70$\times10^{30}$ erg s$^{-1}$\\
         X-ray luminosity  && 1.58$\times10^{30}$ erg s$^{-1}$\\
\\
         Disk mass  && 0.015 $M_{\odot}$\\
         Disk surface density  && 100 (R/AU)$^{-1}$g cm$^{-2}$\\
         Disk inner radius  && 0.1 AU\\
         Disk outer radius  && 150 AU\\
\\
         Dust to gas ratio && 0.01\\
         Dust size distribution    && $n(a)\propto a^{-3.5}$\\
         Dust minimum size  && 0.005 $\mu$m\\
         Dust maximum size  && 1 mm\\
\\
         Wind open angle && $70^{\circ}$\\
         Magnetic lever arm $R_A/R_{0}$ &&  $min\{10.0,~3.2\times[log(R/AU)+1.01]^{-0.7}\}$\\
\\
        Element abundances   &&\\
        H   && 1.0\\
        He   && 0.1\\
        C   && 1.4$\times10^{-4}$\\
        O   && 3.2$\times10^{-4}$\\
        Ne   && 1.2$\times10^{-4}$\\
        Si   && 1.7$\times10^{-6}$\\
        S   && 2.8$\times10^{-5}$\\
        Ar   && 6.3$\times10^{-6}$\\
        Fe   && 1.7$\times10^{-7}$\\
           \hline
    \end{tabular}
\end{table}

\clearpage

% FIGURE VELOCITY UNCERTAINTY
\begin{figure*}[]
\plotone{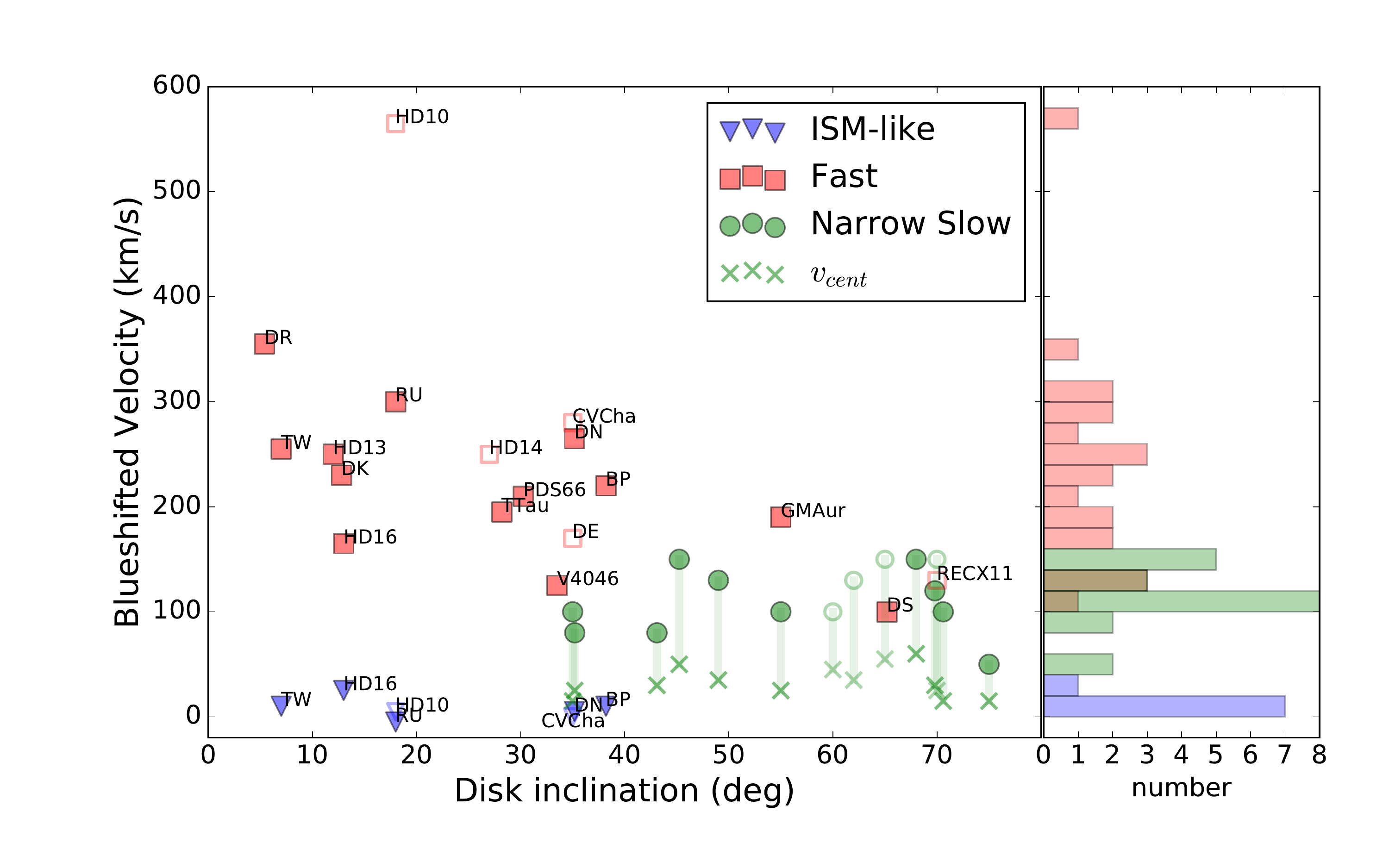}
\caption{\label{fig:velerror} Similar to Figure \ref{fig:windvel}, but with $v_{cent}$ for narrow slow winds presented by green crosses. The uncertainty of velocity measurement for narrow slow winds due to different velocity indicators of $v_{max}$ and $v_{cent}$ is $\lesssim$ 100 \kms.}
\end{figure*}
\clearpage

% FIGURE MODEL MORE
\begin{figure*}[]
\plottwo{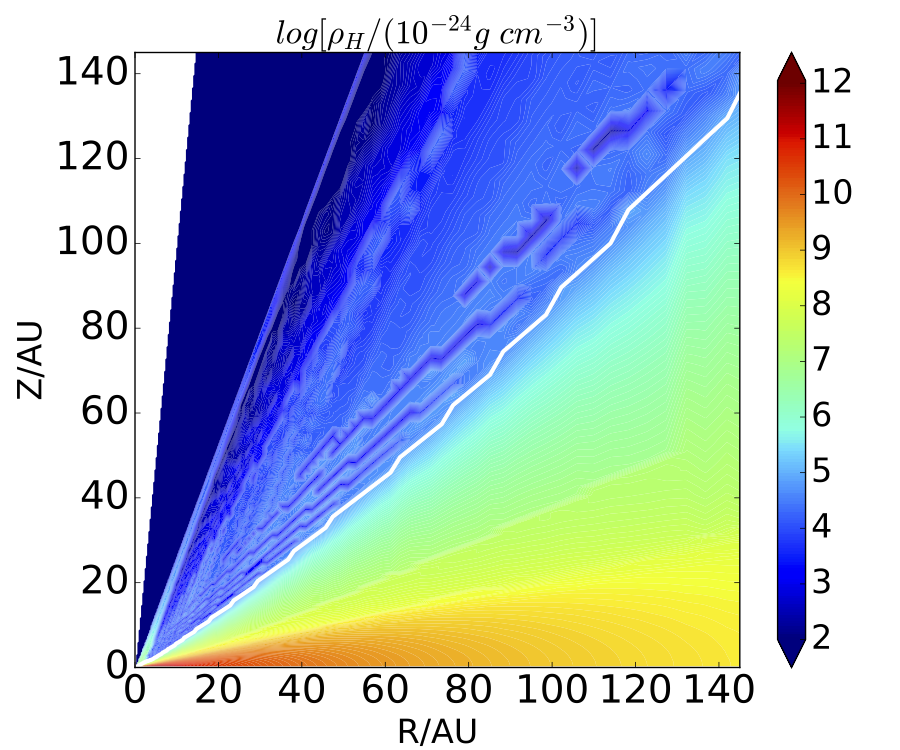}{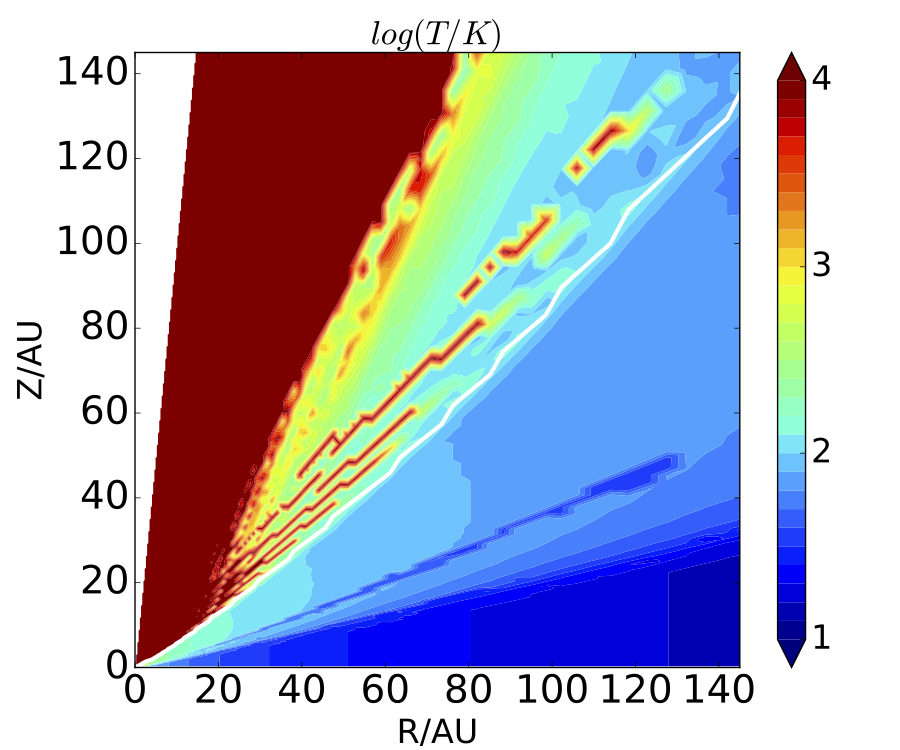}
\plottwo{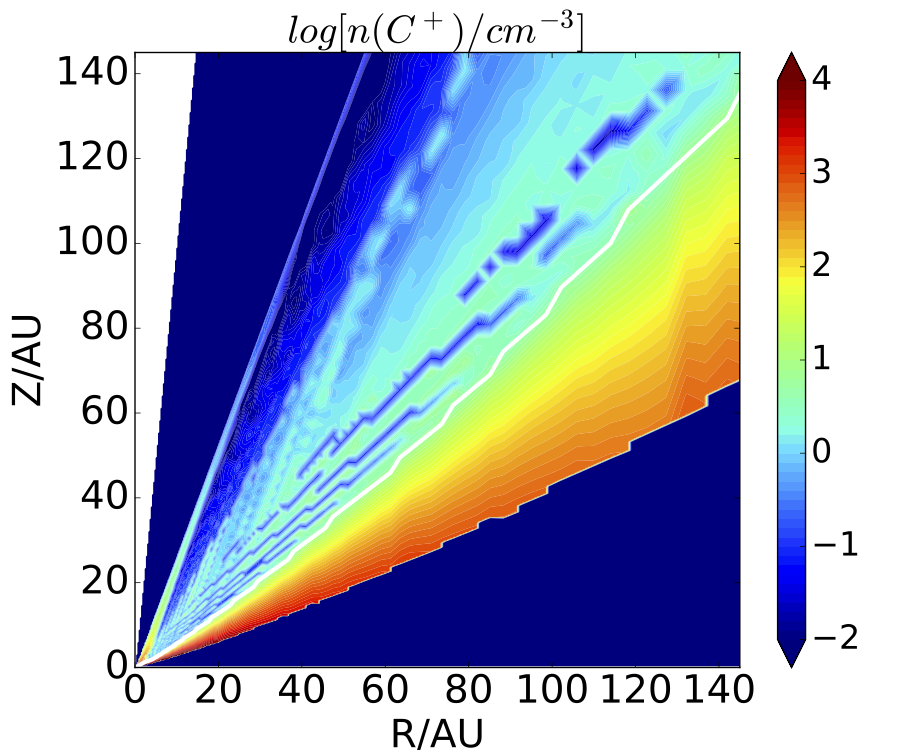}{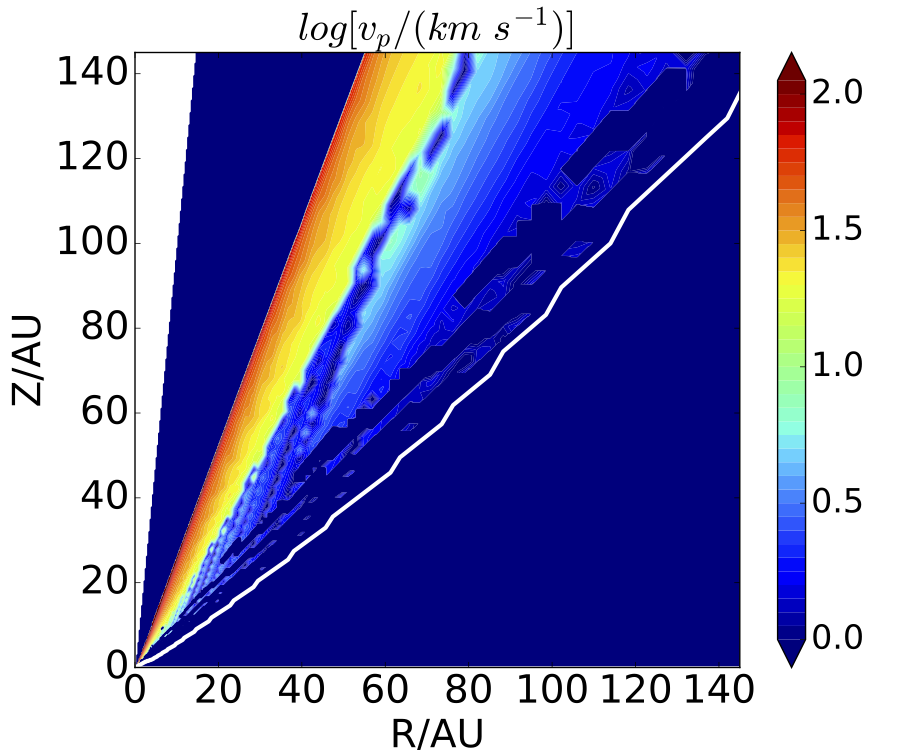}
\plottwo{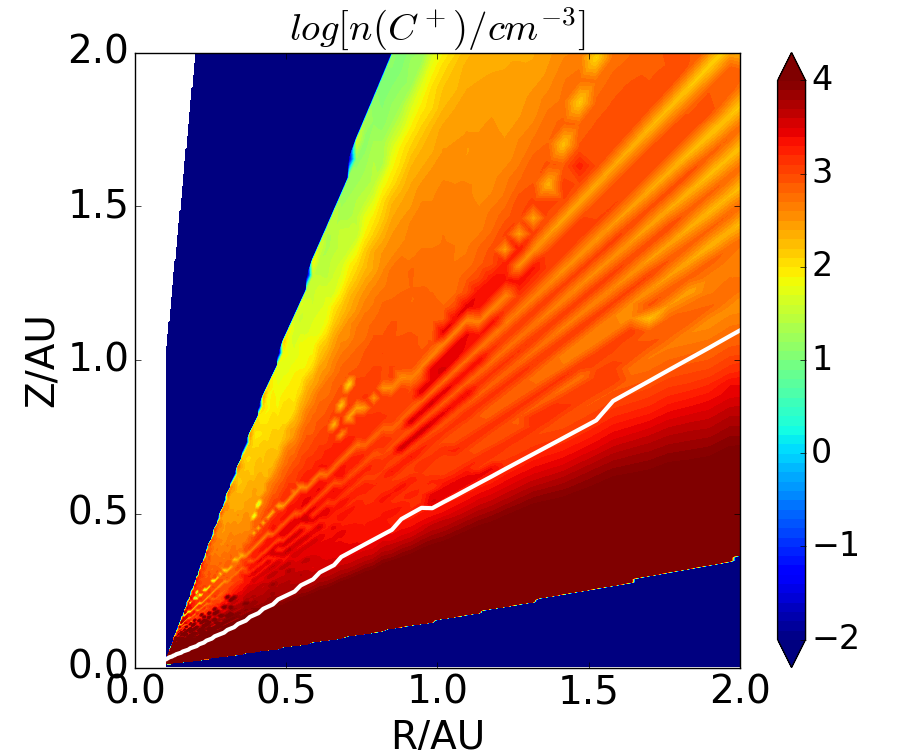}{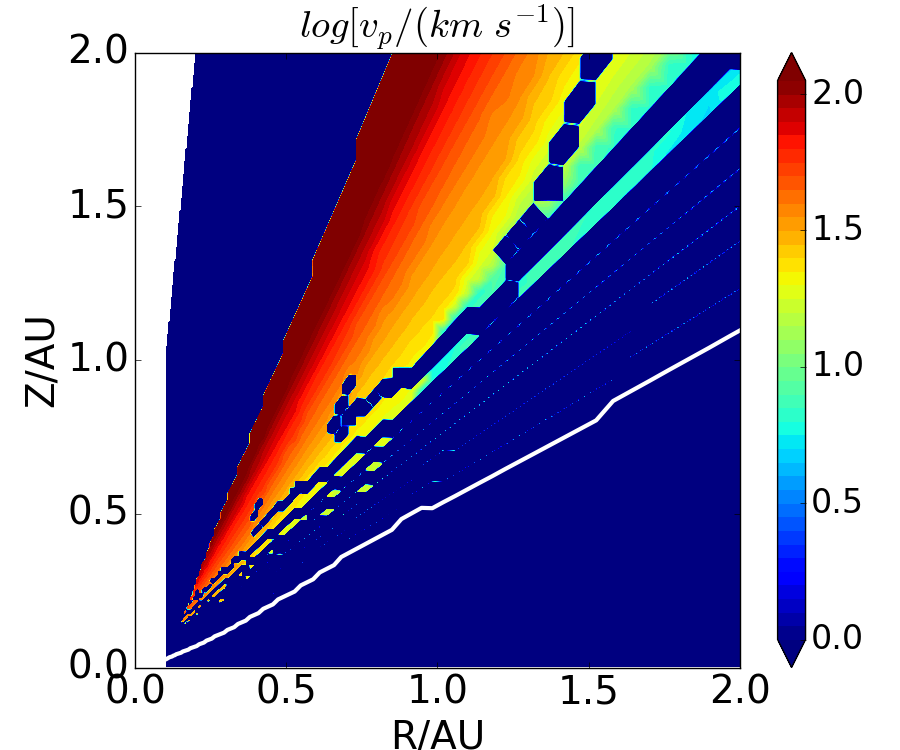}
\caption{\label{fig:fiducial} The disk wind structure predicted by the fiducial model. The flow base is marked by the white solid line, which is slightly higher than the layer of $A_V=1$, and marks the boundary between the disk atmosphere and the wind. Top row: hydrogen mass density in $10^{-24}$~g~cm$^{-1}$ and gas temperature in Kelvin. Middle row: C$^{+}$ number density in cm$^{-3}$ and flow poloidal velocity in \kms. {Bottom row: zoomed-in structure of C$^{+}$ number density (in cm$^{-3}$) and flow poloidal velocity (in \kms) within $R=2AU$.}}
\end{figure*}

% if anyone's here - thank you for going through this, dinner's on me! -ZX 20210723

\end{appendix}
\end{CJK*}
\end{document}